\documentclass[aps,pra,showpacs,twoside,10pt,floatfix,nofootinbib,longbibliography,twocolumn]{revtex4-1}
\usepackage[colorlinks=true, citecolor=red, urlcolor=blue ]{hyperref}
\usepackage{epsfig,newlfont,amssymb,amsfonts,amsmath,bm,subfigure,palatino,mathtools,amsthm,braket,soul,enumitem,color,times,comment,geometry,xcolor}
\usepackage[normalem]{ulem}
\definecolor{iitcolor}{HTML}{F7A600}
\definecolor{checkcolor}{HTML}{7541C0}
\begin{document}

\title{Simultaneous cooling of qubits via a quantum absorption refrigerator and beyond}
\author{Jithin G. Krishnan, Chandrima B. Pushpan,  Amit Kumar Pal}
\affiliation{Department of Physics, Indian Institute of Technology Palakkad, Palakkad 678 623, India}
\date{\today}

\begin{abstract}
We design a quantum thermal device that can simultaneously and dynamically cool multiple target qubits. Using a setup with three bosonic heat baths, we propose an engineering of interaction Hamiltonian using operators on different subspaces of the full Hilbert space of the system labelled by different magnetizations. We demonstrate, using the local as well as global quantum master equations, that a set of target qubits can be cooled  simultaneously using these interaction Hamiltonians, while equal cooling of all target qubits is possible only when the local quantum master equation is used. However, the amount of cooling obtained from different magnetization subspaces, as quantified by a distance-based measure of qubit-local steady-state temperatures, may vary. We also investigate cooling of a set of target qubits when the interaction Hamiltonian has different magnetization components, and when the design of the quantum thermal device involves two heat baths instead of three.  Further, we demonstrate, using local quantum master equation, that during providing cooling to the target qubits, the designed device  operates only as a quantum absorption refrigerator. In contrast, use of the global quantum master equation indicates cooling of the target qubits even when the device works outside the operation regime of a quantum absorption refrigerator. We also extend the design to a star network of qubits interacting via Heisenberg interaction among each other, kept in contact with either three, or two heat baths, and discuss cooling of a set of target qubits using this device.      
\end{abstract}

\maketitle

\section{Introduction}

Investigations of the design and implementation of  thermal devices~\cite{Bhattacharjee_2021} using quantum systems as working fluid, which include quantum refrigerators ~\cite{popescu2010,brunner2014,*skrzypczyk2011,*brask2015,levy2012,correa2014,*wang2015,*Mu_2017,*Nimmrichter2017,*nimmrichter2018,*mitchison2019,mitchison2015,das2019,Ahana2021,hewgill2020,konarghosh22,konar2023,Yan2021,*Yan2022,Konar2022,*Ghosh2024},
batteries~\cite{Allahverdyan_2004,*correa2013,*alicki2013,*Campaioli2018,*shi2022} and transistors~\cite{joulain2016,*mandarino2021}, have now become a large and integral part of quantum thermodynamics~\cite{goold2016,*vinjanampathy2016,*kosloff2013,*Myers2022} in the last decades. While the applications are still far from the envisioned control in power for quantum computing devices~\cite{ikonen2017}, these quantum thermal machines (QTMs) have contributed immensely in achieving a deeper understanding of thermodynamics in the quantum mechanical level~\cite{goold2016,*vinjanampathy2016,*Myers2022,allahverdyan2000,*brandao2015,*gardas2015}, in establishing links between quantum information theory and thermodynamic concepts~\cite{goold2016,*vinjanampathy2016,*Myers2022,*deffner2019,gour2015}, in understanding the interplay between quantum correlations in a quantum system and the work extractable from them when used as working fluids~\cite{goold2016,*vinjanampathy2016,*Myers2022,huber2015,*lostaglio2015,*hewgill2018,*francica2020,*salvia2022,shi2022}, and in investigating the possibility of beating the classical bounds of the efficiency of thermal engines~\cite{goold2016,*vinjanampathy2016,*Myers2022,geva1992,*feldmann2000,*niedenzu2018,*xu2018}. Motivated by this, several experimental realizations of these QTMs using platforms such as trapped ions~\cite{abah2012,*Johannes2016}, nuclear magnetic resonance systems~\cite{peterson2019}, superconducting~\cite{karimi2016,*hardal2017,*manikandan2019} and mesoscopic systems~\cite{giazotto2006} have also been proposed.   

Among the QTMs that have been in focus, quantum refrigerators~\cite{popescu2010, brunner2014,*skrzypczyk2011,*brask2015,levy2012,correa2014,*wang2015,*Mu_2017,*Nimmrichter2017,*nimmrichter2018,*mitchison2019,mitchison2015,das2019,Ahana2021,hewgill2020,konarghosh22,konar2023,Yan2021,*Yan2022,Konar2022,*Ghosh2024} have arguably attracted most of the attention. Theoretical research in quantum refrigerators has branched out mainly in two different directions. In one, a set of $d$-level systems are repeatedly entangled with an ancillary system by global unitary operations generated by specifically engineered Hamiltonian(s), and then disentangled by a subsequent measurement of the ancilla~\cite{Yan2021,*Yan2022,Konar2022,*Ghosh2024}. An appropriate choice of the measurement operator, and a strategic post-selection of the measurement outcomes lead to bringing all the \emph{target} $d$-level systems to their respective ground states, thereby achieving refrigeration in a measurement-based protocol.  In the second approach, a number of $d$-level quantum systems are strategically kept in contact with heat baths at different temperatures, and the full system is evolved by introducing a carefully designed interaction Hamiltonian~\cite{popescu2010,brunner2014,*skrzypczyk2011,*brask2015,levy2012,correa2014,*wang2015,*Mu_2017,*Nimmrichter2017,*nimmrichter2018,*mitchison2019,mitchison2015,das2019,Ahana2021,hewgill2020,konarghosh22,konar2023}. Upon attainment of the steady state of the system, if the population in the ground state of the $d$-level system in contact with the \emph{coldest} bath increases from its initial value, then a cooling is said to be achieved~\cite{popescu2010,das2019,Ahana2021,konarghosh22} via a dynamical process. Experimental realisations of quantum refrigerators have been achieved using nuclear magnetic resonance technique~\cite{huang2024} and trapped ions~\cite{maslennikov2019}, while proposed implementations involve the use of rf-SQUID qubits~\cite{chen2012}, quantum dots~\cite{venturelli2013}, Josephson junctions~\cite{hofer2016}, and atom-cavity systems~\cite{mitchison2016}. 

A large number of studies have explored the latter avenue of refrigeration via dynamics~\cite{popescu2010,brunner2014,*skrzypczyk2011,*brask2015,levy2012,correa2014,*wang2015,*Mu_2017,*Nimmrichter2017,*nimmrichter2018,*mitchison2019,mitchison2015,das2019,Ahana2021,hewgill2020,konarghosh22,konar2023}, and a plethora of designs of such refrigerators exist. Investigations have been carried out to create the \emph{smallest} (in terms of the dimension of the system Hilbert space) possible QTM made of only three qubits ($d=2$), or a qubit and a qutrit ($d=3$)~\cite{popescu2010,brunner2014,*skrzypczyk2011,*brask2015}, which can operate as a quantum absorption refrigerator (QAR) extracting heat from a \emph{hot} and a \emph{cold} bath, and depositing the heat into a third bath at the room temperature, cooling one qubit in the process.  The possibility of obtaining a temperature lower than the steady-state temperature in the transient regime of the dynamics~\cite{mitchison2015,das2019}, and the effect of the presence of coherence in the initial state of the system on the amount of cooling have also been explored~\cite{mitchison2015}. Further, possible designs of QTMs for cooling targeted systems such as a qudit has been explored going beyond the restriction on the size of the fridge and the requirement of the operation of the QTM as a QAR. For example, models of quantum refrigerators using two heat baths instead of three~\cite{Ahana2021}, and also using quantum spin Hamiltonian instead of engineered interaction between the $d$-level systems constructing the refrigerator~\cite{hewgill2020,konarghosh22,konar2023}. Investigations have been also extended to the robustness of these refrigerators against presence of disorder in the interactions~\cite{konarghosh22} as well as to extending the models to systems constituted of subsystems with higher Hilbert space dimensions~\cite{konar2023}.

\begin{figure*}
    \centering
    \includegraphics[width=0.9\linewidth]{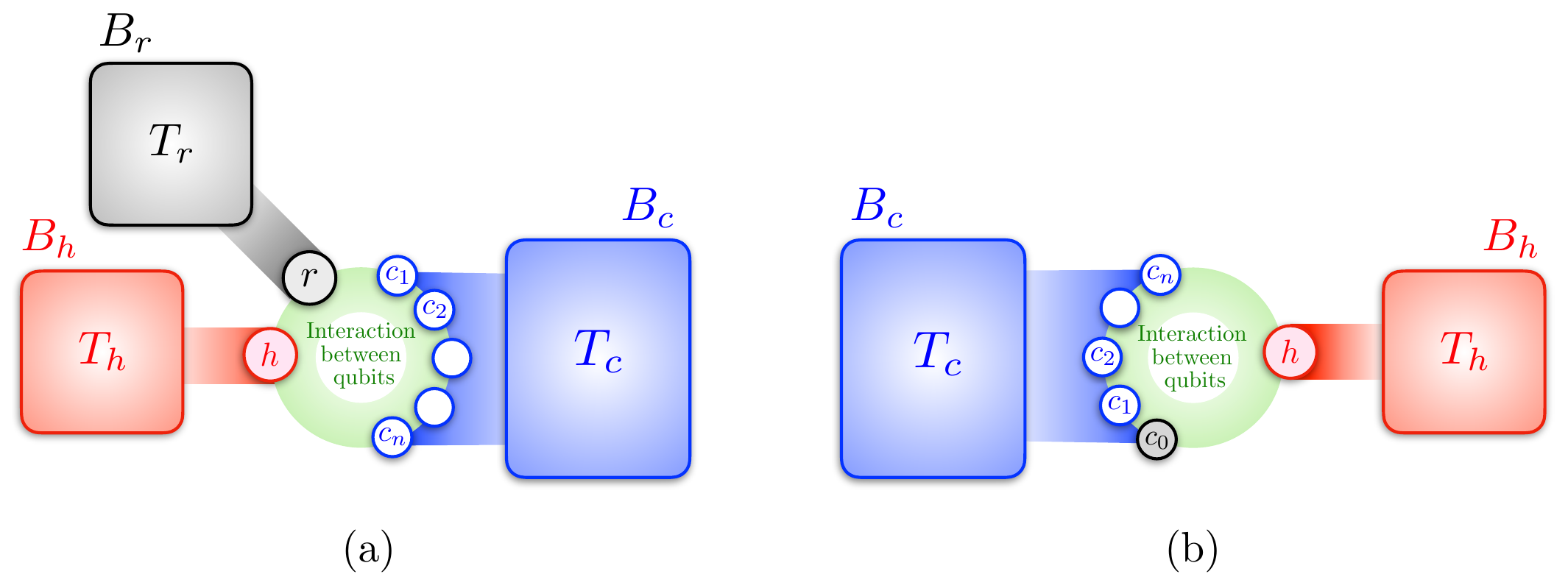}
    \caption{The system-bath setup for simultaneously \emph{cooling} $n$ qubits using (a) three (see Sec.~\ref{sec:refrigerating_multiple_qubits}), and (b) two heat baths (see Sec.~\ref{subsec:two_baths}).}
    \label{fig:schematics}
\end{figure*}

At the wake of these detailed studies on quantum refrigerators, it is natural to explore whether there exists any restriction on the number of  quantum objects to be cooled using the refrigerator of the current designs. While the measurement-based approach of refrigeration~\cite{Yan2021,*Yan2022,Konar2022,*Ghosh2024} is showed to achieve simultaneous cooling of multiple qubits~\cite{Konar2022,*Ghosh2024}, the dynamical route to this problem is hurdled with the challenges of appropriately designing the interaction Hamiltonian for the refrigerator. Investigations have been carried out to generalize the three-qubit \emph{self-contained} quantum refrigerator model~\cite{popescu2010} to one where there can be either $n$ \emph{hot} qubits, or $n$ \emph{cold} qubits, or $n$ qubits at room temperature~\cite{Ghanavati2014}, focusing on the efficiency of the QTM with the goal of approaching Carnot efficiency. Further, from the perspective of designing a quantum refrigerator using a quantum spin Hamiltonian, it has been argued that cooling a multi-qubit system is possible by preparing a refrigerant in the form of a star network of spins and  with \emph{Ising-type} interactions~\cite{Arisoy2021}. However, a complete and comprehensive study on the intricacies of achieving \emph{equal amount of simultaneous cooling} for an arbitrary number, $n$, of  qubits using the dynamical approach is yet to be made. Given a fixed number, $n$, of the \emph{target} qubits to be simultaneously and equally cooled,  our aim is to design the refrigerator with a minimal number of \emph{additional} qubits, which we refer to as a \emph{multi-qubit quantum refrigerator} (MQQR). Specifically, we ask 
\begin{enumerate}
    \item[(a)] how a MQQR can be systematically designed and operated as a QAR in order to simultaneously cool $n$ qubits by an equal amount, and
    \item[(b)] whether a MQQR can operate outside the QAR regime, and still achieve the simultaneous and equal cooling of multiple qubits. 
\end{enumerate}

In this paper, we address the above questions from two different directions. In one, we take a total of $n+2$ qubits in contact with three heat baths, where the $n$ target qubits are \emph{all} in contact with a single \emph{coldest} bath, and systematically design interactions between the qubits in order to model a MQQR working as a QAR. We demonstrate how the design of the interaction Hamiltonian required for this model can be broken down into employing different subspaces of the $(n+2)$-qubit Hilbert space labelled by different \emph{magnetizations}, which are chosen from the motivation of the workings of a QAR. Using local quantum master equation (LQME), and defining local temperature for individual qubits using a distance-based approach, we show that interaction Hamiltonian involving different magnetization sector of the Hilbert space may result in different amount of maximum cooling for the target qubits. Further, for interaction Hamiltonians designed in a single magnetization sector, we show that the QTM can be designed to provide simultaneous and equal cooling for the target qubits while operating only as a QAR . As the number of target qubits increases, within the scope of the parameter regime investigated in this paper, the maximum amount of cooling obtained from a specific magnetization sector deceases monotonically, along with the dependence of the amount of cooling on the magnetization of the transport subspace diminishing. We also consider involving all \emph{allowed} magnetization sectors in defining the interaction Hamiltonian, which provides a temperature landscape that is the resultant of the temperature landscapes corresponding to the individual magnetization sectors.

Given the current discussions on the validity of thermodynamic laws while using the LQME~\cite{Landi2022,levy2012,Wichterich2007,*Barra2015,*Strasberg2017,*Motz2017}, we also employ the global quantum master equation (GQME) to determine the dynamics and the steady state of the system, and carry out the investigation in to the design of the QTM for simultaneously cooling the target qubits. We find that the observations in the case of LQME  qualitatively remains unchanged even when GQME is employed, except the dependence of the maximum cooling on the specific values of the magnetizations corresponding to the transport subspace for a fixed size of the set of target qubits, which, in contrast to the case of the LQME,  becomes more prominent as the number of target qubits increases. Further, unlike the case of the LQME, our analysis suggests that within the scope of the parameter space considered in this paper, along with operating as a QAR, the QTM designed with the engineered interaction Hamiltonian may also operate outside the QAR regime, and the target qubits may still exhibit non-zero cooling as quantified by the distance-based definition of local temperature. To check if simultaneous and equal cooling of $n$ target qubits can be achieved even when the MQQR is not operating as a QAR, we model a QTM with two baths also, and find the observations made with the three-bath QTM with the LQME and GQME to qualitatively remain valid.

Next we consider naturally occurring spin-spin interaction, namely, the Heisenberg interaction~\cite{korepin_bogoliubov_izergin_1993,*Giamarchi2004,*Franchini2017,*fisher1964}, for designing the QTM.  We employ a star network~\cite{Richter_1994,Hutton2004,*Anza_2010,*Militello2011,*haddadi2021,*Karlova2023} among the qubits in contact with the three baths, where the qubits in contact with the hot and the room bath (the cold bath) are considered to be the \emph{center} (\emph{periphery}) of the star network, and a qubit in the center interacts only with all \emph{target} qubits in the periphery, and with no qubits in the center. The permutation symmetry in the peripheral qubits aids in the  simultaneous cooling in all target qubits. However, the challenge lies in determining the Lindblad operators corresponding to the GQME in situations where degeneracy is present in the spectrum of the system Hamiltonian. In order to avoid such situations, the energies of the target qubits are assumed to be different, such that no degeneracy appears in the spectrum of the system Hamiltonian. However, in such cases, the cooling of the thermal qubits  achieved from the QTM designed with the star network of spins are \emph{simultaneous, but not exactly equal}. Also, in contrast to the case of the QTM designed with the engineered interaction Hamiltonian and described with the GQME, the cooling amount with increasing size of the set of target qubits either increases (in the case of three-bath QTM), or remains almost constant (when the QTM has two baths). Further, in the present case, cooling of target qubits quantified in terms of the local temperature is observed \emph{only} in situations where the QTM is not operating as a QAR. 

The rest of the paper is organized as follows. In Sec.~\ref{sec:refrigerating_multiple_qubits}, we discuss the setup of the QTM for simultaneously cooling multiple qubits using three baths. Sec.~\ref{subsec:interaction_hamiltonian} presents the construction of an interaction Hamiltonian designed using different transport subspaces of the full Hilbert space with specific magnetization, while  the necessary details on the quantum master equation describing the dynamics of the system are given in Sec.~\ref{subsec:qme}. The performance of the designed QTM using the engineered interaction Hamiltonian from different magnetization sectors as well as using contributions from multiple magnetization sectors is discussed in Sec.~\ref{subsec:performance}. In Sec.~\ref{subsec:two_baths}, we present a scenario with a two-bath setup of the QTM using which simultaneous cooling of target qubits can be achieved. The extension of the design using a star network of spins with a discussion on the intricacies of the design, and detailed comments on the performance of the QTM are provided in Sec.~\ref{sec:spin_star}. Sec.~\ref{sec:conclude} contains the concluding remarks and outlook.

\section{A three-bath refrigerator for multiple qubits}
\label{sec:refrigerating_multiple_qubits}

\begin{figure*}
    \centering
    \includegraphics[width=\linewidth]{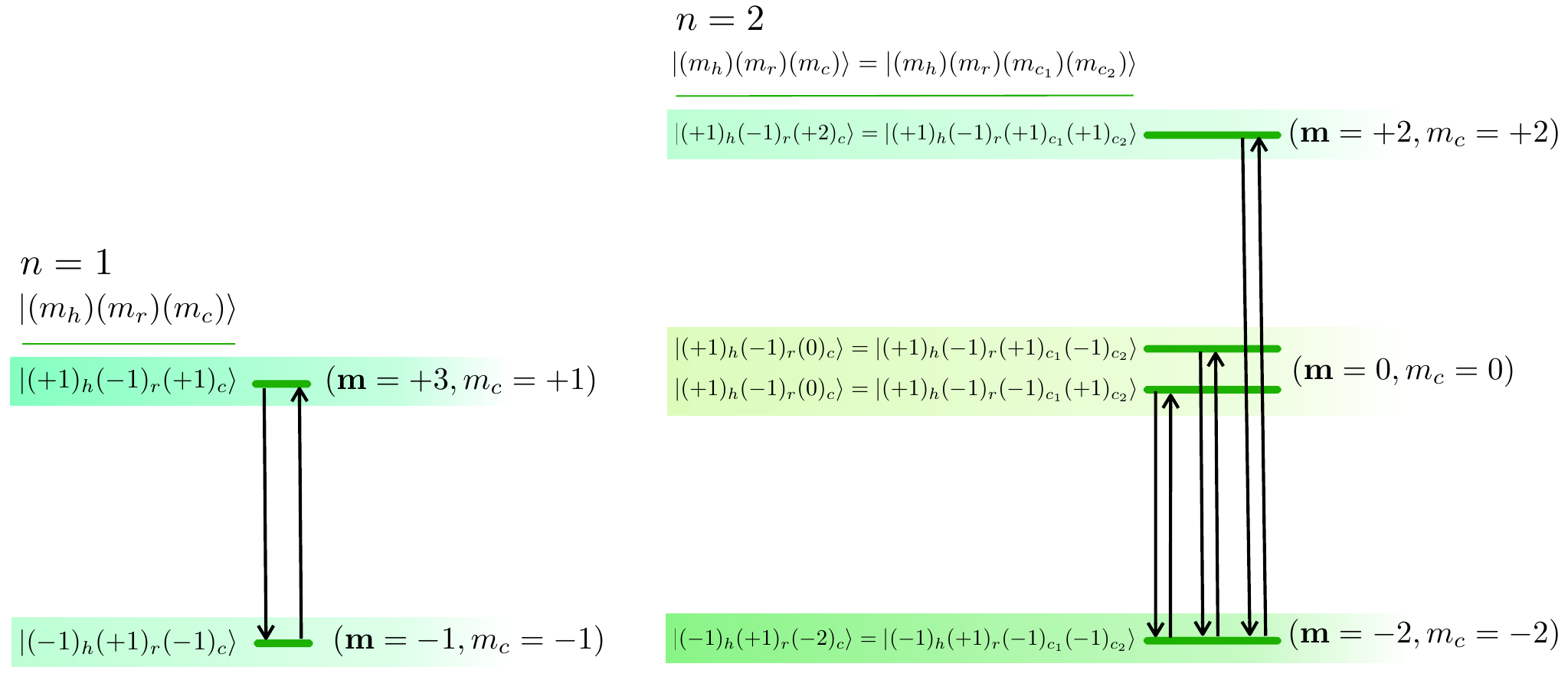}
    \caption{Illustration of different magnetization sectors of $\mathcal{H}$ corresponding to an $(n+2)$-qubit system with $n=1$ and $n=2$. The  levels $\ket{\mathbf{m}_{(m_c)}}$ and $\ket{\mathbf{m}_{(-n)}}$ are marked, using which the interaction Hamiltonians are constructed according to Eq.~(\ref{eq:interaction_fixed_magnetization}). See discussions in Sec.~\ref{subsec:interaction_hamiltonian} for details.}
    \label{fig:hilbert_space}
\end{figure*}

In this section, we model a QTM that can simultaneously lower the temperatures of  multiple qubits by an equal amount.  Let us consider a system of $n+2$ qubits, labelled as $h,r,c_1,c_2,\cdots,c_n$, described by the Hamiltonian 
\begin{eqnarray}
    H_{S}=H_{\text{loc}}+gH_{\text{int}}.
    \label{eq:system_hamiltonian}
\end{eqnarray}
Here,  
\begin{eqnarray}
    H_{\text{loc}}&=& H_h+H_r+\sum_{i=1}^n H_{c_i}
    \label{eq:local_hamiltonian}
\end{eqnarray}
represents the \emph{local} Hamiltonian describing the individual qubits, given by 
\begin{eqnarray}
    H_\alpha=\frac{E_\alpha}{2}\sigma^z_\alpha,\alpha=h,r,c_1,\cdots,c_n,
    \label{eq:single_qubit_hamiltonian}
\end{eqnarray}
where $E_\alpha$ is the energy corresponding to the qubit $\alpha$, and 
\begin{eqnarray}
    H_{\text{int}}=\sum_{j}g_jH_{\text{int}}^{(j)}
\end{eqnarray}
represents the interactions between the qubits, with interaction strengths $g_j$ corresponding to the interaction term $H_{\text{int}}^{(j)}$ generally involving all, or a subset of the qubits $\alpha=h,r,c_1,\cdots,c_n$. The index $j$ signifying different interaction terms may distinguish between $H_{\text{int}}^{(j)}$ with different physical origins, or with different characteristics. We assume control over the interactions $\{g_j\}$ to turn them on or off as per requirement, and  discuss the details on the interaction Hamiltonian $H_{\text{int}}$ in Sec.~\ref{subsec:interaction_hamiltonian}. 

We further assume each of the qubits in the $(n+2)$-qubit system to be in contact with a thermal reservoir. Specifically, we consider a  situation where qubits $h$ and $r$ are respectively in contact with a heat bath $B_h$ and $B_r$, kept respectively at an absolute temperature $T_h$, and $T_r$, while all the qubits in the set $c\equiv\{c_1,\cdots,c_n\}$ are in contact with a \emph{common} bath $B_c$ at an absolute temperature $T_c$ (see Fig.~\ref{fig:schematics}(a) for a demonstration with $n=2$). For reasons that will be clearer in Sec.~\ref{subsec:performance}, from now onward, we refer to the baths $B_h$ and $B_r$ by the names \emph{hot bath} and the \emph{room bath} respectively,  and the bath(s) $B_c$ by the \emph{cold bath}, where the temperature constraint 
\begin{eqnarray}
T_h> T_r\geq T_c 
\label{eq:temperature_ordering}
\end{eqnarray}
on the temperatures of the  baths is implicit.

At time $t=0$, we set the interaction strengths $g_j=0$ $\forall j$ (see Eq.~(\ref{eq:system_hamiltonian})). Also, we assume that each qubit is in thermal equilibrium with its respective bath, such that the state of the system is given by
\begin{eqnarray}
    \rho_S(0)=\bigotimes_{\alpha}\rho_{\alpha}(0),\alpha=h,r,c_1,\cdots,c_n,
\end{eqnarray}
with  
\begin{eqnarray}
    \rho_{\alpha}(0)=\exp -(\beta_{\alpha} H_\alpha)/\text{Tr}\left[\exp -(\beta_\alpha H_\alpha)\right],
\end{eqnarray}
where $\beta_\alpha=(k_BT_\alpha)^{-1}$ and $T_{c_i}=T_c\forall i=1,\cdots,n$, with $k_B$ being the Boltzmann constant. At $t>0$, all or a subset of the interaction terms $\{H_{\text{int}}^{(j)}\}$ are turned on (i.e., $g_j$ set to $>0$ for all, or a subset of $\{j\}$), which takes the system out of equilibrium, followed by a time-evolution of the system according to a quantum master equation (see Sec.~\ref{subsec:qme}) and eventual attainment of the steady state $\rho_S^\infty=\rho_S(t=\infty)$. We consider the performance of the machine once the steady state has been attained by the system.

Our aim is to design the above machine to work as a refrigerator in the steady state for the set of qubits in $c$. We consider the qubits in $c$ to be \emph{simultaneously cooled} if the local density matrices 
\begin{eqnarray}
\rho_{\alpha}^\infty=\text{Tr}_{\underset{\alpha^\prime\neq \alpha}{\{\alpha^\prime\}}}\rho_S^\infty   
\end{eqnarray}
for each of  the qubits $\alpha=c_1,c_2,\cdots,c_n$ corresponds to a \emph{local} temperature $\tau_\alpha^\infty$ such that
\begin{eqnarray}
    \tau_{c_i}^\infty\leq T_c\;\forall i=1,2,\cdots,n, 
\end{eqnarray}
while the amount of cooling for the qubit $c_i$ given by 
\begin{eqnarray}
    \Delta_{c_i}^\infty= T_{c_i}-\tau_{c_i}^\infty.
    \label{eq:cooling_amount}
\end{eqnarray}
The above definition of cooling warrants for an appropriate quantification of the qubit-local temperature. We discuss specific quantification in Sec.~\ref{subsec:performance}.

\subsection{Engineering the interaction Hamiltonian}
\label{subsec:interaction_hamiltonian}

Let us denote the Hilbert space associated to the $(n+2)$-qubit system by $\mathcal{H}=\mathcal{H}_h\otimes\mathcal{H}_r\otimes\mathcal{H}_{c}$, where $\mathcal{H}_h$ and $\mathcal{H}_r$ are two-dimensional  Hilbert spaces corresponding respectively to the qubits $h$ and $r$, and $\mathcal{H}_c=\bigotimes_{i=1}^n\mathcal{H}_{c_i}$ is the joint Hilbert space for the $n$-qubit subsystem $c\equiv\{c_1,c_2,\cdots,c_n\}$,  $\mathcal{H}_{c_i}$ being the individual two-dimensional qubit Hilbert spaces. We find it convenient to work with the eigenbasis of the magnetization operators $M^z_\alpha$, $\alpha=h,r,c$, as the chosen basis for the Hilbert spaces $\mathcal{H}_\alpha$, with
\begin{eqnarray}
    M^z_{h}=\sigma^z_h,M^z_r=\sigma^z_r,M^z_c=\sum_{i=1}^nM^z_{c_i}=\sum_{i=1}^n\sigma^z_{c_i},
\end{eqnarray}
where $M^z_{c_i}=\sigma^z_{c_i}$, such that the basis for individual qubits is the computational basis:
\begin{eqnarray}
    M^z_{\alpha}\ket{m_{\alpha}}=m_\alpha\ket{m_\alpha},\alpha=h,r,c_1,c_2,\cdots,c_n,
\end{eqnarray}
with $m_{\alpha}=\pm 1$, and 
\begin{eqnarray}
    M^z_{c}\ket{m_c}=m_c\ket{m_c}, 
    \label{eq:computational_basis}
\end{eqnarray}
where $\ket{m_c}=\bigotimes_{i=1}^n\ket{m_{c_i}}$, and $m_c=\sum_{i=1}^nm_{c_i}$, which can take $(n+1)$ values given by $m_c=-n,-n+2,\cdots,n-2,n$. One may further define a \emph{total} magnetization $\mathbf{M}^z=M^z_h+M^z_r+M^z_c$ of the $(n+2)$ qubit system, with its eigenbasis $\{\ket{\mathbf{m}}\}$, given by 
\begin{eqnarray}
    \mathbf{M}^z\ket{\mathbf{m}}=\mathbf{m}\ket{\mathbf{m}},
\end{eqnarray}
representing the basis of the $(n+2)$-qubit system, where $\ket{\mathbf{m}}=\ket{m_h}\ket{m_r}\ket{m_c}$, and $\mathbf{m}=m_h+m_r+m_c$. Note that $\mathbf{m}$ can take $n+3$ values, given by $\mathbf{m}=-(n+2),-(n+2)+2,\cdots,n,n+2$. See Fig.~\ref{fig:hilbert_space} for an illustration in the cases of $n=1,2$.

For reasons that will be clear shortly, we look closely into the subsystem $c\equiv\{c_1,c_2,\cdots,c_n\}$, represented by the basis $\{\ket{m_c}\}$. Assuming $n_c^0$ of the $n$ qubits in $\ket{m_c}$ to be in the \emph{excited} state $\ket{m_{c_i}=+1}$ (see Eq.~(\ref{eq:computational_basis})) and the rest ($n_c^1=n-n_c^0$) to be in the \emph{ground} state $\ket{m_{c_i}}=-1$, $m_c$ is expressed as $m_c=2n_c^0-n$, where each value of $m_c$ has a multiplicity $N_{(m_c)}=\genfrac(){0pt}{1}{n}{n_c^0}$. For a fixed value of $m_c$, we denote the $N_{(m_c)}$ basis states by introducing an index $k$ as $\ket{m_c^{(k)}}$, where $k=1,2,\cdots,N_{(m_c)}$, such that a state $\ket{\psi_{(m_c)}}$ of the form 
\begin{eqnarray}
    \ket{\psi_{(m_c)}}=N_{(m_c)}^{-1/2}\sum_{k=1}^{N_{(m_c)}}\ket{m_c^{(k)}},
\end{eqnarray}
which we identify as the permutationally invariant Dicke state~\cite{Dicke1954}, satisfies 
\begin{eqnarray}
    M^z_c\ket{\psi_{(m_c)}}=m_c\ket{\psi_{(m_c)}}. 
\end{eqnarray}
Note that the sector with $m_c=-n$ corresponds to each of qubits $c_i$ being in the \emph{ground} state $\ket{m_{c_i}=-1}$, having a multiplicity $N_{(-n)}=1$, such that $\ket{\psi_{(m_c=-n)}}=\ket{m_c=-n}$.

We systematically design the interaction Hamiltonian $H_{\text{int}}$ by assuming that the different terms $H_{\text{int}}^{(j)}$ labelled by $j$ corresponds to different magnetization sectors of the Hilbert space $\mathcal{H}$ of the $(n+2)$-qubit system, such that $j$ only takes the allowed values of $\mathbf{m}$. To see which of the values of $\mathbf{m}$ can be chosen, note that we would like the QTM to operate as a QAR in the steady state, which is determined by the steady-state heat currents (SSHCs) for the qubits $h$, $r$, and the set of qubits $c\equiv\{c_1,c_2,\cdots,c_n\}$,  defined as the amount of heat flowing from/to the qubit(s) to/from the bath $B_\alpha$ per unit time, when the $(n+2)$-qubit system has attained the state $\rho_S^\infty$. An extraction (deposition) of heat from (to) the bath corresponds to a positive (negative) SSHC. The QTM operates as a QAR in the steady state if  it extracts heat from $B_h$ and $B_{c}$, and deposits heat  to $B_r$, cooling (heating) the set of qubits in $c$ \emph{overall}, and the qubit $h$ (the qubit $r$).  This implies  all qubits $h,c_1,c_2,\cdots,c_n$ \emph{simultaneously} going to the ground state by releasing heat,  while the qubit $r$ going to the excited state by absorbing it. Therefore, one may consider transitions between energy levels having the form $\ket{m_h=+1}\ket{m_r=-1}\ket{m_c\neq -n}$ and $\ket{m_h=-1}\ket{m_r=+1}\ket{m_c=-n}$, which may facilitate such heat flow. Therefore, we consider $\mathbf{m}=m_c$, $m_c=-n,-n+2,-n+4,\cdots,n-2,n$, and write the form of the interaction Hamiltonian corresponding to a specific magnetization $m_c\neq -n$  sector as  
\begin{eqnarray}
    H_{\text{int}}^{(m_c)} &=&g_{(m_c)}\left[\ket{\mathbf{m}_{(m_c)}}\bra{\mathbf{m}_{(-n)}}+\text{h. c.}\right],\label{eq:interaction_fixed_magnetization}
\end{eqnarray}
with an interaction strength $g_{(m_c)}$, where 
\begin{eqnarray}
    \ket{\mathbf{m}_{(m_c)}}&=&N_{(m_c)}^{1/2}\ket{m_h=+1}\ket{m_r=-1}\ket{\psi_{(m_c)}},\nonumber\\
    \ket{\mathbf{m}_{(-n)}}&=&\ket{m_h=-1}\ket{m_r=+1}\ket{m_c=-n},
\end{eqnarray}
defines the \emph{transport subspace},  and we have identified $j$ as $m_c$ (see Fig.~\ref{fig:hilbert_space} for an illustration with $n=1$ and $n=2$). Combining $H_{\text{int}}^{m_c}$ corresponding to all magnetization sectors $m_c=-n+2,-n+4,\cdots,n-2,n$, the full interaction Hamiltonian is given by 
\begin{eqnarray}
    H_{\text{int}} &=& \sum_{m_c=-n+2,\cdots}^{n}g_{(m_c)}H_{\text{int}}^{(m_c)}.
    \label{eq:full_interaction_Hamiltonian}
\end{eqnarray}

\begin{figure*}
    \includegraphics[width=0.8\linewidth]{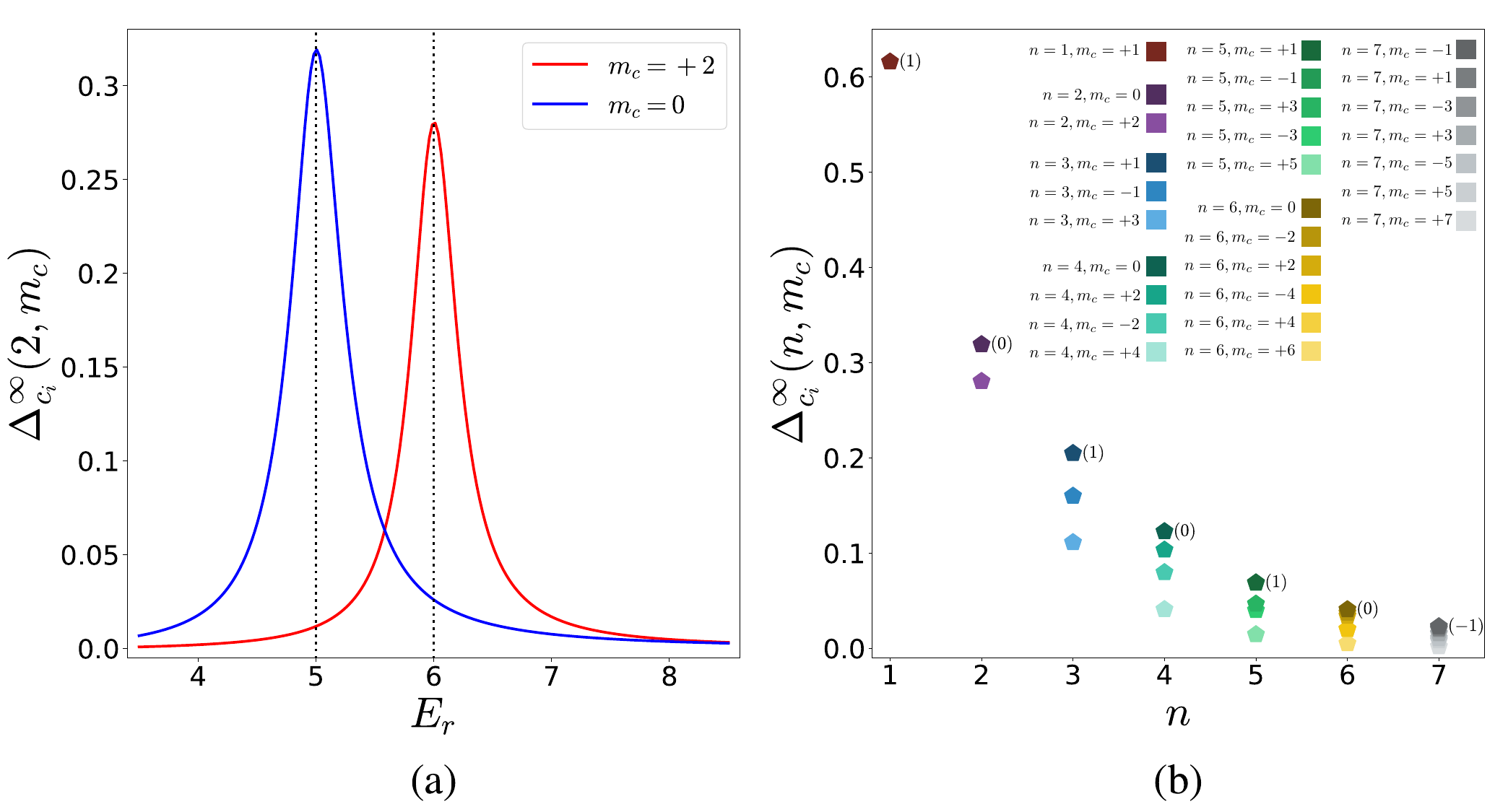}
    \caption{(a) Variations of $\Delta_{c_i}^\infty(2,m_c)$, $i=1,2$, as functions of $E_r$ when interaction terms correspond to $m_c=0,+2$. The relevant parameters are fixed at $E_h=4$, $E_c=1$, $T_h=10$, $T_r=2$, $T_c=2$, $f_{h,r,c}=0.01$, $\Omega_{h,r,c}=10^3$, and $g_{(m_c)}=0.05$ for both $m_c=0,+2$. The maximum in each graph occurs at the value of  $E_r$ for which Eq.~(\ref{eq:self_contained}) is satisfied. (b) Variations of $\Delta_{c_i}^\infty(n,m_c)$ with increasing $n$ for all possible values of $m_c$ corresponding to each value of $n$. The numbers in brackets indicate the magnetization $m_c$ corresponding to which the maximum value of $\Delta_{c_i}^\infty(n,m_c)$ among the allowed values of $m_c$ takes place for a specific $n$. All parameters are same as in (a), while $E_r$ takes a value such that Eq.~(\ref{eq:self_contained}) is obeyed for all points in (b).  All quantities plotted in both (a) and (b) are dimensionless.}
    \label{fig:self_contained_temp_local}
\end{figure*}

We point out the following regarding the interaction Hamiltonian. 
\begin{enumerate}
    \item[(a)] Although $H_{\text{int}}$ is designed with the motivation of constituting a QAR, satisfying (\ref{eq:qar_condition}) at all points in the space of the system parameters given by $\{E_{\alpha},T_{\alpha}\}$, $\alpha=h,r,c_1,\cdots,c_n$, and the system-bath interaction parameters $\{f_\alpha\}$ is not guaranteed. We elaborate on this in Sec.~\ref{subsec:performance}. 
    \item[(b)] In the case of $n=1$, Eq.~(\ref{eq:interaction_fixed_magnetization}) corresponds to the interaction Hamiltonian considered in~\cite{popescu2010,brunner2014,*skrzypczyk2011,*brask2015,levy2012,correa2014,*wang2015,*Mu_2017,*Nimmrichter2017,*nimmrichter2018,*mitchison2019,mitchison2015,das2019,Ahana2021}. See also Fig.~\ref{fig:hilbert_space}. 
\end{enumerate}

\subsection{Quantum master equation}
\label{subsec:qme}

In order to determine the time-evolution of the $(n+2)$-qubit system,
in this paper, we assume the baths to be Bosonic in nature, consisting of infinite set of harmonic oscillators each. The bath Hamiltonian is given by 
\begin{eqnarray}
    H_B=\sum_\alpha H_{B_\alpha},
    \label{eq:bath_hamiltonian}
\end{eqnarray}
with
\begin{eqnarray}
H_{B_\alpha} &=& \sum_{k} \omega^k_\alpha d_{(\alpha,k)}^{\dagger}d_{(\alpha,k)}.
\label{eq:bosonic_bath_hamiltonian}
\end{eqnarray}
Here, $\omega^k_\alpha$ denotes the frequency of the $k$'th mode of the bath $B_{\alpha}$, $d^{\dagger}_{(\alpha,k)}$ and $d_{(\alpha,k)}$ respectively are the creation and annihilation operators for mode $k$ corresponding to the bath $B_\alpha$, $\alpha=h,r,c$. The interaction between the qubits and baths are modelled by the system-bath Hamiltonian $H_{SB}$, given by
\begin{eqnarray}
    H_{SB}&=&H_{SB}^h+H_{SB}^r+H_{SB}^c
    \label{eq:system_bath_interaction}
\end{eqnarray}
with \small 
\begin{eqnarray}
\label{eq:system_bath_interaction_ab}
    H_{SB}^\alpha &=& \sqrt{f_\alpha} \sum_{k}\left[\sigma^{+}_{\alpha} d_{(\alpha,k)}+\sigma_{\alpha}^-d^{\dagger}_{(\alpha,k)}\right],\alpha=h,r,\\
\label{eq:system_bath_interaction_c}
    H_{\text{SB}}^c&=& \sqrt{f_{c}}\sum_k\Bigg[\sum_{i=1}^{n} \left\{\sigma^{+}_{c_i} d_{(c,k)}+\sigma_{c_i}^-d^{\dagger}_{(c,k)}\right\}\nonumber\\&&+\kappa\sum_{\underset{i< j}{\{(c_i,c_j)\}}}\left\{\sigma_{c_i}^+\sigma_{c_{j}}^- d_{(c,k)}+\sigma_{c_i}^-\sigma_{c_{j}}^+ d_{(c,k)}^{\dagger}\right\}\Bigg],
\end{eqnarray}\normalsize 
where $f_h$, $f_r$, and $f_c$  are real parameters determining the strength of the system-bath interaction, and $\kappa$ is the relative strength of the three-body interaction term w.r.t. the two-body interaction term in the case of the common bath $B_c$ attached to all qubits $c_i,i=1,2,\cdots,n$ (cf.~\cite{Ahana2021}). 

Once $H_{\text{int}}$ is turned on at $t>0$, the system-bath duo evolves due to the total Hamiltonian $H=H_S+H_B+H_{SB}$. Assuming Markovian system-bath interactions~\cite{breuer2002,Rivas2010}, the evolution of the system alone is given by the QME (see Appendix~\ref{app:qme} for a derivation) 
\begin{eqnarray}
\frac{d\rho_S(t)}{d t}&=& -\text{i} \left[H_S,\rho_S(t)\right]+ \sum_{\alpha=h,r,c}\mathcal{D}_\alpha(\rho_S(t)),
\label{eq:qme}
\end{eqnarray}
where the second term on the r.h.s. is the dissipative term. For $\alpha=h,r$, $\mathcal{D}(.)$ is given by  
\begin{eqnarray}
    \mathcal{D}_\alpha(\varrho)&=&\sum_{\mathcal{E}>0}\sum_{\ell=1}^2\eta_\alpha^\ell(\mathcal{E})\mathcal{L}_{(\alpha,\ell)}^{\mathcal{E}}(\varrho),
    \label{eq:dissipative_term_main}
\end{eqnarray}
with 
\begin{eqnarray}
    \mathcal{L}_{(\alpha,\ell)}^{\mathcal{E}}(\varrho)&=&\delta_{\ell,1}\left[\mathcal{A}_\alpha^{\dagger}(\mathcal{E})\varrho \mathcal{A}_\alpha(\mathcal{E}) - \frac{1}{2}\left\{\mathcal{A}_\alpha(\mathcal{E}) \mathcal{A}_\alpha^{\dagger}(\mathcal{E}),\varrho \right\}\right]\nonumber\\
    &+&\delta_{\ell,2}\left[\mathcal{A}_\alpha(\mathcal{E})\varrho\mathcal{A}_\alpha^{\dagger}(\mathcal{E}) - \frac{1}{2}\left\{\mathcal{A}_\alpha^{\dagger}(\mathcal{E})\mathcal{A}_\alpha(\mathcal{E}),\varrho \right\}\right],\nonumber\\ 
    \label{eq:lindblad_form}
\end{eqnarray} 
where $\{\mathcal{A}_\alpha(\mathcal{E})\}$ are the Lindblad operators, $\varrho$ is a general density matrix of the $(n+2)$-qubit system, and $\delta_{\ell,\ell^\prime}=1(0)$ for $\ell=\ell^\prime$ ($\ell\neq\ell^\prime$). In terms of the spectrum of the system Hamiltonian $H_S$, where we write
\begin{eqnarray}
H_{S}=\sum_{E}EP(E)=\sum_E E\ket{E}\bra{E},
\end{eqnarray}
the Lindblad operators are given by
\begin{eqnarray}
\mathcal{A}_\alpha(\mathcal{E}) &=&\sum_{\underset{\mathcal{E}=E-E^\prime}{E,E^\prime}} P (E) \left[\sigma_\alpha^{+}+\sigma_\alpha^{-}\right] P(E^\prime).
\label{eq:lindblad_operators_ab}
\end{eqnarray}
The transition rates $\eta_\alpha^\ell(\mathcal{E})$, $\ell=1,2$, corresponding to a jump of an energy $\mathcal{E}$, are given by 
\begin{eqnarray}
    \eta_\alpha^1(\mathcal{E}) &=& \frac{2f_\alpha \mathcal{E} \exp\left(-\mathcal{E}/\Omega_\alpha\right)\exp\left(\mathcal{E}/T_\alpha\right)}{\exp\left(\mathcal{E}/T_\alpha\right)-1},\nonumber\\
    \eta_\alpha^2(\mathcal{E}) &=&\frac{2f_\alpha \mathcal{E} \exp\left(-\mathcal{E}/\Omega_\alpha\right)}{\exp\left(\mathcal{E}/T_\alpha\right)-1},
    \label{eq:transition_probabilities}
\end{eqnarray}
with $\Omega_\alpha$ being the cut-off frequency corresponding to the bath-type $\alpha$, where we have assumed an Ohmic spectral function for all the baths (see Appendix~\ref{app:qme}). 

On the other hand,   $\mathcal{D}_c(\varrho)$ is given by 
\begin{eqnarray}
    \mathcal{D}_c(\varrho)&=&\sum_{\mathcal{E}>0}\sum_{\ell=1}^2\eta_\alpha^\ell(\mathcal{E})\left[\mathcal{L}_{(\alpha,\ell)}^{\mathcal{E}}(\varrho)+\kappa^2\mathcal{K}_{(\alpha,\ell)}^{\mathcal{E}}(\varrho)\right],
\label{eq:dissipative_term_c}
\end{eqnarray}
where $\eta_{\alpha}^\ell(\mathcal{E})$ are as given in Eq.~(\ref{eq:transition_probabilities}), both $\mathcal{L}_{\alpha,\ell}^{\mathcal{E}}(.)$ and $\mathcal{K}_{\alpha,\ell}^{\mathcal{E}}(.)$ have the form as given in Eq.~(\ref{eq:lindblad_form}),  and the operators $\mathcal{A}_\alpha(\mathcal{E})$ corresponding to $\mathcal{L}_\alpha^{\mathcal{E}}(.)$ and $\mathcal{K}_\alpha^{\mathcal{E}}(.)$ are respectively given by 
\begin{eqnarray}
\label{eq:lindblad_operators_c_two_body}
\mathcal{A}_c(\mathcal{E}) &=& \sum_{\underset{\mathcal{E}=E-E^\prime}{E,E^\prime}} P (E) \left[\sum_{i=1}^{n}(\sigma_{c_i}^{+}+\sigma_{c_i}^{-})\right] P(E^\prime),
\end{eqnarray}
and \small 
\begin{eqnarray}
\label{eq:lindblad_operators_c_three_body}
\mathcal{A}_c(\mathcal{E})&=&\sum_{\underset{\mathcal{E}=E-E^\prime}{E,E^\prime}} P (E) \left[\sum_{\underset{i< j}{\{(c_i,c_j)\}}}(\sigma_{c_i}^{+}\sigma_{c_{j}}^{-}+\sigma_{c_i}^{-}\sigma_{c_{j}}^{+})\right] P(E^\prime).\nonumber\\
\end{eqnarray} \normalsize

\begin{figure*}
    \includegraphics[width=\linewidth]{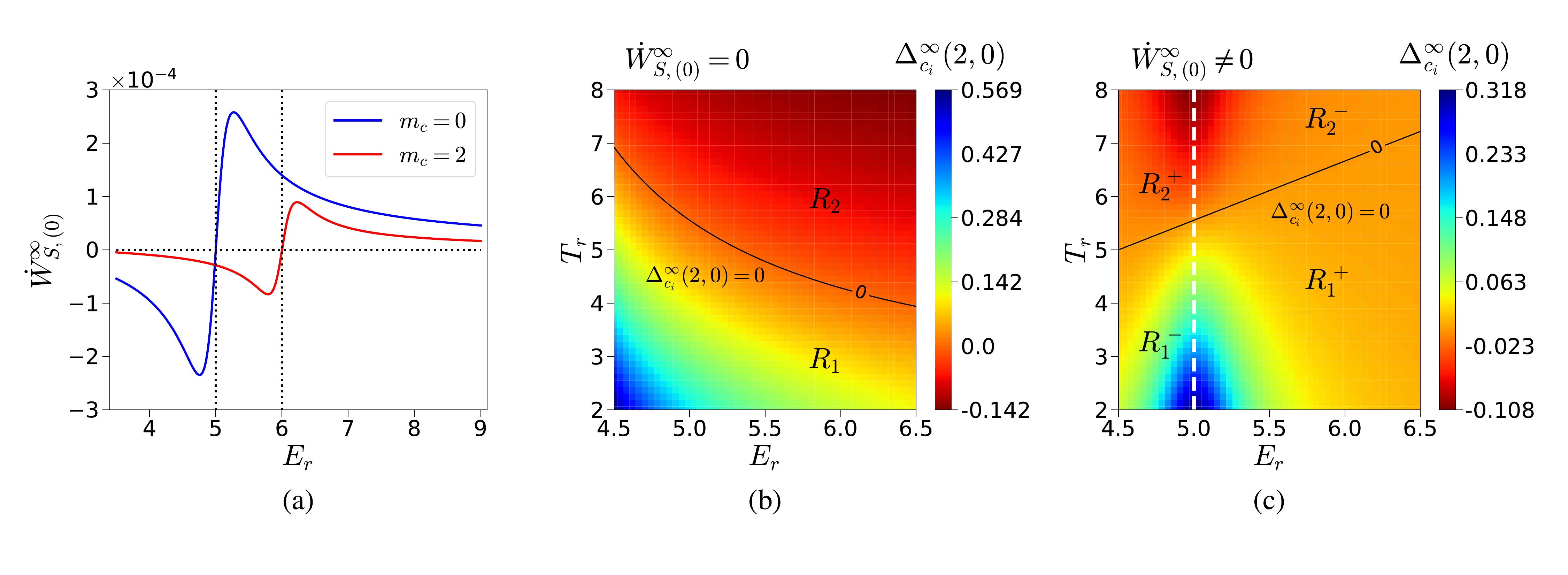}
    \caption{(a) Variations of $\dot{W}_S^\infty$ (Eq.~(\ref{eq:work_imbalence})) as a function of $E_r$ when $n=2$, and the interaction terms correspond to $m_c=0,+2$. The relevant parameters are fixed at $E_h=4$, $E_c=1$, $T_h=10$, $T_r=2$, $T_c=2$, $f_{h,r,c}=0.01$, $\Omega_{h,r,c}=10^3$, and $g_{(m_c)}=0.05$ for both $m_c=0,+2$. The graph for $\dot{W}_S^\infty$ crosses zero at the value of $E_r$ for which Eq.~(\ref{eq:self_contained}) is satisfied. (b) Variations of $\Delta_{c_i}^\infty(2,0)$ as a function of $E_r$ and $T_r$, where $E_h=4$, and the value of $E_c$ is fixed such that Eq.~(\ref{eq:self_contained}) is satisfied on each point of the $E_r-T_r$ plane. The rest of the parameters are fixed as in (a). The contour corresponds to $\Delta_{c_i}^\infty(2,0)=0$, marking the boundary between the region where cooling occurs with $\dot{Q}_{\text{loc},h}^\infty>0$, $\dot{Q}_{\text{loc},r}^\infty<0$, $\dot{Q}_{\text{loc},c}^\infty>0$, and the region where qubits in $c$ get heated, with $\dot{Q}_{\text{loc},h}^\infty<0$, $\dot{Q}_{\text{loc},r}^\infty>0$, $\dot{Q}_{\text{loc},c}^\infty<0$.  All quantities plotted in both (a) and (b) are dimensionless.}
    \label{fig:self_contained_q_local}
\end{figure*}

\subsection{Operation regimes  and performance}
\label{subsec:performance}

We now set the premises for investigating the performance of the QTM, and the underlying thermodynamics.  Note from Eq.~(\ref{eq:qme}) that~\cite{Landi2022,Wichterich2007,*Barra2015,*Strasberg2017,*Motz2017,potts2019}
\begin{eqnarray}
    \frac{d U_S(t)}{dt} &=&\sum_{\alpha=h,r,c}\dot{Q}_\alpha(t), 
\end{eqnarray}
with 
\begin{eqnarray}
    \dot{Q}_{\alpha}(t) &=& \text{Tr}\left[\mathcal{D}_\alpha\left(\rho_S(t)\right)H_S\right], \alpha=h,r,c,
    \label{eq:global_heat_current}
\end{eqnarray}
where one identifies $U_S(t)=\text{Tr}\left[\rho_S(t)H_S\right]$ as the \emph{internal energy}, and $\dot{Q}_S(t)=\sum_{\alpha=h,r,c}\dot{Q}_\alpha(t)$ as the total rate of heat flow (total \emph{heat current}) for the three baths from the \emph{first law of thermodynamics}  $\dot{U}_S(t)=\dot{Q}_S(t)+\dot{W}_S(t)$, where the rate of work (power) $\dot{W}_S(t)=\text{Tr}\left[\rho_S(t)\left(dH_S/dt\right)\right]$ vanishes for a time-independent system Hamiltonian $H_S$. In the steady state, $\dot{U}_S(t=\infty)=0$, leading the following version of the first law:
\begin{eqnarray}
    \label{eq:first_law}
    \sum_{\alpha=h,r,c}\dot{Q}_\alpha^\infty=0, 
\end{eqnarray}
where we have written the SSHC (see discussions preceeding Eq.~(\ref{eq:interaction_fixed_magnetization})) $\dot{Q}_\alpha(t=\infty)=Q_\alpha^\infty$. Further, defining entropy as $S(t)=-\text{Tr}\left[\rho_S(t)\ln\rho_S(t)\right]$~\cite{Landi2022}, the \emph{sum of entropy production due to all processes} is given by $\dot{S}(t)-\sum_{\alpha=h,r,c}\beta_\alpha\dot{Q}_\alpha(t)$, where 
\begin{eqnarray}
    \dot{S}(t)=-\text{Tr}\left[\sum_{\alpha=h,r,c} \mathcal{D}_{\alpha}\left(\rho_S(t)\right) \ln \left(\rho_S(t)\right)\right]. 
\end{eqnarray}
The \emph{second law of thermodynamics} signifies the positivity of entropy production in the steady state, given by ~\cite{Landi2022} 
\begin{eqnarray}
    \dot{S}^\infty-\sum_{\alpha=h,r,c}\beta_\alpha\dot{Q}_\alpha^\infty \geq 0,
    \label{eq:second_law}
\end{eqnarray}
where we write $\dot{S}(t=\infty)=\dot{S}^\infty$. 

In the formulation of the QME and the laws of thermodynamics, no restrictions on the possible values of $\{g_{(m_c)}\}$ are imposed, and the Lindblad operators corresponding to the QME are worked out using the spectrum of the full system Hamiltonian $H_S=H_{\text{loc}}+H_{\text{int}}$ (Eq.~(\ref{eq:system_hamiltonian})). Such QMEs are referred to as the \emph{global} QMEs (GQMEs). In the designed QTM described by the GQME, we expect Eqs.~(\ref{eq:first_law}) and (\ref{eq:second_law}) to be satisfied at all chosen points in the space of the system and the system-bath interaction parameters.

\subsubsection{Transport subspace with a single magnetization}
\label{subsubsec:single_magnetization}

Let  the interaction Hamiltonian have the form (\ref{eq:interaction_fixed_magnetization}) with a specific $m_c$,  which is \emph{energy-conserving}, i.e., $[H_{\text{loc}},H_{\text{int}}^{(m_c)}]=0$~\cite{mitchison2015}   if 
\begin{eqnarray}
    E_h+\left(\frac{m_c+n}{2}\right) E_c=E_r,
    \label{eq:self_contained}
\end{eqnarray}
where we have assumed $E_{c_i}=E_c\;\forall i=1,\cdots,n$, ensuring that $H_{\text{loc}}$ and $H_{\text{int}}$ shares the same eigenspace. This further suggests that  the states $\ket{\mathbf{m}_{(m_c)}}$ and  $\ket{\mathbf{m}_{(-n)}}$ have the same energy, i.e., 
\begin{eqnarray}
\langle\mathbf{m}_{(m_c)}|H_{\text{loc}}|\mathbf{m}_{(m_c)}\rangle=\langle\mathbf{m}_{(-n)}|H_{\text{loc}}|\mathbf{m}_{(-n)}\rangle,
\end{eqnarray}
making the QTM \emph{self-contained}~\cite{popescu2010}, such that no external input of energy is required for transitions between $\ket{\mathbf{m}_{(m_c)}}$ and $\ket{\mathbf{m}_{(-n)}}$. For the cooling of the qubits in $c$ to take place, one would require a higher transition rate corresponding to $\ket{\mathbf{m}_{(m_c)}}\rightarrow\ket{\mathbf{m}_{(-n)}}$ than  $\ket{\mathbf{m}_{(m_c)}}\leftarrow\ket{\mathbf{m}_{(-n)}}$, which is fulfilled by the condition 
\begin{eqnarray}
    \left(\frac{m_c+n}{2}\right)\frac{E_c}{T_c}+\frac{E_h}{T_h}<\frac{E_r}{T_r}.
    \label{eq:population_condition}
\end{eqnarray}

\begin{figure*}
    \includegraphics[width=.8\linewidth]{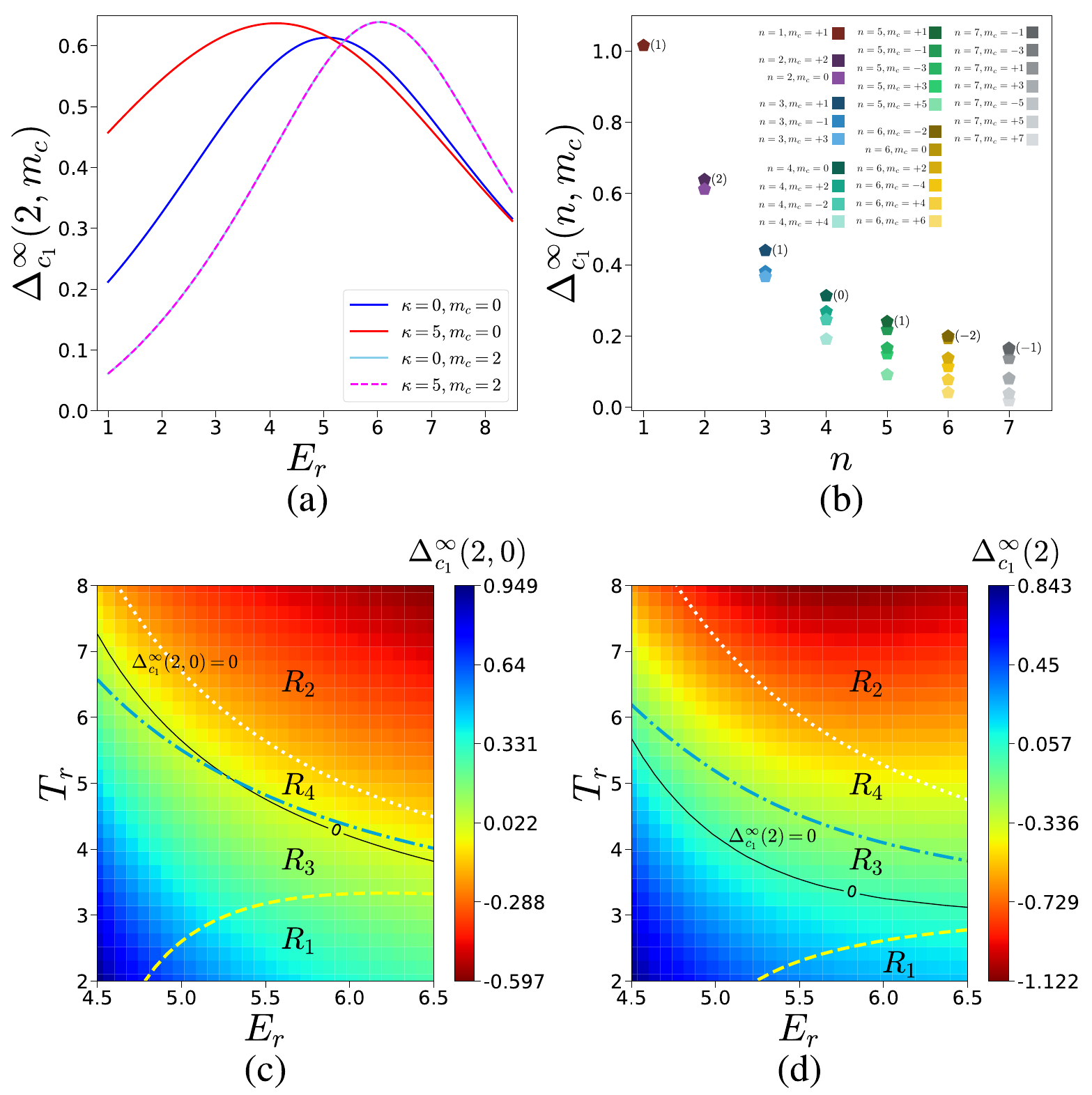}
    \caption{(a) Variations of $\Delta_{c_1}^\infty(2,m_c)$,  as functions of $E_r$, when interaction terms correspond to $m_c=0,+2$, GQME is used for attaining the steady state, and $\kappa$ is set to either $\kappa=0$, or $\kappa=5$. The relevant parameters are fixed at $E_h=4$, $E_{c_1}=1$, $E_{c_2}=1.01$, $T_h=10$, $T_r=2$, $T_c=2$, $f_{h,r,c}=0.01$, $\Omega_{h,r,c}=10^3$, and $g_{(m_c)}=1$ for both $m_c=0,+2$.  (b) Variations of $\Delta_{c_1}^\infty(n,m_c)$ with increasing $n$ for all possible values of $m_c$ corresponding to each value of $n$. The numbers in brackets indicate the magnetization $m_c$ corresponding to which the maximum value of $\Delta_{c_1}^\infty(n,m_c)$ among the allowed values of $m_c$ takes place for a specific $n$. The variations of $\Delta_{c_1}^\infty(2,0)$ and $\Delta_{c_1}^\infty(2)$ as a function of $E_r$ and $T_r$ are shown in (c) and (d) respectively.  We enforce Eq.~(\ref{eq:self_contained}) on the values of $E_{r,h,c}$  for (b), (c), and (d), while the parameters other than $E_{r,h,c}$ are set as in (a). All quantities plotted in (a), (b), (c), and (d) are dimensionless. }
    \label{fig:self_contained_global}
\end{figure*}

\noindent\textbf{Local master equation.} Let us first consider a situation with an interaction Hamiltonian $H_{\text{int}}^{(m_c)}$ corresponding to a specific transition subspace with magnetization $m_c$,  where  $g_{(m_c)}$ is negligible, such that $H_S\approx H_{\text{loc}}$. In this scenario,  derivation of the Lindblad operators can be done using the spectrum of $H_{\text{loc}}$ only, instead of $H_S$, and the corresponding  QME is referred to as the local QME (LQME)~\cite{Landi2022}. We point out here that in the qubit-bath interaction of the $c$-qubits (Eq.~(\ref{eq:system_bath_interaction_c})), $\kappa$ is chosen to be zero while LQME is used. Note further that the qubit-part of the qubit bath interaction (Eqs.~(\ref{eq:system_bath_interaction_ab}), (\ref{eq:system_bath_interaction_c})) facilitates jumps between different levels with different energies.

\paragraph{Local temperature.} To assess whether the QTM designed above is serving its purpose, one needs to appropriately quantify the \emph{local temperature} $\tau_{c_i}^{\infty}$ (see discussions preceding Eq.~(\ref{eq:cooling_amount})). Following~\cite{Tanoy2023}, we assume a \emph{distance-based approach} to quantify the local temperature for a qubit in the state $\rho_\alpha(t)=\text{Tr}_{\underset{\alpha^\prime\neq \alpha}{\{\alpha^\prime\}}}\rho_S(t)$, where the system is in the state $\rho_S(t)$.  We define $\tau_{c_i}$ to be the absolute temperature that minimizes the trace distance~\cite{nielsen2010}
\begin{eqnarray}
    \mathbb{D}(\rho_\alpha(t),\varrho_\alpha)=\text{Tr}\sqrt{(\rho_\alpha(t)-\varrho_\alpha)^{\dagger}(\rho_\alpha(t)-\varrho_\alpha)}
\end{eqnarray}
between $\rho_\alpha(t)$ and a single-qubit \emph{canonical} thermal state 
\begin{eqnarray}
    \varrho_\alpha&=& \exp -( H_\alpha/k_B\tau_\alpha)/\text{Tr}\left[\exp -(H_\alpha/k_B\tau_\alpha)\right],
\end{eqnarray}
where $H_\alpha$ are as defined in Eq.~(\ref{eq:single_qubit_hamiltonian}). In the case of an interaction Hamiltonian $H_{\text{int}}$ that does not lead to coherence generation in $\rho_\alpha(t)$ (i.e., $\rho_\alpha(t)$ are diagonal), the above definition of $\tau_\alpha$ is equivalent to the definition of local temperature considered in~\cite{popescu2010,das2019,Ahana2021,konar2023}, where a cooling of a qubit is indicated by an increase in its ground-state population. The case of the LQME with $H_{\text{int}}^{(m_c)}$ being the interaction Hamiltonian is an example. In such situations, corresponding to the interaction Hamiltonian $H_{\text{int}}^{(m_c)}$ designed for an $n$-qubit cooling, we write $\tau_\alpha$ for $\rho_S(t=\infty)$ to be $\tau_\alpha^\infty(n,m_c)$, which is given by~\cite{popescu2010,mitchison2015,das2019,Ahana2021,konar2023}
\begin{eqnarray}
    \tau_\alpha^{\infty}(n,m_c) &=&  \frac{E_\alpha}{\ln\left[\left(\langle0|\rho_\alpha^\infty|0\rangle\right)^{-1}-1\right]}.
\end{eqnarray}
The corresponding amount of cooling is given by 
\begin{eqnarray}
    \Delta_{\alpha}^\infty(n,m_c)=T_{\alpha}-\tau_\alpha^\infty(n,m_c), 
\end{eqnarray}
where $\alpha=c_1,c_2,\cdots,c_n$. Note that we have introduced the number of target qubits and the magnetization sector in the notation in order to clarify which magnetization sector is facilitating the cooling.  In the situation where we design the interaction Hamiltonian as $H_{\text{int}}=\sum_{m_c}g_{m_c}H_{\text{int}}^{(m_c)}$ for a fixed $n$, we denote the amount of cooling by $\Delta_\alpha^\infty(n)$.

A high value of $\Delta_\alpha^{\infty}(n,m_c)$ depends on a judicious choice of the points in the space of the system parameters on which the QTM operates. In Fig.~\ref{fig:self_contained_temp_local}(a), we plot the variations of $\Delta_{c_i}^\infty(2,m_c)$, $i=1,2$, as  functions of $E_r$ for all allowed values of $m_c$. Observe, for all combinations of $n$ and $m_c$, that for a specific pair of fixed values of $E_h$ and $E_c$, and for a specific value of $m_c$ given a fixed $n$, the variation of $\Delta_{c_i}^\infty(n,m_c)$ attains a maximum at the value of  $E_r$ governed by Eq.~(\ref{eq:self_contained}), while the amount of cooling dies out as one deviates from Eq.~(\ref{eq:self_contained}). In Fig.~\ref{fig:self_contained_temp_local}(b), we present the scatter plot of the values of $\Delta_{c_i}^\infty(n,m_c)$ corresponding to different values of $n$ and the corresponding $m_c$, where the energies of the qubits $h$, $r$, $c_i$, $i=1,2,\cdots,n$ satisfy Eq.~(\ref{eq:self_contained}), indicating a decrease in the amount of cooling as $n$ increases. Further, the maximum amount of cooling that can be obtained from different magnetization sectors of the Hilbert space loses the $m_c$-dependence as one increases $n$, which is clear from the points on the scatter plot approaching each other for a fixed $n$, when $n$ is increasing.

\begin{figure*}
    \includegraphics[width=.8\linewidth]{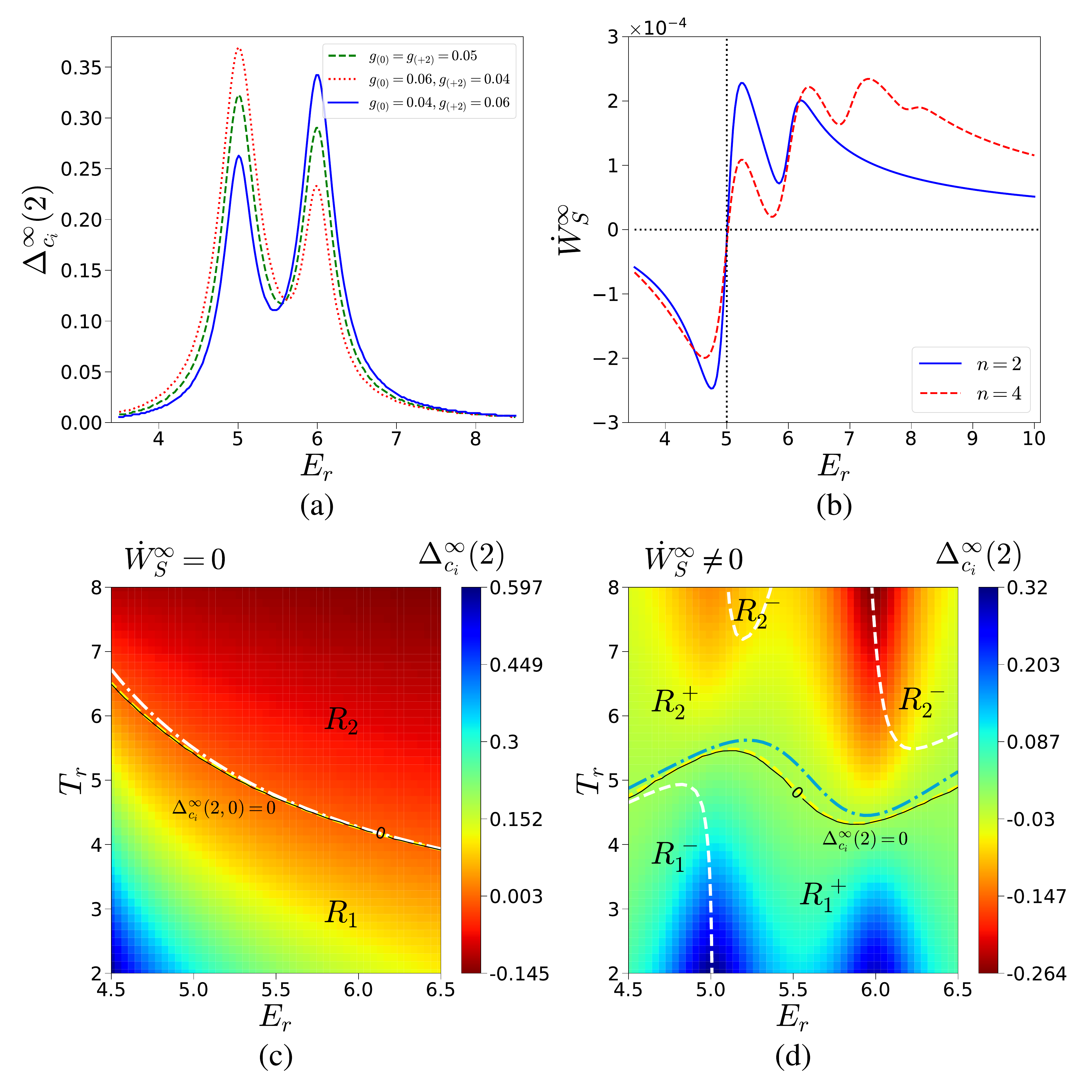}
    \caption{(a) Variations of $\Delta_{c_i}^{\infty}(2)$ as a function of $E_r$ with $g_{(0)}=g_{(+2)}$, $g_{(0)}>g_{(+2)}$, and $g_{(0)}<g_{(+2)}$, keeping $E_h=4$. (b) $\dot{W}_{S}^\infty$ (see Eq.~(\ref{eq:full_work_imbalance})) as a function of $E_r$ for different values of $n$, crossing zero at $E_r\approx E_h+E_c$ for all $n$. Variations of $\Delta_{c_i}^\infty(2)$ as a function of $E_r$ and $T_r$ are depicted in (c) and (d). In (c), $\dot{W}_S^\infty\approx0$ is ensured at each point on the $E_r-T_r$ plane, while in (d), $\dot{W}_S^\infty\neq 0$. All quantities plotted are dimensionless.}
    \label{fig:all_mag_local}
\end{figure*}

\paragraph{Operating regimes.} We now discuss the operation of the QTM in terms of the LSSHCs. With LQME in the use, Eq.~(\ref{eq:global_heat_current}) at $t\rightarrow\infty$ becomes~\cite{hewgill2020}
\begin{eqnarray}
    \sum_{\alpha=h,r,c}\dot{Q}_{\text{loc},\alpha}^\infty=-\dot{W}^\infty_{S},
\end{eqnarray}
where 
\begin{eqnarray}
    \dot{Q}_{\text{loc},\alpha}^\infty &=& \text{Tr}\left[\mathcal{D}_\alpha\left(\rho_S^\infty\right)H_{\text{loc}}\right], \alpha=h,r,c,
    \label{eq:local_heat_current}
\end{eqnarray} 
is the \emph{local} SSHC (LSSHC) corresponding to the bath $\alpha$, and 
\begin{eqnarray}
    \dot{W}^\infty_{S,(m_c)}&=&\text{Tr}\left[\left\{\sum_{\alpha=h,r,c}\mathcal{D}_\alpha(\rho_S^\infty)\right\}H_{\text{int}}^{(m_c)}\right]
    \label{eq:work_imbalence}
\end{eqnarray}
is the contribution to the power due to the interaction Hamiltonian originating from the transport subspace labelled by $m_c$. An extraction (deposition) of heat from (to) the heat bath $B_\alpha$ is designated by $\dot{Q}_{\text{loc},\alpha}^\infty >0$ ($\dot{Q}_{\text{loc},\alpha}^\infty <0$). 

Note that a small, but non-zero $\dot{W}_{S,(m_c)}^\infty$ is the result of the use of LQME, more specifically, the determination of the Lindblad operators using the local spectrum despite the presence of a small, yet not-zero value of $g_{(m_c)}$, which appears as a violation of the first law: $\sum_{\alpha=h,r,c}\dot{Q}_{\text{loc},\alpha}^\infty\neq 0$. However, one may choose the interaction Hamiltonian and the system parameters carefully to ensure $\dot{W}_{S,(m_c)}^\infty=0$. For example, at the values of $E_r$ where $\Delta_{c_i}^{\infty}(n,m_c)$ is maximum (i.e., Eq.~(\ref{eq:self_contained}) is satisfied), $\dot{W}_{S,(m_c)}^\infty=0$, while on both sides of the maximum, $\dot{W}_{S,(m_c)}^\infty\neq 0$ and have opposite signs. This is demonstrated in Fig.~\ref{fig:self_contained_q_local}(a) in the case of $n=2$. Note further that in the parameter regimes considered in this paper, $\dot{S}=-\text{Tr}(\sum_{\alpha} \mathcal{D}_{\alpha}(\rho_S^\infty) \ln (\rho_S^\infty))=0$ while $\sum_\alpha\beta_\alpha\dot{Q}_\alpha<0$,  ensuring the validity of (\ref{eq:second_law}) throughout.

In Fig.~\ref{fig:self_contained_q_local}(b), we plot $\Delta_{c_i}^\infty(2,0)$ as a function of $E_r$ and $T_r$, ensuring that on all points of the $E_r-T_r$ plane, Eq.~(\ref{eq:self_contained}) is valid, thereby guaranteeing 
\begin{eqnarray}
    \sum_{\alpha=h,r,c}\dot{Q}_{\text{loc},\alpha}^\infty=\dot{W}_{S,(m_c)}^\infty=0,
\end{eqnarray}
and further verifying the validity of Eq.~(\ref{eq:second_law}). We find that both cooling as well as heating, as indicated by the variations of $\Delta_{c_i}^\infty(2,0)$, is possible for the qubits in $c$ with the setup of the QTM. Therefore, the designed QTM with the restriction (\ref{eq:self_contained}) on the system parameters may operate as a \emph{refrigerator} as well as a \emph{heater},  the line $\Delta_{c_i}^\infty(2,0)=0$ distinguishing between the two operation regimes, $R_1$ (below the $\Delta_{c_i}^\infty(2,0)=0$ line) and $R_2$ (above the $\Delta_{c_i}^\infty(2,0)=0$ line), on the $E_r-T_r$ plane. Our numerical analysis suggests that 
\begin{eqnarray}
    \dot{Q}^\infty_{\text{loc},h}>0, \dot{Q}^\infty_{\text{loc},r}<0, \dot{Q}^\infty_{\text{loc},c}>0
    \label{eq:qar_condition}
\end{eqnarray}
in $R_1$, when the qubits in $c$ are cooled, thereby the QTM operating as a QAR, as indicated by the signs of the LSSHCs. On the other hand, in region $R_2$ where the qubits in $c$ are heated,
\begin{eqnarray}
    \dot{Q}^\infty_{\text{loc},h}<0, \dot{Q}^\infty_{\text{loc},r}>0, \dot{Q}^\infty_{\text{loc},c}<0.
    \label{eq:heater_condition}
\end{eqnarray}
In Fig.~\ref{fig:self_contained_q_local}(c), we deviate from Eq.~(\ref{eq:self_contained}), and choose a set of values of $E_{h,r,c}$ for which $\dot{W}_{S,(m_c)}^\infty\neq 0$, to find four operating regimes labelled by $R_1^{\pm}$ and $R_2^\pm$, where in $R_1^\pm$ ($R_2^\pm$), (\ref{eq:qar_condition}) is valid ((\ref{eq:heater_condition}) is valid) while the signs $\pm$ signify whether $\dot{W}_{S,(m_c)}^\infty$ is $>0$ or $<0$. For different values of $n$ and with different magnetization sectors constituting the interaction Hamiltonian $H_{\text{int}}^{(m_c)}$, the qualitative features of the  variations of $\Delta_{c_i}^\infty(n,m_c)$, and the operating regimes  are the same as demonstrated in Fig.~\ref{fig:self_contained_q_local}(b) and (c). We point out here that operating the QTM as a QAR does not guarantee a substantial cooling of all the qubits $c_1,c_2,\cdots,c_n$ in $c$, which is determined by the choice of the point in the space of the system parameters where the QTM is running.

\noindent\textbf{Global master equation.} We now focus on the global master equation for the time evolution of the $(n+2)$-qubit system, and determine the Lindblad operators from the spectrum of the full system Hamiltonian, $H_S$. Note here that unlike the LQME, the qubit-part of the qubit-bath interactions (Eqs.~(\ref{eq:system_bath_interaction_ab}), (\ref{eq:system_bath_interaction_c})) can, in principle, facilitate transitions within degenerate energy levels. To avoid this, we incorporate disorder among the energies in the $c$-qubits, such that  $E_{c_i}\neq E_{c_j}$ for $i\neq j$ and $c_i,c_j\in c$, where  all $E_{c_i}$ are fixed along with $E_h$ and $E_r$. While this sacrifices the possibility of equal cooling in $c$-qubits, one can still achieve \emph{almost} equal cooling by fixing $|E_{c_i}-E_c|$ to be small $\forall i$, where $E_c$ is a judiciously chosen energy for the $c$-qubits. Note also that in the case of GQME, Eqs.~(\ref{eq:first_law})-(\ref{eq:second_law}) hold irrespective of the chosen values of $E_h,E_r$, and $E_c$.

For ease of discussion, we consider the case of interaction Hamiltonian constructed from a transport subspace with a specific magnetization, and consider $n=2$ for example. In Fig.~\ref{fig:self_contained_global}(a), $\Delta_{c_1}^\infty(2,m_c)$ is plotted as a function of $E_r$ for different allowed values of $m_c$ and for different values of $\kappa$ ($\kappa=0$ and $\kappa\neq 0$). While  $\Delta_{c_1}^\infty(2,m_c)$ attains a maximum value at a specific $E_r$ as in the case of LQME (see Fig.~\ref{fig:self_contained_temp_local}(a)), the peaks of  $\Delta_{c_1}^\infty(2,m_c)$ are broader, and the position of the peaks on the $E_r$ axis depends on the chosen values of $\kappa$. The variations of $\Delta_{c_1}^\infty(n,m_c)$ with $E_r$ are qualitatively similar for all values of $n$ and the corresponding all values of $m_c$. In Fig.~\ref{fig:self_contained_global}(b), we depict the scatter plot of the values of $\Delta_{c_1}^\infty(n,m_c)$ with varying $n$ and $m_c$, while obeying (\ref{eq:self_contained}) for the sake of comparison with Fig.~\ref{fig:self_contained_temp_local}(b). Note that the overall decrease in the values of  $\Delta_{c_1}^\infty(n,m_c)$ with increasing $n$ is similar to the case of the LQME, although the dependence of $\Delta_{c_1}^\infty(n,m_c)$ on the values of $m_c$, for a fixed $n$, becomes prominent when $n$ is large, which is in contrast to the case of LQME, as evident from Fig.~\ref{fig:self_contained_temp_local}(b).

\begin{table}[]
    \centering
    \begin{tabular}{ccc}
    \hline
    \hline
        Operating Region & Signs of SSHCs & $\dot{W}_S^\infty$ \\
        \hline
        $R_1$ &  $\dot{Q}_h^\infty>0,\dot{Q}_r^\infty<0, \dot{Q}_c^\infty>0$ & 0 \\
        $R_2$ & $\dot{Q}_h^\infty<0,\dot{Q}_r^\infty>0, \dot{Q}_c^\infty<0$ & 0 \\
        $R_3$ & $\dot{Q}_h^\infty>0,\dot{Q}_r^\infty<0, \dot{Q}_c^\infty<0$ & 0\\
        $R_4$ & $\dot{Q}_h^\infty>0,\dot{Q}_r^\infty>0, \dot{Q}_c^\infty<0$ & 0\\        
    \end{tabular}
    \caption{Signs of SSHCs in different regions $R_1-R_4$ on the $E_r-T_r$ plane depicted in Fig.~\ref{fig:self_contained_global}(c). All quantities listed are dimensionless.}
    \label{tab:operating_regions}
\end{table}

In Fig.~\ref{fig:self_contained_global}(c), we plot $\Delta_{c_1}^\infty(2,0)$ as a function of $E_r$ and $T_r$, and for the sake of comparison with Fig.~\ref{fig:self_contained_q_local}(b) (note that $\dot{W}_S^\infty=0$ in the case of the GQME irrespective of the values of $E_{r,h,c}$, see discussions preceding Eqs.~(\ref{eq:first_law})), we ensure the satisfaction of Eq.~(\ref{eq:self_contained})  on all points of the $E_r-T_r$ plane considered in Fig.~\ref{fig:self_contained_global}(c) by choosing similar values of $E_h$ and $E_c$ as in Fig.~\ref{fig:self_contained_q_local}(b). The contour $\Delta_{c_1}^{\infty}(2,0)=0$ marks the boundary between the regions where $\Delta_{c_1}^{\infty}(2,0)>0$ (cooling, below the contour) and $<0$ (heating, above the contour). The different configurations of the SSHCs on the $E_r-T_r$ plane shown in Fig.~\ref{fig:self_contained_global} are given in Table~\ref{tab:operating_regions}, while the dashed, dot-dashed, and dotted lines on the $E_r-T_r$ plane mark the boundaries between the operation regions $R_1$ and $R_3$, $R_3$, and $R_4$, and $R_4$ and $R_2$ respectively. Note that for ease of discussion and comparison with the case of the LQME, we have defined the regions $R_1$ and $R_2$ as they are defined with the LSSHCs in Eqs.~(\ref{eq:qar_condition}) and (\ref{eq:heater_condition}). Note further that Fig.~\ref{fig:self_contained_global}(c) suggests the existence of operation regions of the designed QTM where the machine does not work as a QAR (for example, $R_3$ and $R_4$), and yet a cooling of the target qubits in $c$, as quantified by $\Delta_{c_1}^\infty(n,m_c)$, takes place. While the demonstration is done with the example of $n=2$ and $m_c=0$, qualitatively similar results are found in the cases of higher $n$ and for all corresponding allowed values of $m_c$. Further, note that the operating regions of the designed QTM are consistent with the operating regimes of QTMs designed with harmonic baths with the requirement $\dot{W}_S^\infty=0$, as pointed out in~\cite{hewgill2020}.    

\begin{figure*}
    \includegraphics[width=\linewidth]{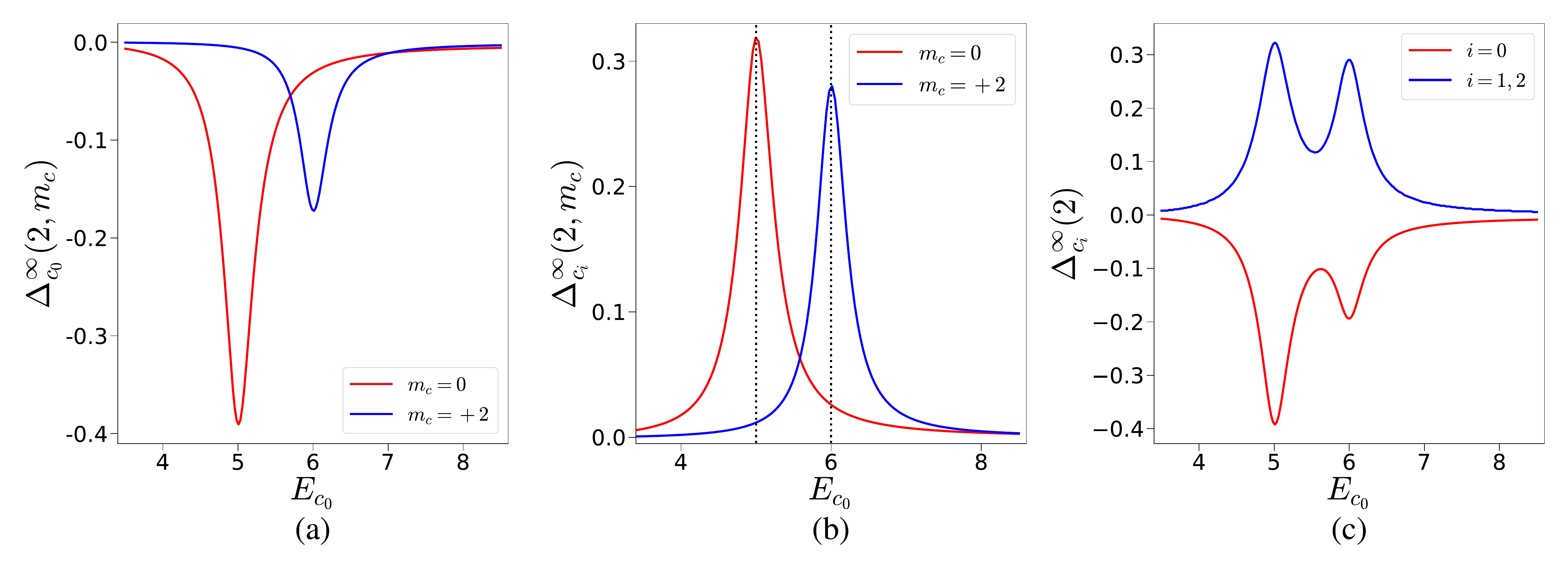}
    \caption{Variations of (a) $\Delta_{c_0}^\infty(2,m_c)$ and (b) $\Delta_{c_i}^\infty(2,m_c)$ as functions of $E_{c_0}$ in the case of a two-bath QTM. When contributions from both values of $m_c=0,+2$ are present in the interaction Hamiltonian, the variations of $\Delta_{c_0}^{\infty}$ and $\Delta_{c_i}^{\infty}$, $i=1,2$, with $E_{c_0}$ are shown in (c). All quantities plotted are dimensionless.}
    \label{fig:two-baths}
\end{figure*}

\subsubsection{Transport subspace with multiple magnetization sectors}
\label{subsubsec:multiple_magnetization}

It is now logical to ask how the above features of the QTM and the simultaneous cooling of the target qubits by it are modified when the interaction Hamiltonian $H_{\text{int}}$ has contribution from different magnetization sectors (see Eq.~(\ref{eq:full_interaction_Hamiltonian})). We assume that for a given $n$, all, or a subset of $g_{(m_c)}$s are non-zero and, in general, not equal, but negligible so that the LQME can be employed. Note that demanding $[H_{\text{loc}},H_{\text{int}}]=0$ requires \small 
\begin{eqnarray}
    \sum_{m_c}g_{(m_c)}f_{(m_c)}\left[\ket{\mathbf{m}_{(m_c)}}\bra{\mathbf{m}_{(-n)}}
    -\ket{\mathbf{m}_{(-n)}}\bra{\mathbf{m}_{(m_c)}}\right]=0,
\end{eqnarray}\normalsize 
with 
\begin{eqnarray}
    f_{(m_c)}&=& E_h-E_r+\frac{m_c+n}{2}E_c, 
\end{eqnarray}
which, in turn, requires $f_{(m_c)}=0$ for all magnetization sectors for which $g_{(m_c)}\neq 0$. This can not be achieved for all magnetization sectors \emph{simultaneously} by fixing $E_{h,r,c}$, as for a fixed $n$, 
\begin{eqnarray}
    1\leq \frac{m_c+n}{2}\leq n,
\end{eqnarray}
with the value of $(m_c+n)/2$ increasing by $1$ from $1$ to $n$, while $-n+2\leq m_c\leq n$, and there is no degeneracy in the values of $m_c$. However, one can fix the values of $E_{r,h,c}$ such that for a transport sector corresponding to a specific $m_c=m_c^0$, $f_{(m_c^0)}=0$, i.e., 
\begin{eqnarray}
    \left[H_{\text{loc}},H_{\text{int}}\right] &=& \sum_{m_c\neq m_c^0}\left[H_{\text{loc}},H_{\text{int}}^{(m_c)}\right].
\end{eqnarray}
Note further, from Eq.~(\ref{eq:work_imbalence}), that the contribution to power due to $H_{\text{int}}$ having the form (\ref{eq:full_interaction_Hamiltonian}) is  
\begin{eqnarray}
    \dot{W}_S^{\infty}&=&\sum_{m_c}\dot{W}_{S,(m_c)}^\infty.
    \label{eq:full_work_imbalance}
\end{eqnarray}
Also, in this scenario, we denote the amount of cooling for the $n$-qubit subsystem by $\Delta_{c_i}^{\infty}(n)$.

To discuss the effect of $g_{(m_c)}$ on the cooling of the target qubits, we consider the case of $n=2$ for demonstration. In Fig.~\ref{fig:all_mag_local}(a), we plot $\Delta_{c_i}^\infty(2)$ as a function of $E_r$ for the case of $g_{(0)}=g_{(+2)}$, $g_{(0)}>g_{(+2)}$, and $g_{(0)}<g_{(+2)}$. It is clear from Figs.~\ref{fig:all_mag_local}(a) and \ref{fig:self_contained_temp_local}(a) that $\Delta_{c_i}^\infty(2)$ attains local maximums at the values of $E_r$ corresponding to Eq.~(\ref{eq:self_contained}) for the two allowed values of $m_c$, and the hierarchy of the values of $\Delta_{c_i}^\infty(2)$ at the local maximums can be altered by tuning the values of $g_{(0)}$ and $g_{(+2)}$. For instance, with $g_{(0)}\geq g_{(+2)}$, the peak of $\Delta_{c_i}^{\infty}(2)$ corresponding to $m_c=0$, at $E_r=E_h+E_c$, is higher than the same corresponding to $m_c=+2$, occurring at $E_r=E_h+2E_c$. However, as $g_{(0)}$ ($g_{(+2)}$) is further reduced (increased) keeping $g_{(0)}<g_{(+2)}$, the peak  $\Delta_{c_i}^{\infty}(2)$ corresponding to $m_c=0$, at $E_r=E_h+E_c$, diminishes below the same corresponding to $m_c=+2$, occurring at $E_r=E_h+2E_c$, which is clear from Fig.~\ref{fig:all_mag_local}(a). Such tuning of the local peaks can be done in the cases of all $n$, although a large value of $n$ allows a larger number of values of $m_c$, implying a more complicated hierarchy of the values of $g_{(m_c)}$, and in turn, the values of $\Delta_{c_i}^\infty(n)$ at the local peaks.         

In Fig.~\ref{fig:all_mag_local}(b), we plot $\dot{W}_S^\infty$ as a function of $E_r$ and find $\dot{W}_S^\infty=0$ to occur always at $f_{(m_{c}=-n+2)}\approx 0$ (i.e., $m_c^0=-n+2$), irrespective of the value of $n$, indicating $\sum_{\alpha=h,r,c}\dot{Q}_{\text{loc},\alpha}^\infty\approx 0$ ($\neq 0$) for $E_r\approx E_h+E_c$ ($E_r=E_h+(m_c+n)E_c/2$ with $m_c>-n+2$)\footnote{Here, we have assumed $g_{(m_c)}$ to be equal for all $m_c$. In the cases where $g_{(m_c)}$ is different for different magnetization sectors, $\dot{W}_S^\infty$ may vanish at other values of $E_r$.}. In Fig.~\ref{fig:all_mag_local}(c), we plot $\Delta_{c_i}^{\infty}(2)$ as a function of $E_r$ and $T_r$, ensuring $E_{r}=E_h+E_c$ (and therefore $\dot{W}_S^\infty\approx 0$) at every point on the $E_r-T_r$ plane. Similar to Fig.~\ref{fig:self_contained_q_local}(b), the line $\Delta^\infty_{c_i}(2)=0$ marks the boundary between the regions on the $E_r-T_r$ plane where cooling ($\Delta_{c_i}^\infty(2)>0$) and heating ($\Delta_{c_i}^\infty(2)<0$) of the qubits in $c$ occur. However, unlike the case of interaction Hamiltonian constituted of transport subspace with a single magnetization, $\Delta_{c_i}^\infty(2)=0$ only approximately marks the boundary between $R_2$ and $R_1$ according to the signs of the LSSHCs (see Eqs.~(\ref{eq:qar_condition}) and (\ref{eq:heater_condition})), as evident from the difference between the lines representing $\Delta_{c_i}^\infty(2)=0$ and the boundary between $R_1$ and $R_2$. A similar observation is made in the case of the variation of $\Delta_{c_i}^\infty(2)$  with $E_r$ and $T_r$ where $\dot{W}_S^\infty\neq 0$ on all points considered, as shown in Fig.~\ref{fig:all_mag_local}(d). Note, however, that the regions $R_{1}$ and $R_2$ are further split into regions $R_1^\pm$ and $R_2^\pm$, according to whether $\dot{W}_S^\infty$ is positive or negative, as in the case of Fig.~\ref{fig:self_contained_q_local}(c).

When GQME is used for determining the dynamics, the local peaks observed in the variation of $\Delta_{c_i}^\infty(2,m_c)$ against $E_r$ corresponding to different allowed values of $m_c$, as observed in the case of LQME, vanish, and similar to the situation described in Fig.~\ref{fig:self_contained_global}(a),  $\Delta_{c_i}^\infty$ exhibits a single maximum corresponding to a specific value of $E_r$. Further, similar to the case presented in Fig.~\ref{fig:self_contained_global}(c),  the landscape of $\Delta_{c_i}^\infty(2)$ on the $E_r-T_r$ plane exhibits both cooling and heating of the qubits $c_i$, $i=1,2$, with the cooling taking place mostly in regions $R_1$ and $R_3$, where $R_1$ corresponds to the heat currents satisfying conditions for a QAR (see Table~\ref{tab:operating_regions}). For higher $n$, these trends are found to remain unchanged. Also, the variation of $\Delta_{c_i}^\infty(n)$ on the $E_r-T_r$ plane suggests that the QTM, described by the GQME, may operate in working regimes other than a QAR, and yet indicate a cooling when probed via the definition of local temperature provides in Sec.~\ref{subsubsec:single_magnetization} -- a trend similar to the cases where the interaction Hamiltonian is designed out of a single magnetization sector.

\begin{figure*}
    \includegraphics[width=0.9\linewidth]{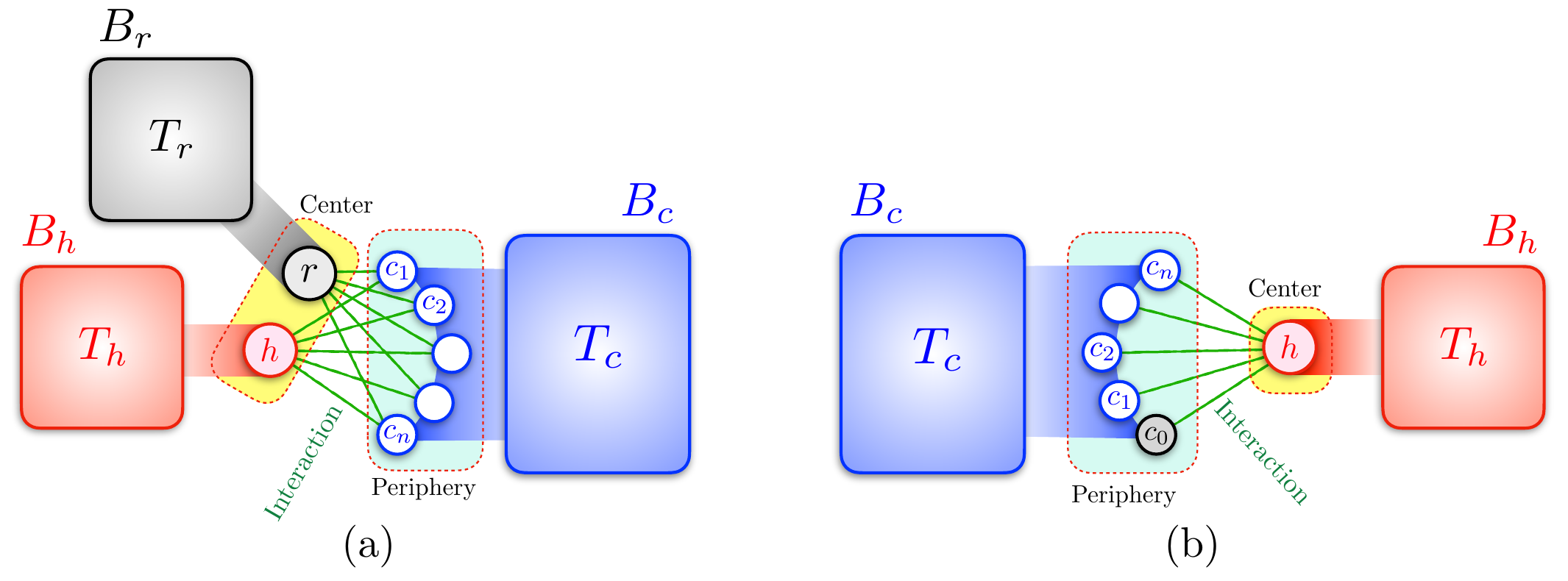}
    \caption{Setups of QTMs with (a) three and (b) two baths, using star network of qubits. }
    \label{fig:spin_star}
\end{figure*}

\subsection{Multi-qubit cooling with two baths}
\label{subsec:two_baths}

We now explore the possibility of cooling multiple qubits by setting up the QTM using two bosonic heat baths -- a hot bath $B_h$, and the cold bath $B_c$ -- where the \emph{room} qubit $r$ is also kept in contact with the cold bath (see Fig.~\ref{fig:schematics}(b)). For ease of representation, we re-label the qubit $r$ with $c_0$,  and denote the set of $(n+1)$ qubits $c_0,c_1,c_2,\cdots,c_n$ collectively by $c$, such that the Hamiltonian $H_{SB}^c$ is now given by Eq.~(\ref{eq:system_bath_interaction_c}), $\mathcal{D}_h(.)$ has the form as given in Eq.~(\ref{eq:dissipative_term_main}), while $\mathcal{D}_c(.)$ reads as Eq.~(\ref{eq:dissipative_term_c}), with the Lindblad operators taking the form as given in Eq.~(\ref{eq:lindblad_operators_c_two_body}) and (\ref{eq:lindblad_operators_c_three_body}). The interaction Hamiltonian $H_{\text{int}}$ is designed as per the steps given in Sec.~\ref{subsec:interaction_hamiltonian}, and in the same fashion as described in Sec.~\ref{subsec:performance}, one can either take a local, or a global approach in determining the Lindblad operators in the QME.

We consider the LQME for the sake of discussion, and work with energy conserving interaction Hamiltonian ($[H_{\text{loc}},H_{\text{int}}^{(m_c)}]=0$) corresponding to a transport subspace with a fixed magnetization.   Note that with the setup described above, $T_r=T_c$, and Eq.~(\ref{eq:population_condition}), along with Eq.~(\ref{eq:self_contained}), leads to $T_h>T_c$ to be the required condition for a higher transition rate corresponding to $\ket{\mathbf{m}_{(m_c)}}\rightarrow\ket{\mathbf{m}_{(-n)}}$ than  $\ket{\mathbf{m}_{(m_c)}}\leftarrow\ket{\mathbf{m}_{(-n)}}$. Note here that unlike the three-bath setup, $E_{c_0}\neq E_{c_i}=E_c$ $\forall i=1,2,\cdots,n$, and one would expect different dynamics for the qubit $c_0$ and the qubits $c_i,i=1,2,\cdots,n$, thereby different behaviours of the steady-state temperature of the qubits $c_0$ and $c_i$. The variations of $\Delta_{c_0}^\infty(2,m_c)$ and $\Delta_{c_i}^\infty(2,m_c)$, $i=1,2$, as functions of $E_{c_0}$ ($\equiv E_r$), are demonstrated in Figs.~\ref{fig:two-baths}(a) and (b) respectively for $H_{\text{int}}^{(m_c)}$ corresponding to different $m_c$, exhibiting respectively a heating and a cooling. Similar observation sustains as one constructs the interaction Hamiltonian with contributions from different magnetization sectors, as shown in Fig.~\ref{fig:two-baths}(c), with the local maximums (minimums) of $\Delta_{c_i}^\infty(2,m_c)$ ($\Delta_{c_0}^\infty(2,m_c)$) corresponding to different $m_c$ values prominent (see Figs.~\ref{fig:all_mag_local}(a) for a comparison). Variations of $\Delta_{c_i}^{\infty}(2,m_c)$  as functions of $E_{c_0}$ and $T_h$ exhibit both cooling and heating when $\dot{W}_S^\infty$ is made to vanish (via satisfaction of (\ref{eq:self_contained}), the boundary marked by the line $T_h=T_c$. For $T_h<T_c$ ($T_h>T_c$), $\Delta_{c_i}^\infty(2,m_c)<0$ ($>0$) for all $m_c$, indicating a heating (cooling), with $\dot{Q}_{\text{loc},c}^\infty>0$, $\dot{Q}_{\text{loc},h}^\infty<0$, ($\dot{Q}_{\text{loc},c}^\infty<0$, $\dot{Q}_{\text{loc},h}^\infty>0$), which we represent by $R_1$ ($R_2$).  When $\dot{W}_S^\infty\neq 0$, the $E_{c_0}-T_h$ plane is divided into four regions, represented by $R_1^\pm$ and $R_2^\pm$, as in the case of the Fig.~\ref{fig:self_contained_q_local}(c). Similarly, with $H_{\text{int}}$ involving different sectors of magnetization, landscape of $\Delta_{c_i}^\infty(2)$ on the $E_{c_0}-T_h$ plane is qualitatively similar to those shown in Figs.~\ref{fig:all_mag_local}(c) and (d), with majority of cooling occurring in $R_2^+$. 

One may also consider the GQME, in which case $\dot{W}_S^\infty=0$ throughout the parameter space. The results obtained are qualitatively similar to the same for the three-bath model, as described in Sec.~\ref{subsec:performance}, and demonstrated in Figs.~\ref{fig:self_contained_global}(a) and (c).

\section{Design with a star-network of spins}
\label{sec:spin_star}

So far, we have used strategically designed forms of $H_{\text{int}}$ systematically involving different subspaces of $\mathcal{H}$ for designing the QTM. It is now logical to ask whether the operations and performance of the QTM change if one takes into account the full $\mathcal{H}$ for designing the $H_{\text{int}}$. A convenient way to do so is to write $H_{\text{int}}$ in terms of well-known spin interactions, which has been explored extensively in literature. In this paper, we focus on the \emph{star network} of qubits ~\cite{Richter_1994,Hutton2004,*Anza_2010,*Militello2011,*haddadi2021,*Karlova2023}, where each qubit in a set of \emph{central} qubits interact with each qubit in a set of \emph{peripheral} qubits, while none of the central (peripheral) qubits interact with each other. Further, given the setups of QTMs discussed in Sec.~\ref{sec:refrigerating_multiple_qubits}, we consider two specific scenarios, where the set of central qubits is constituted of only (a) two spins when three baths are used for the design, and (b) one spin if two baths are used. The motivation behind the choice of the star-network of qubits and the corresponding $H_{\text{int}}$ lies in the permutation symmetry in the peripheral qubits, which is in alignment with the requirement of simultaneous and equal cooling of all qubits when they are chosen from the periphery. 

First, let us consider the three baths setup, where the qubits $h$ and $r$ constitute the center, while the qubits $c\equiv\{c_1,c_2,\cdots,c_n\}$ belong to the periphery (see Fig.~\ref{fig:spin_star}(a)). The interaction Hamiltonian is given by ~\cite{korepin_bogoliubov_izergin_1993,*Giamarchi2004,*Franchini2017,*fisher1964}
\begin{eqnarray}
    H_{\text{int}} &=& \sum_{\alpha=h,r}\left[\frac{J_\alpha}{4}\sum_{i=1}^n\left(\sigma^x_\alpha\sigma^x_{c_i}+\sigma^y_\alpha\sigma^y_{c_i}+\sigma^z_\alpha\sigma^z_{c_i}\right)\right],
\end{eqnarray}
where $J_\alpha$ is the exchange interaction strength between the qubit $\alpha$ ($\alpha=h,r$) in the center, and any one of the qubits $c_i$ ($i=1,\cdots,n$) in the periphery. In all our calculations in this paper, we assume ferromagnetic interaction among the qubits, i.e., $J_\alpha<0$ $\forall \alpha$. Also, in view of the fact that LQME leads to negligible cooling of the target qubits, we employ GQME for determining the dynamics of the system, and for obtaining the steady state. 

\begin{figure*}
    \centering
    \includegraphics[width=\linewidth]{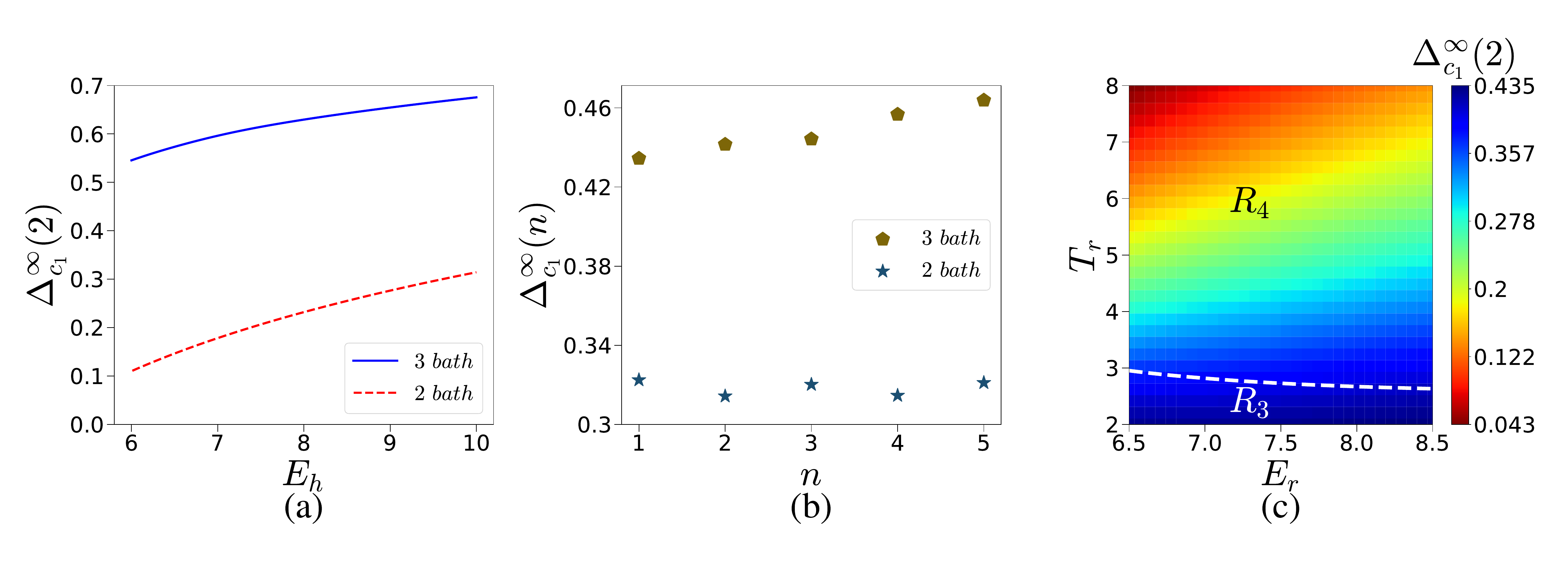}
    \caption{(a) Variations of $\Delta_{c_1}^\infty(2)$ as a function of $E_h$ in the case of the QTM designed with a star network of spins using three and two baths. The system paramteres are fixed as $J=-1$, $E_r=10$ (for the case of 3 bath), $E_{c_1}=1$, $E_{c_2}=1.01$, $E_{c_0}=0.99$ (for the case of 2 bath). (b) Variations of $\Delta_{c_1}^\infty(n)$ as a function of $n$ in the cases of three- and two-bath QTM designed to cool $n$ qubits. For 3 bath case, $E_h=4$, $E_r=10$ where as $E_h=10$ for the case of 2 baths. $E_{c_i}$'s are chosen similar to (a) and also following the discussion in Sec.~\ref{sec:spin_star}. (c) Variation of $\Delta_{c_1}^\infty(2)$ as a function of $E_r$ and $T_r$ in the case of the three-bath QTM. The white dashed line indicates the boundary between $\dot{Q}_{r}^\infty<0$ (below the line) and $>0$ (above the line, see also Table~\ref{tab:operating_regions} for definitions of $R_3$ and $R_4$). The unspecified parameters takes the value from Fig.~\ref{fig:self_contained_temp_local}. All quantities plotted are dimensionless. }
    \label{fig:spin_star_data}
\end{figure*}

To ensure there is no transition among degenerate eigen states,  we choose system parameters appropriately to avoid degeneracy in the spectrum of $H_S$. Specifically, we allow presence of disorder in the values of $E_{c_i}$, and fix them such that $E_{c_i}\neq E_{c_j}$ for $i\neq j$ and $c_i,c_j\in c$, with $|E_{c_i}-E_c|$ is small $\forall i$, where $E_c$ is fixed along with $E_h$ and $E_r$ (see Sec.~\ref{subsec:performance}). Further, we ensure that with the chosen parameter values, Eqs.~(\ref{eq:first_law}) and (\ref{eq:second_law}) remain valid.  In contrast to the case of engineered interaction Hamiltonian discussed in Sec.~\ref{subsec:performance}, in the present case, the variation of $\Delta_{c_i}^\infty(n)$ with varying $E_r$ is found to be negligible, while $E_h$ has a more prominent effect on the amount of cooling, as demonstrated for $n=2$ and $c_1$ in Fig.~\ref{fig:spin_star_data}(a) for the setup with three baths. Additionally, for all $c_i\in c$, $\Delta_{c_i}^\infty(n)$ is found to slowly increase with increasing $n$ in the case of three-bath setup as shown in Fig.~\ref{fig:spin_star_data}(b) for $c_1$ and $n=2$. This is advantageous compared to the results corresponding to the GQME with engineered interaction. Fig.~\ref{fig:spin_star_data}(c) plots $\Delta_{c_1}^{\infty}(n)$ as a function of $E_r$ and $T_r$ corresponding to the device with three baths, to reveal that a cooling for the qubit $c_1$, as indicated in terms of the local temperature defined in Sec.~\ref{subsec:performance}, is obtained over the entire $E_r-T_r$ plane, where the QTM works in two different regimes, given by $R_3$ and $R_4$ (see Table~\ref{tab:operating_regions}). This is in contrast to the case of the engineered interaction Hamiltonian, where in all cases considered so far, an operation regime $R_1$ is found where the QTM operates as a QAR and cooling of the qubits in $c$ are obtained as indicated by the local temperatures.

We further consider a  scenario where the center of the star network hosts only one qubit, labelled by $h$, while the qubits $c_0\equiv r$, $c_1,c_2,\cdots,c_n$ are now parts of the periphery. Further, the central qubit $h$ is in contact with the heat bath $B_h$, while the peripheral qubits $c_0,c_1,\cdots,c_n$  are in contact with $B_c$. The interaction Hamiltonian is given by ~\cite{korepin_bogoliubov_izergin_1993,*Giamarchi2004,*Franchini2017,*fisher1964}
\begin{eqnarray}
    H_{\text{int}} &=& \frac{J}{4}\sum_{i=0}^n\left(\sigma^x_{h}\sigma^x_{c_i}+\sigma^y_{h}\sigma^y_{c_i}+ \sigma^z_{h}\sigma^z_{c_i}\right),
\end{eqnarray}
where $J$ is the strength of the exchange interaction between the qubit $h$ and all qubits $c_i$, $i=0,1,2,\cdots,n$ including $c_0$. This is a setup comparable to the design discussed in Sec.~\ref{subsec:two_baths}, and the results regarding the variations of $\Delta_{c_i}^\infty(n)$ with  $E_h$, the landscape of  $\Delta_{c_i}^\infty(n)$ on the $E_h-T_h$ plane, and the operating regions of the designed QTM remains qualitatively the same as in the case of the three-bath model discussed above. However, in contrast to the three-bath model, $\Delta_{c_i}^\infty(n)$ for a fixed $c_i$ remains almost constant with $n$ for a fixed point in the parameter space, as illustrated in Fig.~\ref{fig:spin_star_data}(b).

\section{Conclusion}
\label{sec:conclude}

In this paper, we present the  design of a quantum thermal machine that can be used to simultaneously and dynamically cool a number of target qubits using an interaction Hamiltonian that is designed using different subspaces of the full Hilbert space of the system having specific magnetization. More specifically, we use a $(n+2)$-qubit system in contact with three bosonic heat baths in such a fashion that turning on the designed  interaction Hamiltonian makes the system evolve in time and attain a steady state in which the $n$ target qubits attains a temperature lower than their initial temperature, which is the temperature of the coldest available bath.  Using both local as well as global quantum master equations, we show that the amount of cooling obtained for the target qubits depend on the magnetization of the transport subspace using which the interaction Hamiltonian is constructed,  and there is a decrease in cooling corresponding to a specific magnetization sector with increasing the number of target qubits. We demonstrate that within the parameter space considered in this paper, use of the local quantum master equation suggests a cooling of the target qubits only when the quantum thermal machines operates as a quantum absorption refrigerator. However, when global quantum master equation is employed, our analysis reveals the possibility of the quantum thermal machine to operate outside the quantum absorption refrigerator regime, and still provide cooling of the target qubits as quantified by the distance-based definition of qubit-local temperature.  We further investigate a two-bath design of the quantum thermal machine, and find these features to remain unchanged. We also investigate the possibility to achieve simultaneous cooling of target qubits via a quantum thermal machine designed using naturally occurring spin-spin interactions such as the Heisenberg interaction, and achieve this exploiting the permutation symmetry of the peripheral spins in a star network of spins. However, unlike the engineered interaction Hamiltonian providing cooling in the operation regime of a quantum absorption refrigerator with the use of LQME, in the case of the star network of spins, use of LQME provides negligible cooling. We also demonstrate that equal cooling of the target qubits is not possible with the star network of spins, and the cooling takes place in an operating regime out of the quantum absorption  refrigerator regime.   

Our work gives rise to a number of interesting questions. The interaction Hamiltonian proposed in this paper are motivated from the development of a quantum absorption refrigerator. Outside the scope of the absorption refrigerator, in order for cooling $n$ qubits constituting the subsystem $c$, it is logical to ask whether an interaction Hamiltonian of the form $H_{\text{int}}=\mathcal{O}_{(h,r)}\otimes\left[\ket{\mathbf{0}}\bra{\mathbf{1}}+\text{h.c.}\right]_c$, with $\mathcal{O}_{(h,r)}$ being the operator having support on the qubits $h$ and $r$, and $\ket{\mathbf{0}}=\otimes_{i=1}^n\ket{0_i}$, and $\ket{\mathbf{1}}=\otimes_{i=1}^n\ket{1_i}$, can be designed for the said purpose such that the thermodynamic laws remain valid during the operation of the QTM, and what would be the possible restrictions on $\mathcal{O}_{(h,r)}$. Also, within the scope of using naturally occurring quantum spin Hamiltonian to design a refrigerator simultaneously cooling multiple qubits, it is logical to ask whether an interaction Hamiltonian can be found which provides an equal amount of cooling along with providing the cooling simultaneously.  Further, while designs for refrigerating $d$-dimensional systems have been proposed using quantum spin Hamiltonian, the hurdles in developing the architecture of quantum thermal devices for simultaneously cooling $n$ $d$-dimensional systems are yet to be addressed.    

\acknowledgements

The Authors acknowledge the use of \href{https://github.com/titaschanda/QIClib}{QIClib} -- a modern C++ library for general purpose quantum information processing and quantum computing. A.K.P acknowledges the support from the Anusandhan National Research Foundation (ANRF) of the Department of Science and Technology (DST), India, through the Core Research Grant (CRG) (File No. CRG/2023/001217, Sanction Date 16 May 2024).

\onecolumngrid 

\appendix

\section{Determination of the quantum master equation}
\label{app:qme}

Here we derive the quantum master equation corresponding to the model discussed in Sec.~\ref{sec:refrigerating_multiple_qubits}, where the qubits are designated by $\alpha=h,r,c_1\cdots c_n$. The qubits $\alpha=h,r$ are coupled respectively to the baths $B_h, B_r$, while the set of qubits $c\equiv\{c_1,c_2,\cdots,c_n\}$ are attached to the bath $B_c$ (see Fig.~\ref{fig:schematics}(a)). At $t=0$, the system-bath Hamiltonian is given by   $H_{\text{loc}}+H_{B}$, while the system is in the state $ \rho_S(0)=\bigotimes_{\alpha}\rho_{\alpha}(0),\alpha=h,r,c_1,\cdots,c_n,$ with $\rho_{\alpha}(0)=\exp -(\beta_{\alpha} H_\alpha)/\text{Tr}\left[\exp -(\beta_\alpha H_\alpha)\right]$  and $\beta_\alpha=(k_BT_\alpha)^{-1}$, $k_B$ being the Boltzmann constant and $T_\alpha$ being the absolute temperature of the bath to which the qubit is attached. The baths $B_h$, $B_r$ and $B_c$ are initialized at their thermal states  $\rho_{\text{bath}}^h$, $\rho_{\text{bath}}^r$ and $\rho_{\text{bath}}^c$ respectively. At $t>0$, $H_{\text{int}}$ describing the qubit-qubit interactions is turned on, and the von Neumann equation of motion for the closed system with Hamiltonian $H_{\text{tot}}=H_{S}+H_{B}+H_{SB}$ is given by
\begin{eqnarray}
    \frac{d\rho(t)}{dt} = -i [H_{\text{tot}},\rho(t)],
    \label{eq:vne}
\end{eqnarray}
whose solution, $\rho(t)$, is the state of the system-bath duo. Here, $H_B$ and $H_{SB}$ respectively represent the Hamiltonians describing the baths, and the system-bath interactions (see Eq.~(\ref{eq:bath_hamiltonian}) and (\ref{eq:system_bath_interaction})). We work in the interaction picture via the transformation
\begin{eqnarray}
    \mathcal{O}(t) \rightarrow \mathcal{O}^{I}(t)&=& e^{\text{i}(H_{S}+H_{B})t}  \mathcal{O}(t) e^{-\text{i}(H_{S}+H_{B})t}
    \label{eq:int_pic}
\end{eqnarray}
for all operators $\mathcal{O}$, with which the von Neumann equation assumes the form
\begin{eqnarray}\label{eq:int von neu}
    \frac{d\rho^{I}(t)}{dt} = -\text{i} \left[H_{SB}^{I},\rho^{I}(t)\right],
    \label{eq:vne_int}
\end{eqnarray}
having the solution 
\begin{eqnarray}\label{eq:soln}
    \rho^{I}(t)=\rho^{I}(0)-\text{i}\int_{0}^{t} ds \left[H^{I}_{SB}(s),\rho^{I}(s)\right].
    \label{eq:vne_int_sol}
\end{eqnarray}
Plugging the solution back into Eq.~(\ref{eq:vne_int}) results in 
\begin{eqnarray}
    \frac{d\rho^{I}(t)}{d t}= -\text{i}[H^{I}_{SB},\rho^{I}(0)] - \int_{0}^{t} ds \left[H^{I}_{SB}(t),\left[H^{I}_{SB}(s),\rho^{I}(s)\right]\right],
\end{eqnarray}
which, upon partial trace over the bath degrees of freedom, leads to  
\begin{eqnarray}\label{eq: integral DE}
    \frac{d\rho^{I}_{S}(t)}{dt}= -\text{i}\text{Tr}_{B}\left[H^{I}_{SB},\rho^{I}(0)\right] - \int_{0}^{t} ds \text{Tr}_{B}\left[H^{I}_{SB}(t),\left[H^{I}_{SB}(s),\rho^{I}(s)\right]\right].
    \label{eq:vne_int_1}
\end{eqnarray}
The first term on the R.H.S can be made zero with suitable rearrangements of $H_{\text{tot}}$. Further, we consider the initial ($t=0$) state of the system-bath duo to be $\rho^{I}(0)=\rho^{I}_{S}(0)\otimes \rho^{h,I}_{\text{bath}}(0)\otimes \rho^{r,I}_{\text{bath}}(0)\otimes \rho^{c,I}_{\text{bath}}(0)$, and a weak system-bath coupling. Assuming  $\rho^{I}(t)=\rho^{I}_{S}(t)\otimes \rho^{h,I}_{\text{bath}}(0)\otimes \rho^{r,I}_{\text{bath}}(0)\otimes \rho^{c,I}_{\text{bath}}(0)=\rho^{I}_{S}(t)\otimes \rho^{h,I}_{\text{bath}}\otimes \rho^{r,I}_{\text{bath}}\otimes \rho^{c,I}_{\text{bath}}$ (since the state of the bath is not changing with time) for $t>0$ by the Born approximation, and a quick decay of the bath correlation functions over time leading to a replacement of $\rho_{S}^{I}(s)$ with  $\rho_{S}^{I}(t)$ within the integral via the Markovian approximation, a transformation of the variables via $\tau=t-s$ changes Eq.~(\ref{eq:vne_int_1}) to the Born-Markov master equation  
\begin{eqnarray}
\label{eq:Born_Markov_eqn}
    \frac{d\rho^{I}_{S}(t)}{d t}&=& - \int_{0}^{\infty} d\tau \text{Tr}_{B}[H^I_{SB}(t),[H^I_{SB}(t-\tau),\rho^{I}_{S}(t)\otimes \rho^{I}_{\text{bath}}]],
\end{eqnarray}
where we have written $\rho^{I}_{\text{bath}}=\rho^{h,I}_{\text{bath}}\otimes \rho^{r,I}_{\text{bath}}\otimes \rho^{c,I}_{\text{bath}}$. The upper limit of the integral is set to infinity as the integral remains unchanged on account of Markovian approximation.

Let us now focus on the qubits $\alpha=h,r$ which are attached to one bath each, and define the spectrum of the system Hamiltonian as $H_{S}=\sum_{E}EP(E)$ with $P(E)=\ket{E}\bra{E}$, 
and the eigenoperators $\{\mathcal{A}(\mathcal{E})\}$ of the Hamiltonian $H_{S}$ as   
\begin{eqnarray}
\mathcal{A}(\mathcal{E})=\sum_{\underset{\mathcal{E}=E-E^\prime}{E,E^\prime}} P (E) (\sigma_{\alpha}^{+}+\sigma_{\alpha}^{-}) P(E^\prime)
\label{eq:lindblad_single_qubit}
\end{eqnarray}
satisfying $[H_{S},\mathcal{A}(\mathcal{E})] = \mathcal{E} \mathcal{A}(\mathcal{E})$, $[H_{S},\mathcal{A}^{\dagger}(\mathcal{E})] = -\mathcal{E} \mathcal{A}^{\dagger}(\mathcal{E})$ and $[H_{S},\mathcal{A}(\mathcal{E})\mathcal{A}^{\dagger}(\mathcal{E})]=0$. Note that this corresponds to a two-body interaction between system and bath (see Eq.~(\ref{eq:system_bath_interaction_ab})).  For demonstration, let us work out the quantum master equation corresponding to the situation where \emph{only} the hot qubit $h$ is connected to the hot bath $B_h$.  The system-bath interaction Hamiltonian (Eq.~(\ref{eq:system_bath_interaction_ab})), represented in terms of the eigenoperators, reads
\begin{eqnarray}\label{eq: interaction hamiltonian}
    H^h_{SB} = \sum\limits_
    {\mathcal{E} > 0}\left[\mathcal{A}_h(\mathcal{E})\otimes \mathcal{B}_h + \mathcal{A}_h^{\dagger}(\mathcal{E}) \otimes \mathcal{B}_h^{\dagger}\right],
\end{eqnarray}
where $\mathcal{B}_h=\sqrt{f_h}\sum_k d_{h,k}$, and one can determine  the eigenoperator $\mathcal{A}_h(\mathcal{E})$ by the spectral decomposition of $\sigma_h^+ + \sigma_h^-$. In the interaction picture,
\begin{eqnarray}\label{eq:int_int}
    H^{I,h}_{SB}(t) = \sum_{\mathcal{E}>0} \left[\text{e}^{\text{i}\mathcal{E}t} \mathcal{A}_h(\mathcal{E})\otimes \mathcal{B}_h(t) + \text{e}^{-\text{i} \mathcal{E} t} \mathcal{A}_h^{\dagger}(\mathcal{E}) \otimes \mathcal{B}_h^{\dagger}(t)\right], 
\end{eqnarray}
with $\mathcal{B}_h(t)=\sqrt{f_h}\sum_{k} \text{e}^{-\text{i} \omega^k_h t}  d_{h,k}$. Expanding the Born-Markov master equation (Eq.~(\ref{eq:Born_Markov_eqn})) as
\begin{eqnarray}\label{eq:born_markov_expanded_1}
    \frac{d\rho^{I}_{S}(t)}{d t}&=& \int_{0}^{\infty} d\tau \text{Tr}_{B}\left[H^{h,I}_{SB}(t-\tau)(\rho^{I}_{S}(t)\otimes \rho^{h,I}_{\text{bath}}) H^{h,I}_{SB}(t) - H^{h,I}_{SB}(t)H^{h,I}_{SB}(t-\tau ) (\rho^{I}_{S}(t)\otimes \rho^{h,I}_{\text{bath}})\right]+ h.c.,
\end{eqnarray}
and substituting for $H^{h,I}_{SB}(t)$ (see Eq.~(\ref{eq:int_int})) leads to 
\begin{eqnarray}\label{eq:born_markov expanded_2}
    \frac{d\rho^{I}_{S}(t)}{d t}&=&\sum_{\mathcal{E},\mathcal{E}^{\prime}}\int_{0}^{\infty} d\tau  \text{e}^{\text{i} (\mathcal{E}^{\prime}- \mathcal{E})t}\bigg[\mathcal{A}_h^{\dagger}(\mathcal{E})\rho^{I}_{S}(t) \mathcal{A}_h(\mathcal{E}^{\prime}) \otimes \text{e}^{\text{i} \mathcal{E} \tau}\text{Tr}\big[\mathcal{B}_h^{\dagger}(t-\tau) \rho^{h,I}_{\text{bath}}\mathcal{B}_h(t)\big]
   \nonumber\\&& -\mathcal{A}_h(\mathcal{E}^{\prime}) \mathcal{A}_h^{\dagger}(\mathcal{E})\rho^{I}_{S}(t)  \otimes \text{e}^{\text{i} \mathcal{E} \tau} \text{Tr}\big[\mathcal{B}_h(t)\mathcal{B}_h^{\dagger}(t-\tau) \rho^{h,I}_{\text{bath}}\big]\bigg]  \nonumber\\  &&+  \text{e}^{-\text{i} (\mathcal{E}^{\prime}- \mathcal{E})t}\bigg[\mathcal{A}_h(\mathcal{E})\rho^{I}_{S}(t) \mathcal{A}_h^{\dagger}(\mathcal{E}^{\prime}) \otimes \text{e}^{-\text{i} \mathcal{E} \tau}\text{Tr}\big[\mathcal{B}_h(t-\tau) \rho^{h,I}_{\text{bath}}\mathcal{B}_h^{\dagger}(t)\big] \nonumber\\ && -\mathcal{A}_h^{\dagger}(\mathcal{E}^{\prime}) \mathcal{A}_h(\mathcal{E})\rho^{I}_{S}(t)  \otimes \text{e}^{-\text{i} \mathcal{E} \tau} \text{Tr}\big[\mathcal{B}_h^{\dagger}(t)\mathcal{B}_h(t-\tau) \rho^{h,I}_{\text{bath}}\big]\bigg]   + h.c.
\end{eqnarray}
The term $\text{Tr}\left[\mathcal{B}_h(t)\mathcal{B}_h^{\dagger}(t-\tau) \rho^{h,I}_{\text{bath}}\right]$ contributes to the bath correlation function,  which can be evaluated as 
\begin{eqnarray}
    \Gamma(\mathcal{E}) &=& \int_{0}^{\infty} d \tau \text{e}^{\text{i} \mathcal{E} \tau} \text{Tr}\left[\mathcal{B}_h(t)\mathcal{B}_h^{\dagger}(t-\tau) \rho^{h}_{\text{bath}}\right]
    = f_h\sum_{k} \int_{0}^{\infty} d \tau \left[\text{e}^{\text{i} (\mathcal{E}-\omega^k_h)\tau}(1+n_h(\omega^k_h)\right],
\end{eqnarray}
where $\rho^{h,I}_{\text{bath}}=\rho^{h}_{\text{bath}}$, and $\langle d_{h,k} d^{\dagger}_{h,k^{\prime}} \rangle=\delta_{kk^{\prime}} (1+n_h(\omega^k_h))$ with $n_h(\omega^k_h)=1/\left[\exp\left(\omega^k_h/T_h\right)-1\right]$. Here, $T_h$ is the absolute temperature of the bath $h$, and $\langle d_{h,k} d^{\dagger}_{h,k^{\prime}} \rangle=\text{Tr}\left[d_{h,k} d^{\dagger}_{h, k^{\prime}}\rho^{h}_{\text{bath}}\right]$. Considering continuous bath frequency, one obtains 
\begin{eqnarray}
    \Gamma(\mathcal{E}) &=& f_h\int_{0}^{\infty}d \omega_h D( \omega_h) \int_{0}^{\infty} d \tau \text{e}^{\text{i} (\mathcal{E}-\omega_h)\tau}(1+n_h(\omega_h))\nonumber\\
    &=& f_h\int_{0}^{\infty}d \omega_h  D( \omega_h)\left[ \pi \delta(\mathcal{E}-\omega_h)+ \text{i}\frac{\mathcal{P}}{\mathcal{E} - \omega_h}\right]\left[1+n_h(\omega_h)\right],   
\end{eqnarray}
where we have used $\int_{0}^{\infty}d \tau \text{e}^{\text{i} (\mathcal{E}-\omega_h)\tau} = \pi \delta(\mathcal{E}-\omega_h)+ \text{i}\frac{\mathcal{P}}{\mathcal{E} - \omega_h}$ ~\cite{breuer2002,lidar2020lecture}, with $\mathcal{P}$ being the Cauchy principal value contributing to the Lamb shift Hamiltonian, which we neglect throughout the current study. The simplified bath correlation function can be written as 
\begin{eqnarray}
    \Gamma(\mathcal{E}) &=& \int_{0}^{\infty}d \omega_h J_h(\omega_h)(1+n_h(\omega_h))\delta(\mathcal{E}-\omega_h)
    = J_h(\mathcal{E})(1+n_h(\mathcal{E})),
\end{eqnarray}
where $J_h(\mathcal{E})=\pi f_h D(\mathcal{E})$ is the spectral density of the bath, and $n_h(\mathcal{E})=1/\left[\exp\left(\mathcal{E}/T_h\right)-1\right]$. We choose $D(\mathcal{E})$ to be $D(\mathcal{E})=\pi^{-1}\mathcal{E}\exp\left(-\mathcal{E}/\Omega_h\right)$ with $\Omega_h$ as the characteristic (cut-off) frequency of the bath $h$, so that $J_h(\mathcal{E})$ takes the form of the Ohmic spectral function~\cite{wiess2012}, given by  $J_h(\mathcal{E}) = f_h \mathcal{E} \text{e}^{-\mathcal{E}/\Omega_h}$. 
Eq.~(\ref{eq:born_markov expanded_2}) now simplifies to
\begin{eqnarray}
     \frac{d\rho^{I}_{S}(t)}{d t}&=&\sum_{\mathcal{E}, \mathcal{E}^{\prime}} \text{e}^{\text{i} (\mathcal{E}^{\prime}- \mathcal{E})t}\left[\mathcal{A}_h^{\dagger}(\mathcal{E})\rho^{I}_{S}(t) \mathcal{A}_h(\mathcal{E}^{\prime})-\mathcal{A}_h(\mathcal{E}^{\prime}) \mathcal{A}_h^{\dagger}(\mathcal{E})\rho^{I}_{S}(t) \right] J_h(\mathcal{E})\left[1+n_h(\mathcal{E})\right]   \nonumber\\  &&+  \text{e}^{-\text{i} (\mathcal{E}^{\prime}- \mathcal{E})t}\left[\mathcal{A}_h(\mathcal{E})\rho^{I}_{S}(t) \mathcal{A}_h^{\dagger}(\mathcal{E}^{\prime})-\mathcal{A}_h^{\dagger}(\mathcal{E}^{\prime}) \mathcal{A}_h(\mathcal{E})\rho^{I}_{S}(t) \right]J_h(\mathcal{E})n_h(\mathcal{E})  + h.c.,
\end{eqnarray}
which, upon application of rotating wave approximation, keeps only those terms corresponding to  $\mathcal{E}=\mathcal{E}^{\prime}$~\cite{breuer2002,lidar2020lecture} and  becomes 
\begin{eqnarray}
     \frac{d\rho^{I}_{S}(t)}{d t}&=&\sum_{\mathcal{E}>0} J_h(\mathcal{E})\left[1+n_h(\mathcal{E})\right]\left[2 \mathcal{A}_h^{\dagger}(\mathcal{E})\rho^{I}_{S}(t) \mathcal{A}_h(\mathcal{E})- \left\{\mathcal{A}_h(\mathcal{E}) \mathcal{A}_h^{\dagger}(\mathcal{E}),\rho^{I}_{S}(t) \right\} \right]    \nonumber\\  &&+ \sum_{\mathcal{E}>0}J_h(\mathcal{E})n_h(\mathcal{E}) \left[2 \mathcal{A}_h(\mathcal{E})\rho^{I}_{S}(t) \mathcal{A}_h^{\dagger}(\mathcal{E})- \left\{ \mathcal{A}_h^{\dagger}(\mathcal{E}) \mathcal{A}_h(\mathcal{E}),\rho^{I}_{S}(t)\right\} \right].
\end{eqnarray}
Reverting back to the Schrodinger picture, one can now write the Lindblad form of the quantum master equation as
\begin{eqnarray}
     \frac{d\rho_{S}(t)}{d t}
    &=&-\text{i} \left[H_{SB},\rho_{S}(t)\right]+\mathcal{D}_h\left(\rho_{S}(t)\right),
\end{eqnarray}
where the second term on the right hand side represents the dissipative term due to the connection of \emph{only} the qubit $h$ with the hot bath $B_h$, given by 
\begin{eqnarray}
\mathcal{D}_h\left(\varrho\right)&=&\sum_{\mathcal{E}>0}\bigg[ \eta_h^1(\mathcal{E}) \left[\mathcal{A}_h^{\dagger}(\mathcal{E})\varrho \mathcal{A}_h(\mathcal{E}) - \frac{1}{2}\left\{\mathcal{A}_h(\mathcal{E}) \mathcal{A}_h^{\dagger}(\mathcal{E}),\varrho \right\}\right] \nonumber \\&& +  \eta_h^2(\mathcal{E})\left[ \mathcal{A}_h(\mathcal{E})\varrho \mathcal{A}_h^{\dagger}(\mathcal{E}) - \frac{1}{2}\left\{\mathcal{A}_h^{\dagger}(\mathcal{E}) \mathcal{A}_h(\mathcal{E}),\varrho \right\}\right]\bigg]
\label{eq:dissipative_term}
\end{eqnarray}
for an arbitrary density matrix $\varrho$, with $\eta_h^1(\mathcal{E})=2J_h(\mathcal{E})\left[1+n_h(\mathcal{E})\right]$ and $\eta_h^2(\mathcal{E})=2J_h(\mathcal{E})n_h(\mathcal{E})$. 
In a similar fashion, the dissipative term $\mathcal{D}_r(.)$ corresponding to the attachment of the   qubit $r$ with the bath $B_r$ can be determined.

In the case of the set of qubits $c\equiv\{c_1,c_2,\cdots,c_n\}$ attached to the bath $B_c$,  one may consider \emph{many-body} interactions where multiple system variables interact with a single bath mode~\cite{Liao2011,Ahana2021}. Along with the two-body interactions between the system and bath degrees of freedom, in the present case, we consider  a three-body interaction  between the system and bath degrees of freedom (see Eq.~(\ref{eq:system_bath_interaction_c})). Lindblad operators for the set of qubits, $c$, corresponding to the energy gap $\mathcal{E}$ and  are constructed as
\begin{eqnarray}
\mathcal{A}_{(2)}(\mathcal{E})&=&\sum_{\underset{\mathcal{E}=E-E^\prime}{E,E^\prime}} P (E) \left[\sum_{c_i}(\sigma_{c_i}^{+}+\sigma_{c_{i}}^{-})\right] P(E^\prime),
\end{eqnarray}
and 
\begin{eqnarray}
\mathcal{A}_{(3)}(\mathcal{E})&=&\sum_{\underset{\mathcal{E}=E-E^\prime}{E,E^\prime}} P (E) \left[\sum_{\underset{i< j}{\{(c_i,c_j)\}}}(\sigma_{c_i}^{+}\sigma_{c_{j}}^{-}+\sigma_{c_i}^{-}\sigma_{c_{j}}^{+})\right] P(E^\prime),
\end{eqnarray}
for the two- and three-body interaction terms respectively. Following the footsteps of the calculation for the qubit $h$, the dissipative term due to the attachment of the set of qubits $c$ with the bath $B_c$  can be determined as 
\begin{eqnarray}
    \mathcal{D}_c(\varrho)&=&\sum_{\mathcal{E}>0}\bigg[ \eta_c^1(\mathcal{E}) \left[\mathcal{A}_{(2)}^{\dagger}(\mathcal{E})\varrho \mathcal{A}_{(2)}(\mathcal{E}) - \frac{1}{2}\left\{\mathcal{A}_{(2)}(\mathcal{E}) \mathcal{A}_{(2)}^{\dagger}(\mathcal{E}),\varrho \right\}\right] \nonumber \\&& +  \eta_c^2(\mathcal{E})\left[ \mathcal{A}_{(2)}(\mathcal{E})\varrho \mathcal{A}_{(2)}^{\dagger}(\mathcal{E}) - \frac{1}{2}\left\{\mathcal{A}_{(2)}^{\dagger}(\mathcal{E}) \mathcal{A}_{(2)}(\mathcal{E}),\varrho \right\}\right]\bigg]\nonumber\\
    &&+\kappa^2\sum_{\mathcal{E}>0}\bigg[ \eta_c^1(\mathcal{E}) \left[\mathcal{A}_{(3)}^{\dagger}(\mathcal{E})\varrho \mathcal{A}_{(3)}(\mathcal{E}) - \frac{1}{2}\left\{\mathcal{A}_{(3)}(\mathcal{E}) \mathcal{A}_{(3)}^{\dagger}(\mathcal{E}),\varrho \right\}\right] \nonumber \\&& +  \eta_c^2(\mathcal{E})\left[ \mathcal{A}_{(3)}(\mathcal{E})\varrho \mathcal{A}_{(3)}^{\dagger}(\mathcal{E}) - \frac{1}{2}\left\{\mathcal{A}_{(3)}^{\dagger}(\mathcal{E}) \mathcal{A}_{(3)}(\mathcal{E}),\varrho \right\}\right]\bigg],
\end{eqnarray}
where $\kappa$ is the relative strength of the three-body interaction compared to the two-body interaction (see Eq.~(\ref{eq:system_bath_interaction_c})). Upon determining $\mathcal{D}_\alpha(.)$, $\alpha=h,r,c$, one can write the full quantum master equation as  
\begin{eqnarray}\label{app eq: IHB}
      \frac{d\rho_{S}(t)}{d t}&=& -\text{i} \left[H_{S},\rho_{S}(t)\right]+ \sum_{i=0}^{n}\mathcal{D}_i(\rho_{S}(t)).
\end{eqnarray}
We point out here that no restrictions on the strengths of the system parameters are assumed in the derivation of the quantum master equation, which corresponds to a \emph{global} approach. One can also define a local master equation by assuming $H_S\approx H_{\text{loc}}$, and thereby determining the Lindblad operators from the eigenspectrum of $H_{\text{loc}}$. See Sec.~\ref{subsec:performance} for details. 
  
\twocolumngrid

\bibliography{ref}

\begin{thebibliography}{87}%
\makeatletter
\providecommand \@ifxundefined [1]{%
 \@ifx{#1\undefined}
}%
\providecommand \@ifnum [1]{%
 \ifnum #1\expandafter \@firstoftwo
 \else \expandafter \@secondoftwo
 \fi
}%
\providecommand \@ifx [1]{%
 \ifx #1\expandafter \@firstoftwo
 \else \expandafter \@secondoftwo
 \fi
}%
\providecommand \natexlab [1]{#1}%
\providecommand \enquote  [1]{``#1''}%
\providecommand \bibnamefont  [1]{#1}%
\providecommand \bibfnamefont [1]{#1}%
\providecommand \citenamefont [1]{#1}%
\providecommand \href@noop [0]{\@secondoftwo}%
\providecommand \href [0]{\begingroup \@sanitize@url \@href}%
\providecommand \@href[1]{\@@startlink{#1}\@@href}%
\providecommand \@@href[1]{\endgroup#1\@@endlink}%
\providecommand \@sanitize@url [0]{\catcode `\\12\catcode `\$12\catcode
  `\&12\catcode `\#12\catcode `\^12\catcode `\_12\catcode `\%12\relax}%
\providecommand \@@startlink[1]{}%
\providecommand \@@endlink[0]{}%
\providecommand \url  [0]{\begingroup\@sanitize@url \@url }%
\providecommand \@url [1]{\endgroup\@href {#1}{\urlprefix }}%
\providecommand \urlprefix  [0]{URL }%
\providecommand \Eprint [0]{\href }%
\providecommand \doibase [0]{http://dx.doi.org/}%
\providecommand \selectlanguage [0]{\@gobble}%
\providecommand \bibinfo  [0]{\@secondoftwo}%
\providecommand \bibfield  [0]{\@secondoftwo}%
\providecommand \translation [1]{[#1]}%
\providecommand \BibitemOpen [0]{}%
\providecommand \bibitemStop [0]{}%
\providecommand \bibitemNoStop [0]{.\EOS\space}%
\providecommand \EOS [0]{\spacefactor3000\relax}%
\providecommand \BibitemShut  [1]{\csname bibitem#1\endcsname}%
\let\auto@bib@innerbib\@empty
\bibitem [{\citenamefont {Bhattacharjee}\ and\ \citenamefont
  {Dutta}(2021)}]{Bhattacharjee_2021}%
  \BibitemOpen
  \bibfield  {author} {\bibinfo {author} {\bibfnamefont {S.}~\bibnamefont
  {Bhattacharjee}}\ and\ \bibinfo {author} {\bibfnamefont {A.}~\bibnamefont
  {Dutta}},\ }\href {\doibase 10.1140/epjb/s10051-021-00235-3} {\bibfield
  {journal} {\bibinfo  {journal} {The European Physical Journal B}\ }\textbf
  {\bibinfo {volume} {94}},\ \bibinfo {pages} {239} (\bibinfo {year}
  {2021})}\BibitemShut {NoStop}%
\bibitem [{\citenamefont {Linden}\ \emph {et~al.}(2010)\citenamefont {Linden},
  \citenamefont {Popescu},\ and\ \citenamefont {Skrzypczyk}}]{popescu2010}%
  \BibitemOpen
  \bibfield  {author} {\bibinfo {author} {\bibfnamefont {N.}~\bibnamefont
  {Linden}}, \bibinfo {author} {\bibfnamefont {S.}~\bibnamefont {Popescu}}, \
  and\ \bibinfo {author} {\bibfnamefont {P.}~\bibnamefont {Skrzypczyk}},\
  }\href {\doibase 10.1103/PhysRevLett.105.130401} {\bibfield  {journal}
  {\bibinfo  {journal} {Phys. Rev. Lett.}\ }\textbf {\bibinfo {volume} {105}},\
  \bibinfo {pages} {130401} (\bibinfo {year} {2010})}\BibitemShut {NoStop}%
\bibitem [{\citenamefont {Brunner}\ \emph {et~al.}(2014)\citenamefont
  {Brunner}, \citenamefont {Huber}, \citenamefont {Linden}, \citenamefont
  {Popescu}, \citenamefont {Silva},\ and\ \citenamefont
  {Skrzypczyk}}]{brunner2014}%
  \BibitemOpen
  \bibfield  {author} {\bibinfo {author} {\bibfnamefont {N.}~\bibnamefont
  {Brunner}}, \bibinfo {author} {\bibfnamefont {M.}~\bibnamefont {Huber}},
  \bibinfo {author} {\bibfnamefont {N.}~\bibnamefont {Linden}}, \bibinfo
  {author} {\bibfnamefont {S.}~\bibnamefont {Popescu}}, \bibinfo {author}
  {\bibfnamefont {R.}~\bibnamefont {Silva}}, \ and\ \bibinfo {author}
  {\bibfnamefont {P.}~\bibnamefont {Skrzypczyk}},\ }\href {\doibase
  10.1103/PhysRevE.89.032115} {\bibfield  {journal} {\bibinfo  {journal} {Phys.
  Rev. E}\ }\textbf {\bibinfo {volume} {89}},\ \bibinfo {pages} {032115}
  (\bibinfo {year} {2014})}\BibitemShut {NoStop}%
\bibitem [{\citenamefont {Skrzypczyk}\ \emph {et~al.}(2011)\citenamefont
  {Skrzypczyk}, \citenamefont {Brunner}, \citenamefont {Linden},\ and\
  \citenamefont {Popescu}}]{skrzypczyk2011}%
  \BibitemOpen
  \bibfield  {author} {\bibinfo {author} {\bibfnamefont {P.}~\bibnamefont
  {Skrzypczyk}}, \bibinfo {author} {\bibfnamefont {N.}~\bibnamefont {Brunner}},
  \bibinfo {author} {\bibfnamefont {N.}~\bibnamefont {Linden}}, \ and\ \bibinfo
  {author} {\bibfnamefont {S.}~\bibnamefont {Popescu}},\ }\href {\doibase
  10.1088/1751-8113/44/49/492002} {\bibfield  {journal} {\bibinfo  {journal}
  {Journal of Physics A: Mathematical and Theoretical}\ }\textbf {\bibinfo
  {volume} {44}},\ \bibinfo {pages} {492002} (\bibinfo {year}
  {2011})}\BibitemShut {NoStop}%
\bibitem [{\citenamefont {Brask}\ and\ \citenamefont
  {Brunner}(2015)}]{brask2015}%
  \BibitemOpen
  \bibfield  {author} {\bibinfo {author} {\bibfnamefont {J.~B.}\ \bibnamefont
  {Brask}}\ and\ \bibinfo {author} {\bibfnamefont {N.}~\bibnamefont
  {Brunner}},\ }\href {\doibase 10.1103/PhysRevE.92.062101} {\bibfield
  {journal} {\bibinfo  {journal} {Phys. Rev. E}\ }\textbf {\bibinfo {volume}
  {92}},\ \bibinfo {pages} {062101} (\bibinfo {year} {2015})}\BibitemShut
  {NoStop}%
\bibitem [{\citenamefont {Levy}\ and\ \citenamefont
  {Kosloff}(2012)}]{levy2012}%
  \BibitemOpen
  \bibfield  {author} {\bibinfo {author} {\bibfnamefont {A.}~\bibnamefont
  {Levy}}\ and\ \bibinfo {author} {\bibfnamefont {R.}~\bibnamefont {Kosloff}},\
  }\href {\doibase 10.1103/PhysRevLett.108.070604} {\bibfield  {journal}
  {\bibinfo  {journal} {Phys. Rev. Lett.}\ }\textbf {\bibinfo {volume} {108}},\
  \bibinfo {pages} {070604} (\bibinfo {year} {2012})}\BibitemShut {NoStop}%
\bibitem [{\citenamefont {Correa}\ \emph {et~al.}(2014)\citenamefont {Correa},
  \citenamefont {Palao}, \citenamefont {Alonso},\ and\ \citenamefont
  {Adesso}}]{correa2014}%
  \BibitemOpen
  \bibfield  {author} {\bibinfo {author} {\bibfnamefont {L.~A.}\ \bibnamefont
  {Correa}}, \bibinfo {author} {\bibfnamefont {J.}~\bibnamefont {Palao}},
  \bibinfo {author} {\bibfnamefont {D.}~\bibnamefont {Alonso}}, \ and\ \bibinfo
  {author} {\bibfnamefont {G.}~\bibnamefont {Adesso}},\ }\href {\doibase
  10.1038/srep03949} {\bibfield  {journal} {\bibinfo  {journal} {Scientific
  Reports}\ }\textbf {\bibinfo {volume} {4}},\ \bibinfo {pages} {3949}
  (\bibinfo {year} {2014})}\BibitemShut {NoStop}%
\bibitem [{\citenamefont {Wang}\ \emph {et~al.}(2015)\citenamefont {Wang},
  \citenamefont {Lai}, \citenamefont {Ye}, \citenamefont {He}, \citenamefont
  {Ma},\ and\ \citenamefont {Liao}}]{wang2015}%
  \BibitemOpen
  \bibfield  {author} {\bibinfo {author} {\bibfnamefont {J.}~\bibnamefont
  {Wang}}, \bibinfo {author} {\bibfnamefont {Y.}~\bibnamefont {Lai}}, \bibinfo
  {author} {\bibfnamefont {Z.}~\bibnamefont {Ye}}, \bibinfo {author}
  {\bibfnamefont {J.}~\bibnamefont {He}}, \bibinfo {author} {\bibfnamefont
  {Y.}~\bibnamefont {Ma}}, \ and\ \bibinfo {author} {\bibfnamefont
  {Q.}~\bibnamefont {Liao}},\ }\href {\doibase 10.1103/PhysRevE.91.050102}
  {\bibfield  {journal} {\bibinfo  {journal} {Phys. Rev. E}\ }\textbf {\bibinfo
  {volume} {91}},\ \bibinfo {pages} {050102} (\bibinfo {year}
  {2015})}\BibitemShut {NoStop}%
\bibitem [{\citenamefont {Mu}\ \emph {et~al.}(2017)\citenamefont {Mu},
  \citenamefont {Agarwalla}, \citenamefont {Schaller},\ and\ \citenamefont
  {Segal}}]{Mu_2017}%
  \BibitemOpen
  \bibfield  {author} {\bibinfo {author} {\bibfnamefont {A.}~\bibnamefont
  {Mu}}, \bibinfo {author} {\bibfnamefont {B.~K.}\ \bibnamefont {Agarwalla}},
  \bibinfo {author} {\bibfnamefont {G.}~\bibnamefont {Schaller}}, \ and\
  \bibinfo {author} {\bibfnamefont {D.}~\bibnamefont {Segal}},\ }\href
  {\doibase 10.1088/1367-2630/aa9b75} {\bibfield  {journal} {\bibinfo
  {journal} {New Journal of Physics}\ }\textbf {\bibinfo {volume} {19}},\
  \bibinfo {pages} {123034} (\bibinfo {year} {2017})}\BibitemShut {NoStop}%
\bibitem [{\citenamefont {Nimmrichter}\ \emph {et~al.}(2017)\citenamefont
  {Nimmrichter}, \citenamefont {Dai}, \citenamefont {Roulet},\ and\
  \citenamefont {Scarani}}]{Nimmrichter2017}%
  \BibitemOpen
  \bibfield  {author} {\bibinfo {author} {\bibfnamefont {S.}~\bibnamefont
  {Nimmrichter}}, \bibinfo {author} {\bibfnamefont {J.}~\bibnamefont {Dai}},
  \bibinfo {author} {\bibfnamefont {A.}~\bibnamefont {Roulet}}, \ and\ \bibinfo
  {author} {\bibfnamefont {V.}~\bibnamefont {Scarani}},\ }\href {\doibase
  10.22331/q-2017-12-11-37} {\bibfield  {journal} {\bibinfo  {journal}
  {{Quantum}}\ }\textbf {\bibinfo {volume} {1}},\ \bibinfo {pages} {37}
  (\bibinfo {year} {2017})}\BibitemShut {NoStop}%
\bibitem [{\citenamefont {Nimmrichter}\ \emph {et~al.}(2018)\citenamefont
  {Nimmrichter}, \citenamefont {Roulet},\ and\ \citenamefont
  {Scarani}}]{nimmrichter2018}%
  \BibitemOpen
  \bibfield  {author} {\bibinfo {author} {\bibfnamefont {S.}~\bibnamefont
  {Nimmrichter}}, \bibinfo {author} {\bibfnamefont {A.}~\bibnamefont {Roulet}},
  \ and\ \bibinfo {author} {\bibfnamefont {V.}~\bibnamefont {Scarani}},\
  }\enquote {\bibinfo {title} {Quantum rotor engines},}\ in\ \href {\doibase
  10.1007/978-3-319-99046-0_6} {\emph {\bibinfo {booktitle} {Thermodynamics in
  the Quantum Regime: Fundamental Aspects and New Directions}}},\ \bibinfo
  {editor} {edited by\ \bibinfo {editor} {\bibfnamefont {F.}~\bibnamefont
  {Binder}}, \bibinfo {editor} {\bibfnamefont {L.~A.}\ \bibnamefont {Correa}},
  \bibinfo {editor} {\bibfnamefont {C.}~\bibnamefont {Gogolin}}, \bibinfo
  {editor} {\bibfnamefont {J.}~\bibnamefont {Anders}}, \ and\ \bibinfo {editor}
  {\bibfnamefont {G.}~\bibnamefont {Adesso}}}\ (\bibinfo  {publisher} {Springer
  International Publishing},\ \bibinfo {address} {Cham},\ \bibinfo {year}
  {2018})\ pp.\ \bibinfo {pages} {227--245}\BibitemShut {NoStop}%
\bibitem [{\citenamefont {Mitchison}(2019)}]{mitchison2019}%
  \BibitemOpen
  \bibfield  {author} {\bibinfo {author} {\bibfnamefont {M.~T.}\ \bibnamefont
  {Mitchison}},\ }\href {\doibase 10.1080/00107514.2019.1631555} {\bibfield
  {journal} {\bibinfo  {journal} {Contemporary Physics}\ }\textbf {\bibinfo
  {volume} {60}},\ \bibinfo {pages} {164} (\bibinfo {year} {2019})}\BibitemShut
  {NoStop}%
\bibitem [{\citenamefont {Mitchison}\ \emph {et~al.}(2015)\citenamefont
  {Mitchison}, \citenamefont {Woods}, \citenamefont {Prior},\ and\
  \citenamefont {Huber}}]{mitchison2015}%
  \BibitemOpen
  \bibfield  {author} {\bibinfo {author} {\bibfnamefont {M.~T.}\ \bibnamefont
  {Mitchison}}, \bibinfo {author} {\bibfnamefont {M.~P.}\ \bibnamefont
  {Woods}}, \bibinfo {author} {\bibfnamefont {J.}~\bibnamefont {Prior}}, \ and\
  \bibinfo {author} {\bibfnamefont {M.}~\bibnamefont {Huber}},\ }\href
  {\doibase 10.1088/1367-2630/17/11/115013} {\bibfield  {journal} {\bibinfo
  {journal} {New Journal of Physics}\ }\textbf {\bibinfo {volume} {17}},\
  \bibinfo {pages} {115013} (\bibinfo {year} {2015})}\BibitemShut {NoStop}%
\bibitem [{\citenamefont {Das}\ \emph {et~al.}(2019)\citenamefont {Das},
  \citenamefont {Misra}, \citenamefont {Pal}, \citenamefont {Sen(De)},\ and\
  \citenamefont {Sen}}]{das2019}%
  \BibitemOpen
  \bibfield  {author} {\bibinfo {author} {\bibfnamefont {S.}~\bibnamefont
  {Das}}, \bibinfo {author} {\bibfnamefont {A.}~\bibnamefont {Misra}}, \bibinfo
  {author} {\bibfnamefont {A.~K.}\ \bibnamefont {Pal}}, \bibinfo {author}
  {\bibfnamefont {A.}~\bibnamefont {Sen(De)}}, \ and\ \bibinfo {author}
  {\bibfnamefont {U.}~\bibnamefont {Sen}},\ }\href {\doibase
  10.1209/0295-5075/125/20007} {\bibfield  {journal} {\bibinfo  {journal}
  {{EPL} (Europhysics Letters)}\ }\textbf {\bibinfo {volume} {125}},\ \bibinfo
  {pages} {20007} (\bibinfo {year} {2019})}\BibitemShut {NoStop}%
\bibitem [{\citenamefont {Ghoshal}\ \emph {et~al.}(2021)\citenamefont
  {Ghoshal}, \citenamefont {Das}, \citenamefont {Pal}, \citenamefont
  {Sen(De)},\ and\ \citenamefont {Sen}}]{Ahana2021}%
  \BibitemOpen
  \bibfield  {author} {\bibinfo {author} {\bibfnamefont {A.}~\bibnamefont
  {Ghoshal}}, \bibinfo {author} {\bibfnamefont {S.}~\bibnamefont {Das}},
  \bibinfo {author} {\bibfnamefont {A.~K.}\ \bibnamefont {Pal}}, \bibinfo
  {author} {\bibfnamefont {A.}~\bibnamefont {Sen(De)}}, \ and\ \bibinfo
  {author} {\bibfnamefont {U.}~\bibnamefont {Sen}},\ }\href {\doibase
  10.1103/PhysRevA.104.042208} {\bibfield  {journal} {\bibinfo  {journal}
  {Phys. Rev. A}\ }\textbf {\bibinfo {volume} {104}},\ \bibinfo {pages}
  {042208} (\bibinfo {year} {2021})}\BibitemShut {NoStop}%
\bibitem [{\citenamefont {Hewgill}\ \emph {et~al.}(2020)\citenamefont
  {Hewgill}, \citenamefont {Gonz\'alez}, \citenamefont {Palao}, \citenamefont
  {Alonso}, \citenamefont {Ferraro},\ and\ \citenamefont
  {De~Chiara}}]{hewgill2020}%
  \BibitemOpen
  \bibfield  {author} {\bibinfo {author} {\bibfnamefont {A.}~\bibnamefont
  {Hewgill}}, \bibinfo {author} {\bibfnamefont {J.~O.}\ \bibnamefont
  {Gonz\'alez}}, \bibinfo {author} {\bibfnamefont {J.~P.}\ \bibnamefont
  {Palao}}, \bibinfo {author} {\bibfnamefont {D.}~\bibnamefont {Alonso}},
  \bibinfo {author} {\bibfnamefont {A.}~\bibnamefont {Ferraro}}, \ and\
  \bibinfo {author} {\bibfnamefont {G.}~\bibnamefont {De~Chiara}},\ }\href
  {\doibase 10.1103/PhysRevE.101.012109} {\bibfield  {journal} {\bibinfo
  {journal} {Phys. Rev. E}\ }\textbf {\bibinfo {volume} {101}},\ \bibinfo
  {pages} {012109} (\bibinfo {year} {2020})}\BibitemShut {NoStop}%
\bibitem [{\citenamefont {Konar}\ \emph
  {et~al.}(2022{\natexlab{a}})\citenamefont {Konar}, \citenamefont {Ghosh},
  \citenamefont {Pal},\ and\ \citenamefont {Sen(De)}}]{konarghosh22}%
  \BibitemOpen
  \bibfield  {author} {\bibinfo {author} {\bibfnamefont {T.~K.}\ \bibnamefont
  {Konar}}, \bibinfo {author} {\bibfnamefont {S.}~\bibnamefont {Ghosh}},
  \bibinfo {author} {\bibfnamefont {A.~K.}\ \bibnamefont {Pal}}, \ and\
  \bibinfo {author} {\bibfnamefont {A.}~\bibnamefont {Sen(De)}},\ }\href
  {\doibase 10.1103/PhysRevA.105.022214} {\bibfield  {journal} {\bibinfo
  {journal} {Phys. Rev. A}\ }\textbf {\bibinfo {volume} {105}},\ \bibinfo
  {pages} {022214} (\bibinfo {year} {2022}{\natexlab{a}})}\BibitemShut
  {NoStop}%
\bibitem [{\citenamefont {Konar}\ \emph
  {et~al.}(2023{\natexlab{a}})\citenamefont {Konar}, \citenamefont {Ghosh},
  \citenamefont {Pal},\ and\ \citenamefont {Sen(De)}}]{konar2023}%
  \BibitemOpen
  \bibfield  {author} {\bibinfo {author} {\bibfnamefont {T.~K.}\ \bibnamefont
  {Konar}}, \bibinfo {author} {\bibfnamefont {S.}~\bibnamefont {Ghosh}},
  \bibinfo {author} {\bibfnamefont {A.~K.}\ \bibnamefont {Pal}}, \ and\
  \bibinfo {author} {\bibfnamefont {A.}~\bibnamefont {Sen(De)}},\ }\href
  {\doibase 10.1103/PhysRevA.107.032602} {\bibfield  {journal} {\bibinfo
  {journal} {Phys. Rev. A}\ }\textbf {\bibinfo {volume} {107}},\ \bibinfo
  {pages} {032602} (\bibinfo {year} {2023}{\natexlab{a}})}\BibitemShut
  {NoStop}%
\bibitem [{\citenamefont {Yan}\ and\ \citenamefont {Jing}(2021)}]{Yan2021}%
  \BibitemOpen
  \bibfield  {author} {\bibinfo {author} {\bibfnamefont {J.-s.}\ \bibnamefont
  {Yan}}\ and\ \bibinfo {author} {\bibfnamefont {J.}~\bibnamefont {Jing}},\
  }\href {\doibase 10.1103/PhysRevA.104.063105} {\bibfield  {journal} {\bibinfo
   {journal} {Phys. Rev. A}\ }\textbf {\bibinfo {volume} {104}},\ \bibinfo
  {pages} {063105} (\bibinfo {year} {2021})}\BibitemShut {NoStop}%
\bibitem [{\citenamefont {Yan}\ and\ \citenamefont {Jing}(2022)}]{Yan2022}%
  \BibitemOpen
  \bibfield  {author} {\bibinfo {author} {\bibfnamefont {J.-s.}\ \bibnamefont
  {Yan}}\ and\ \bibinfo {author} {\bibfnamefont {J.}~\bibnamefont {Jing}},\
  }\href {\doibase 10.1103/PhysRevA.105.052607} {\bibfield  {journal} {\bibinfo
   {journal} {Phys. Rev. A}\ }\textbf {\bibinfo {volume} {105}},\ \bibinfo
  {pages} {052607} (\bibinfo {year} {2022})}\BibitemShut {NoStop}%
\bibitem [{\citenamefont {Konar}\ \emph
  {et~al.}(2022{\natexlab{b}})\citenamefont {Konar}, \citenamefont {Ghosh},\
  and\ \citenamefont {Sen(De)}}]{Konar2022}%
  \BibitemOpen
  \bibfield  {author} {\bibinfo {author} {\bibfnamefont {T.~K.}\ \bibnamefont
  {Konar}}, \bibinfo {author} {\bibfnamefont {S.}~\bibnamefont {Ghosh}}, \ and\
  \bibinfo {author} {\bibfnamefont {A.}~\bibnamefont {Sen(De)}},\ }\href
  {\doibase 10.1103/PhysRevA.106.022616} {\bibfield  {journal} {\bibinfo
  {journal} {Phys. Rev. A}\ }\textbf {\bibinfo {volume} {106}},\ \bibinfo
  {pages} {022616} (\bibinfo {year} {2022}{\natexlab{b}})}\BibitemShut
  {NoStop}%
\bibitem [{\citenamefont {Ghosh}\ \emph {et~al.}(2024)\citenamefont {Ghosh},
  \citenamefont {Konar},\ and\ \citenamefont {Sen(De)}}]{Ghosh2024}%
  \BibitemOpen
  \bibfield  {author} {\bibinfo {author} {\bibfnamefont {D.}~\bibnamefont
  {Ghosh}}, \bibinfo {author} {\bibfnamefont {T.~K.}\ \bibnamefont {Konar}}, \
  and\ \bibinfo {author} {\bibfnamefont {A.}~\bibnamefont {Sen(De)}},\ }\href
  {https://arxiv.org/abs/2409.08375v1} {\bibfield  {journal} {\bibinfo
  {journal} {arXiv:2409.08375}\ } (\bibinfo {year} {2024})}\BibitemShut
  {NoStop}%
\bibitem [{\citenamefont {Allahverdyan}\ \emph {et~al.}(2004)\citenamefont
  {Allahverdyan}, \citenamefont {Balian},\ and\ \citenamefont
  {Nieuwenhuizen}}]{Allahverdyan_2004}%
  \BibitemOpen
  \bibfield  {author} {\bibinfo {author} {\bibfnamefont {A.~E.}\ \bibnamefont
  {Allahverdyan}}, \bibinfo {author} {\bibfnamefont {R.}~\bibnamefont
  {Balian}}, \ and\ \bibinfo {author} {\bibfnamefont {T.~M.}\ \bibnamefont
  {Nieuwenhuizen}},\ }\href {\doibase 10.1209/epl/i2004-10101-2} {\bibfield
  {journal} {\bibinfo  {journal} {Europhysics Letters}\ }\textbf {\bibinfo
  {volume} {67}},\ \bibinfo {pages} {565} (\bibinfo {year} {2004})}\BibitemShut
  {NoStop}%
\bibitem [{\citenamefont {Correa}\ \emph {et~al.}(2013)\citenamefont {Correa},
  \citenamefont {Palao}, \citenamefont {Adesso},\ and\ \citenamefont
  {Alonso}}]{correa2013}%
  \BibitemOpen
  \bibfield  {author} {\bibinfo {author} {\bibfnamefont {L.~A.}\ \bibnamefont
  {Correa}}, \bibinfo {author} {\bibfnamefont {J.~P.}\ \bibnamefont {Palao}},
  \bibinfo {author} {\bibfnamefont {G.}~\bibnamefont {Adesso}}, \ and\ \bibinfo
  {author} {\bibfnamefont {D.}~\bibnamefont {Alonso}},\ }\href {\doibase
  10.1103/PhysRevE.87.042131} {\bibfield  {journal} {\bibinfo  {journal} {Phys.
  Rev. E}\ }\textbf {\bibinfo {volume} {87}},\ \bibinfo {pages} {042131}
  (\bibinfo {year} {2013})}\BibitemShut {NoStop}%
\bibitem [{\citenamefont {Alicki}\ and\ \citenamefont
  {Fannes}(2013)}]{alicki2013}%
  \BibitemOpen
  \bibfield  {author} {\bibinfo {author} {\bibfnamefont {R.}~\bibnamefont
  {Alicki}}\ and\ \bibinfo {author} {\bibfnamefont {M.}~\bibnamefont
  {Fannes}},\ }\href {\doibase 10.1103/PhysRevE.87.042123} {\bibfield
  {journal} {\bibinfo  {journal} {Phys. Rev. E}\ }\textbf {\bibinfo {volume}
  {87}},\ \bibinfo {pages} {042123} (\bibinfo {year} {2013})}\BibitemShut
  {NoStop}%
\bibitem [{\citenamefont {Campaioli}\ \emph {et~al.}(2018)\citenamefont
  {Campaioli}, \citenamefont {Pollock},\ and\ \citenamefont
  {Vinjanampathy}}]{Campaioli2018}%
  \BibitemOpen
  \bibfield  {author} {\bibinfo {author} {\bibfnamefont {F.}~\bibnamefont
  {Campaioli}}, \bibinfo {author} {\bibfnamefont {F.~A.}\ \bibnamefont
  {Pollock}}, \ and\ \bibinfo {author} {\bibfnamefont {S.}~\bibnamefont
  {Vinjanampathy}},\ }\enquote {\bibinfo {title} {Quantum batteries},}\ in\
  \href {\doibase 10.1007/978-3-319-99046-0_8} {\emph {\bibinfo {booktitle}
  {Thermodynamics in the Quantum Regime: Fundamental Aspects and New
  Directions}}},\ \bibinfo {editor} {edited by\ \bibinfo {editor}
  {\bibfnamefont {F.}~\bibnamefont {Binder}}, \bibinfo {editor} {\bibfnamefont
  {L.~A.}\ \bibnamefont {Correa}}, \bibinfo {editor} {\bibfnamefont
  {C.}~\bibnamefont {Gogolin}}, \bibinfo {editor} {\bibfnamefont
  {J.}~\bibnamefont {Anders}}, \ and\ \bibinfo {editor} {\bibfnamefont
  {G.}~\bibnamefont {Adesso}}}\ (\bibinfo  {publisher} {Springer International
  Publishing},\ \bibinfo {address} {Cham},\ \bibinfo {year} {2018})\ pp.\
  \bibinfo {pages} {207--225}\BibitemShut {NoStop}%
\bibitem [{\citenamefont {Shi}\ \emph {et~al.}(2022)\citenamefont {Shi},
  \citenamefont {Ding}, \citenamefont {Wan}, \citenamefont {Wang},\ and\
  \citenamefont {Yang}}]{shi2022}%
  \BibitemOpen
  \bibfield  {author} {\bibinfo {author} {\bibfnamefont {H.-L.}\ \bibnamefont
  {Shi}}, \bibinfo {author} {\bibfnamefont {S.}~\bibnamefont {Ding}}, \bibinfo
  {author} {\bibfnamefont {Q.-K.}\ \bibnamefont {Wan}}, \bibinfo {author}
  {\bibfnamefont {X.-H.}\ \bibnamefont {Wang}}, \ and\ \bibinfo {author}
  {\bibfnamefont {W.-L.}\ \bibnamefont {Yang}},\ }\href {\doibase
  10.1103/PhysRevLett.129.130602} {\bibfield  {journal} {\bibinfo  {journal}
  {Phys. Rev. Lett.}\ }\textbf {\bibinfo {volume} {129}},\ \bibinfo {pages}
  {130602} (\bibinfo {year} {2022})}\BibitemShut {NoStop}%
\bibitem [{\citenamefont {Joulain}\ \emph {et~al.}(2016)\citenamefont
  {Joulain}, \citenamefont {Drevillon}, \citenamefont {Ezzahri},\ and\
  \citenamefont {Ordonez-Miranda}}]{joulain2016}%
  \BibitemOpen
  \bibfield  {author} {\bibinfo {author} {\bibfnamefont {K.}~\bibnamefont
  {Joulain}}, \bibinfo {author} {\bibfnamefont {J.}~\bibnamefont {Drevillon}},
  \bibinfo {author} {\bibfnamefont {Y.}~\bibnamefont {Ezzahri}}, \ and\
  \bibinfo {author} {\bibfnamefont {J.}~\bibnamefont {Ordonez-Miranda}},\
  }\href {\doibase 10.1103/PhysRevLett.116.200601} {\bibfield  {journal}
  {\bibinfo  {journal} {Phys. Rev. Lett.}\ }\textbf {\bibinfo {volume} {116}},\
  \bibinfo {pages} {200601} (\bibinfo {year} {2016})}\BibitemShut {NoStop}%
\bibitem [{\citenamefont {Mandarino}\ \emph {et~al.}(2021)\citenamefont
  {Mandarino}, \citenamefont {Joulain}, \citenamefont {G\'omez},\ and\
  \citenamefont {Bellomo}}]{mandarino2021}%
  \BibitemOpen
  \bibfield  {author} {\bibinfo {author} {\bibfnamefont {A.}~\bibnamefont
  {Mandarino}}, \bibinfo {author} {\bibfnamefont {K.}~\bibnamefont {Joulain}},
  \bibinfo {author} {\bibfnamefont {M.~D.}\ \bibnamefont {G\'omez}}, \ and\
  \bibinfo {author} {\bibfnamefont {B.}~\bibnamefont {Bellomo}},\ }\href
  {\doibase 10.1103/PhysRevApplied.16.034026} {\bibfield  {journal} {\bibinfo
  {journal} {Phys. Rev. Appl.}\ }\textbf {\bibinfo {volume} {16}},\ \bibinfo
  {pages} {034026} (\bibinfo {year} {2021})}\BibitemShut {NoStop}%
\bibitem [{\citenamefont {Goold}\ \emph {et~al.}(2016)\citenamefont {Goold},
  \citenamefont {Huber}, \citenamefont {Riera}, \citenamefont {del Rio},\ and\
  \citenamefont {Skrzypczyk}}]{goold2016}%
  \BibitemOpen
  \bibfield  {author} {\bibinfo {author} {\bibfnamefont {J.}~\bibnamefont
  {Goold}}, \bibinfo {author} {\bibfnamefont {M.}~\bibnamefont {Huber}},
  \bibinfo {author} {\bibfnamefont {A.}~\bibnamefont {Riera}}, \bibinfo
  {author} {\bibfnamefont {L.}~\bibnamefont {del Rio}}, \ and\ \bibinfo
  {author} {\bibfnamefont {P.}~\bibnamefont {Skrzypczyk}},\ }\href {\doibase
  10.1088/1751-8113/49/14/143001} {\bibfield  {journal} {\bibinfo  {journal}
  {Journal of Physics A: Mathematical and Theoretical}\ }\textbf {\bibinfo
  {volume} {49}},\ \bibinfo {pages} {143001} (\bibinfo {year}
  {2016})}\BibitemShut {NoStop}%
\bibitem [{\citenamefont {Vinjanampathy}\ and\ \citenamefont
  {Anders}(2016)}]{vinjanampathy2016}%
  \BibitemOpen
  \bibfield  {author} {\bibinfo {author} {\bibfnamefont {S.}~\bibnamefont
  {Vinjanampathy}}\ and\ \bibinfo {author} {\bibfnamefont {J.}~\bibnamefont
  {Anders}},\ }\href {\doibase 10.1080/00107514.2016.1201896} {\bibfield
  {journal} {\bibinfo  {journal} {Contemporary Physics}\ }\textbf {\bibinfo
  {volume} {57}},\ \bibinfo {pages} {545} (\bibinfo {year} {2016})}\BibitemShut
  {NoStop}%
\bibitem [{\citenamefont {Kosloff}(2013)}]{kosloff2013}%
  \BibitemOpen
  \bibfield  {author} {\bibinfo {author} {\bibfnamefont {R.}~\bibnamefont
  {Kosloff}},\ }\href {\doibase 10.3390/e15062100} {\bibfield  {journal}
  {\bibinfo  {journal} {Entropy}\ }\textbf {\bibinfo {volume} {15}},\ \bibinfo
  {pages} {2100} (\bibinfo {year} {2013})}\BibitemShut {NoStop}%
\bibitem [{\citenamefont {Myers}\ \emph {et~al.}(2022)\citenamefont {Myers},
  \citenamefont {Abah},\ and\ \citenamefont {Deffner}}]{Myers2022}%
  \BibitemOpen
  \bibfield  {author} {\bibinfo {author} {\bibfnamefont {N.~M.}\ \bibnamefont
  {Myers}}, \bibinfo {author} {\bibfnamefont {O.}~\bibnamefont {Abah}}, \ and\
  \bibinfo {author} {\bibfnamefont {S.}~\bibnamefont {Deffner}},\ }\href
  {\doibase 10.1116/5.0083192} {\bibfield  {journal} {\bibinfo  {journal} {AVS
  Quantum Sci}\ }\textbf {\bibinfo {volume} {4 (2)}},\ \bibinfo {pages}
  {027101} (\bibinfo {year} {2022})}\BibitemShut {NoStop}%
\bibitem [{\citenamefont {Ikonen}\ \emph {et~al.}(2017)\citenamefont {Ikonen},
  \citenamefont {Salmilehto},\ and\ \citenamefont
  {M{\"o}tt{\"o}nen}}]{ikonen2017}%
  \BibitemOpen
  \bibfield  {author} {\bibinfo {author} {\bibfnamefont {J.}~\bibnamefont
  {Ikonen}}, \bibinfo {author} {\bibfnamefont {J.}~\bibnamefont {Salmilehto}},
  \ and\ \bibinfo {author} {\bibfnamefont {M.}~\bibnamefont
  {M{\"o}tt{\"o}nen}},\ }\href {\doibase 10.1038/s41534-017-0015-5} {\bibfield
  {journal} {\bibinfo  {journal} {npj Quantum Information}\ }\textbf {\bibinfo
  {volume} {3}},\ \bibinfo {pages} {17} (\bibinfo {year} {2017})}\BibitemShut
  {NoStop}%
\bibitem [{\citenamefont {Allahverdyan}\ and\ \citenamefont
  {Nieuwenhuizen}(2000)}]{allahverdyan2000}%
  \BibitemOpen
  \bibfield  {author} {\bibinfo {author} {\bibfnamefont {A.~E.}\ \bibnamefont
  {Allahverdyan}}\ and\ \bibinfo {author} {\bibfnamefont {T.~M.}\ \bibnamefont
  {Nieuwenhuizen}},\ }\href {\doibase 10.1103/PhysRevLett.85.1799} {\bibfield
  {journal} {\bibinfo  {journal} {Phys. Rev. Lett.}\ }\textbf {\bibinfo
  {volume} {85}},\ \bibinfo {pages} {1799} (\bibinfo {year}
  {2000})}\BibitemShut {NoStop}%
\bibitem [{\citenamefont {Brand{\~a}o}\ \emph {et~al.}(2015)\citenamefont
  {Brand{\~a}o}, \citenamefont {Horodecki}, \citenamefont {Ng}, \citenamefont
  {Oppenheim},\ and\ \citenamefont {Wehner}}]{brandao2015}%
  \BibitemOpen
  \bibfield  {author} {\bibinfo {author} {\bibfnamefont {F.}~\bibnamefont
  {Brand{\~a}o}}, \bibinfo {author} {\bibfnamefont {M.}~\bibnamefont
  {Horodecki}}, \bibinfo {author} {\bibfnamefont {N.}~\bibnamefont {Ng}},
  \bibinfo {author} {\bibfnamefont {J.}~\bibnamefont {Oppenheim}}, \ and\
  \bibinfo {author} {\bibfnamefont {S.}~\bibnamefont {Wehner}},\ }\href
  {\doibase 10.1073/pnas.1411728112} {\bibfield  {journal} {\bibinfo  {journal}
  {Proceedings of the National Academy of Sciences}\ }\textbf {\bibinfo
  {volume} {112}},\ \bibinfo {pages} {3275} (\bibinfo {year}
  {2015})}\BibitemShut {NoStop}%
\bibitem [{\citenamefont {Gardas}\ and\ \citenamefont
  {Deffner}(2015)}]{gardas2015}%
  \BibitemOpen
  \bibfield  {author} {\bibinfo {author} {\bibfnamefont {B.}~\bibnamefont
  {Gardas}}\ and\ \bibinfo {author} {\bibfnamefont {S.}~\bibnamefont
  {Deffner}},\ }\href {\doibase 10.1103/PhysRevE.92.042126} {\bibfield
  {journal} {\bibinfo  {journal} {Phys. Rev. E}\ }\textbf {\bibinfo {volume}
  {92}},\ \bibinfo {pages} {042126} (\bibinfo {year} {2015})}\BibitemShut
  {NoStop}%
\bibitem [{\citenamefont {Deffner}\ and\ \citenamefont
  {Campbell}(2019)}]{deffner2019}%
  \BibitemOpen
  \bibfield  {author} {\bibinfo {author} {\bibfnamefont {S.}~\bibnamefont
  {Deffner}}\ and\ \bibinfo {author} {\bibfnamefont {S.}~\bibnamefont
  {Campbell}},\ }\href {\doibase 10.1088/2053-2571/ab21c6} {\emph {\bibinfo
  {title} {Quantum Thermodynamics}}},\ 2053-2571\ (\bibinfo  {publisher}
  {Morgan and Claypool Publishers},\ \bibinfo {year} {2019})\BibitemShut
  {NoStop}%
\bibitem [{\citenamefont {Gour}\ \emph {et~al.}(2015)\citenamefont {Gour},
  \citenamefont {M\"{u}ller}, \citenamefont {Narasimhachar}, \citenamefont
  {Spekkens},\ and\ \citenamefont {Halpern}}]{gour2015}%
  \BibitemOpen
  \bibfield  {author} {\bibinfo {author} {\bibfnamefont {G.}~\bibnamefont
  {Gour}}, \bibinfo {author} {\bibfnamefont {M.~P.}\ \bibnamefont
  {M\"{u}ller}}, \bibinfo {author} {\bibfnamefont {V.}~\bibnamefont
  {Narasimhachar}}, \bibinfo {author} {\bibfnamefont {R.~W.}\ \bibnamefont
  {Spekkens}}, \ and\ \bibinfo {author} {\bibfnamefont {N.~Y.}\ \bibnamefont
  {Halpern}},\ }\href {\doibase 10.1016/j.physrep.2015.04.003} {\bibfield
  {journal} {\bibinfo  {journal} {Phys. Rep.}\ }\textbf {\bibinfo {volume}
  {583}},\ \bibinfo {pages} {1} (\bibinfo {year} {2015})}\BibitemShut {NoStop}%
\bibitem [{\citenamefont {Huber}\ \emph {et~al.}(2015)\citenamefont {Huber},
  \citenamefont {Perarnau-Llobet}, \citenamefont {Hovhannisyan}, \citenamefont
  {Skrzypczyk}, \citenamefont {Klöckl}, \citenamefont {Brunner},\ and\
  \citenamefont {Ac{\'{\i}}n}}]{huber2015}%
  \BibitemOpen
  \bibfield  {author} {\bibinfo {author} {\bibfnamefont {M.}~\bibnamefont
  {Huber}}, \bibinfo {author} {\bibfnamefont {M.}~\bibnamefont
  {Perarnau-Llobet}}, \bibinfo {author} {\bibfnamefont {K.~V.}\ \bibnamefont
  {Hovhannisyan}}, \bibinfo {author} {\bibfnamefont {P.}~\bibnamefont
  {Skrzypczyk}}, \bibinfo {author} {\bibfnamefont {C.}~\bibnamefont {Klöckl}},
  \bibinfo {author} {\bibfnamefont {N.}~\bibnamefont {Brunner}}, \ and\
  \bibinfo {author} {\bibfnamefont {A.}~\bibnamefont {Ac{\'{\i}}n}},\ }\href
  {\doibase 10.1088/1367-2630/17/6/065008} {\bibfield  {journal} {\bibinfo
  {journal} {New Journal of Physics}\ }\textbf {\bibinfo {volume} {17}},\
  \bibinfo {pages} {065008} (\bibinfo {year} {2015})}\BibitemShut {NoStop}%
\bibitem [{\citenamefont {Lostaglio}\ \emph {et~al.}(2015)\citenamefont
  {Lostaglio}, \citenamefont {Jennings},\ and\ \citenamefont
  {Rudolph}}]{lostaglio2015}%
  \BibitemOpen
  \bibfield  {author} {\bibinfo {author} {\bibfnamefont {M.}~\bibnamefont
  {Lostaglio}}, \bibinfo {author} {\bibfnamefont {D.}~\bibnamefont {Jennings}},
  \ and\ \bibinfo {author} {\bibfnamefont {T.}~\bibnamefont {Rudolph}},\ }\href
  {\doibase 10.1038/ncomms7383} {\bibfield  {journal} {\bibinfo  {journal}
  {Nature Communications}\ }\textbf {\bibinfo {volume} {6}},\ \bibinfo {pages}
  {6383} (\bibinfo {year} {2015})}\BibitemShut {NoStop}%
\bibitem [{\citenamefont {Hewgill}\ \emph {et~al.}(2018)\citenamefont
  {Hewgill}, \citenamefont {Ferraro},\ and\ \citenamefont
  {De~Chiara}}]{hewgill2018}%
  \BibitemOpen
  \bibfield  {author} {\bibinfo {author} {\bibfnamefont {A.}~\bibnamefont
  {Hewgill}}, \bibinfo {author} {\bibfnamefont {A.}~\bibnamefont {Ferraro}}, \
  and\ \bibinfo {author} {\bibfnamefont {G.}~\bibnamefont {De~Chiara}},\ }\href
  {\doibase 10.1103/PhysRevA.98.042102} {\bibfield  {journal} {\bibinfo
  {journal} {Phys. Rev. A}\ }\textbf {\bibinfo {volume} {98}},\ \bibinfo
  {pages} {042102} (\bibinfo {year} {2018})}\BibitemShut {NoStop}%
\bibitem [{\citenamefont {Francica}\ \emph {et~al.}(2020)\citenamefont
  {Francica}, \citenamefont {Binder}, \citenamefont {Guarnieri}, \citenamefont
  {Mitchison}, \citenamefont {Goold},\ and\ \citenamefont
  {Plastina}}]{francica2020}%
  \BibitemOpen
  \bibfield  {author} {\bibinfo {author} {\bibfnamefont {G.}~\bibnamefont
  {Francica}}, \bibinfo {author} {\bibfnamefont {F.~C.}\ \bibnamefont
  {Binder}}, \bibinfo {author} {\bibfnamefont {G.}~\bibnamefont {Guarnieri}},
  \bibinfo {author} {\bibfnamefont {M.~T.}\ \bibnamefont {Mitchison}}, \bibinfo
  {author} {\bibfnamefont {J.}~\bibnamefont {Goold}}, \ and\ \bibinfo {author}
  {\bibfnamefont {F.}~\bibnamefont {Plastina}},\ }\href {\doibase
  10.1103/PhysRevLett.125.180603} {\bibfield  {journal} {\bibinfo  {journal}
  {Phys. Rev. Lett.}\ }\textbf {\bibinfo {volume} {125}},\ \bibinfo {pages}
  {180603} (\bibinfo {year} {2020})}\BibitemShut {NoStop}%
\bibitem [{\citenamefont {Salvia}\ and\ \citenamefont
  {Giovannetti}(2022)}]{salvia2022}%
  \BibitemOpen
  \bibfield  {author} {\bibinfo {author} {\bibfnamefont {R.}~\bibnamefont
  {Salvia}}\ and\ \bibinfo {author} {\bibfnamefont {V.}~\bibnamefont
  {Giovannetti}},\ }\href {\doibase 10.1103/PhysRevA.105.012414} {\bibfield
  {journal} {\bibinfo  {journal} {Phys. Rev. A}\ }\textbf {\bibinfo {volume}
  {105}},\ \bibinfo {pages} {012414} (\bibinfo {year} {2022})}\BibitemShut
  {NoStop}%
\bibitem [{\citenamefont {Geva}\ and\ \citenamefont
  {Kosloff}(1992)}]{geva1992}%
  \BibitemOpen
  \bibfield  {author} {\bibinfo {author} {\bibfnamefont {E.}~\bibnamefont
  {Geva}}\ and\ \bibinfo {author} {\bibfnamefont {R.}~\bibnamefont {Kosloff}},\
  }\href {\doibase 10.1063/1.463909} {\bibfield  {journal} {\bibinfo  {journal}
  {J. Chem. Phys.}\ }\textbf {\bibinfo {volume} {97}},\ \bibinfo {pages} {4398}
  (\bibinfo {year} {1992})}\BibitemShut {NoStop}%
\bibitem [{\citenamefont {Feldmann}\ and\ \citenamefont
  {Kosloff}(2000)}]{feldmann2000}%
  \BibitemOpen
  \bibfield  {author} {\bibinfo {author} {\bibfnamefont {T.}~\bibnamefont
  {Feldmann}}\ and\ \bibinfo {author} {\bibfnamefont {R.}~\bibnamefont
  {Kosloff}},\ }\href {\doibase 10.1103/PhysRevE.61.4774} {\bibfield  {journal}
  {\bibinfo  {journal} {Phys. Rev. E}\ }\textbf {\bibinfo {volume} {61}},\
  \bibinfo {pages} {4774} (\bibinfo {year} {2000})}\BibitemShut {NoStop}%
\bibitem [{\citenamefont {Niedenzu}\ \emph {et~al.}(2018)\citenamefont
  {Niedenzu}, \citenamefont {Mukherjee}, \citenamefont {Ghosh}, \citenamefont
  {Kofman},\ and\ \citenamefont {Kurizki}}]{niedenzu2018}%
  \BibitemOpen
  \bibfield  {author} {\bibinfo {author} {\bibfnamefont {W.}~\bibnamefont
  {Niedenzu}}, \bibinfo {author} {\bibfnamefont {V.}~\bibnamefont {Mukherjee}},
  \bibinfo {author} {\bibfnamefont {A.}~\bibnamefont {Ghosh}}, \bibinfo
  {author} {\bibfnamefont {A.~G.}\ \bibnamefont {Kofman}}, \ and\ \bibinfo
  {author} {\bibfnamefont {G.}~\bibnamefont {Kurizki}},\ }\href {\doibase
  10.1038/s41467-017-01991-6} {\bibfield  {journal} {\bibinfo  {journal}
  {Nature Communications}\ }\textbf {\bibinfo {volume} {9}},\ \bibinfo {pages}
  {165} (\bibinfo {year} {2018})}\BibitemShut {NoStop}%
\bibitem [{\citenamefont {Xu}\ \emph {et~al.}(2018)\citenamefont {Xu},
  \citenamefont {Chen},\ and\ \citenamefont {Liu}}]{xu2018}%
  \BibitemOpen
  \bibfield  {author} {\bibinfo {author} {\bibfnamefont {Y.~Y.}\ \bibnamefont
  {Xu}}, \bibinfo {author} {\bibfnamefont {B.}~\bibnamefont {Chen}}, \ and\
  \bibinfo {author} {\bibfnamefont {J.}~\bibnamefont {Liu}},\ }\href {\doibase
  10.1103/PhysRevE.97.022130} {\bibfield  {journal} {\bibinfo  {journal} {Phys.
  Rev. E}\ }\textbf {\bibinfo {volume} {97}},\ \bibinfo {pages} {022130}
  (\bibinfo {year} {2018})}\BibitemShut {NoStop}%
\bibitem [{\citenamefont {Abah}\ \emph {et~al.}(2012)\citenamefont {Abah},
  \citenamefont {Ro\ss{}nagel}, \citenamefont {Jacob}, \citenamefont {Deffner},
  \citenamefont {Schmidt-Kaler}, \citenamefont {Singer},\ and\ \citenamefont
  {Lutz}}]{abah2012}%
  \BibitemOpen
  \bibfield  {author} {\bibinfo {author} {\bibfnamefont {O.}~\bibnamefont
  {Abah}}, \bibinfo {author} {\bibfnamefont {J.}~\bibnamefont {Ro\ss{}nagel}},
  \bibinfo {author} {\bibfnamefont {G.}~\bibnamefont {Jacob}}, \bibinfo
  {author} {\bibfnamefont {S.}~\bibnamefont {Deffner}}, \bibinfo {author}
  {\bibfnamefont {F.}~\bibnamefont {Schmidt-Kaler}}, \bibinfo {author}
  {\bibfnamefont {K.}~\bibnamefont {Singer}}, \ and\ \bibinfo {author}
  {\bibfnamefont {E.}~\bibnamefont {Lutz}},\ }\href {\doibase
  10.1103/PhysRevLett.109.203006} {\bibfield  {journal} {\bibinfo  {journal}
  {Phys. Rev. Lett.}\ }\textbf {\bibinfo {volume} {109}},\ \bibinfo {pages}
  {203006} (\bibinfo {year} {2012})}\BibitemShut {NoStop}%
\bibitem [{\citenamefont {Roßnagel}\ \emph {et~al.}(2016)\citenamefont
  {Roßnagel}, \citenamefont {Dawkins}, \citenamefont {Tolazzi}, \citenamefont
  {Abah}, \citenamefont {Lutz}, \citenamefont {Schmidt-Kaler},\ and\
  \citenamefont {Singer}}]{Johannes2016}%
  \BibitemOpen
  \bibfield  {author} {\bibinfo {author} {\bibfnamefont {J.}~\bibnamefont
  {Roßnagel}}, \bibinfo {author} {\bibfnamefont {S.~T.}\ \bibnamefont
  {Dawkins}}, \bibinfo {author} {\bibfnamefont {K.~N.}\ \bibnamefont
  {Tolazzi}}, \bibinfo {author} {\bibfnamefont {O.}~\bibnamefont {Abah}},
  \bibinfo {author} {\bibfnamefont {E.}~\bibnamefont {Lutz}}, \bibinfo {author}
  {\bibfnamefont {F.}~\bibnamefont {Schmidt-Kaler}}, \ and\ \bibinfo {author}
  {\bibfnamefont {K.}~\bibnamefont {Singer}},\ }\href {\doibase
  10.1126/science.aad6320} {\bibfield  {journal} {\bibinfo  {journal}
  {Science}\ }\textbf {\bibinfo {volume} {352}},\ \bibinfo {pages} {325}
  (\bibinfo {year} {2016})}\BibitemShut {NoStop}%
\bibitem [{\citenamefont {Peterson}\ \emph {et~al.}(2019)\citenamefont
  {Peterson}, \citenamefont {Batalh\~ao}, \citenamefont {Herrera},
  \citenamefont {Souza}, \citenamefont {Sarthour}, \citenamefont {Oliveira},\
  and\ \citenamefont {Serra}}]{peterson2019}%
  \BibitemOpen
  \bibfield  {author} {\bibinfo {author} {\bibfnamefont {J.~P.~S.}\
  \bibnamefont {Peterson}}, \bibinfo {author} {\bibfnamefont {T.~B.}\
  \bibnamefont {Batalh\~ao}}, \bibinfo {author} {\bibfnamefont
  {M.}~\bibnamefont {Herrera}}, \bibinfo {author} {\bibfnamefont {A.~M.}\
  \bibnamefont {Souza}}, \bibinfo {author} {\bibfnamefont {R.~S.}\ \bibnamefont
  {Sarthour}}, \bibinfo {author} {\bibfnamefont {I.~S.}\ \bibnamefont
  {Oliveira}}, \ and\ \bibinfo {author} {\bibfnamefont {R.~M.}\ \bibnamefont
  {Serra}},\ }\href {\doibase 10.1103/PhysRevLett.123.240601} {\bibfield
  {journal} {\bibinfo  {journal} {Phys. Rev. Lett.}\ }\textbf {\bibinfo
  {volume} {123}},\ \bibinfo {pages} {240601} (\bibinfo {year}
  {2019})}\BibitemShut {NoStop}%
\bibitem [{\citenamefont {Karimi}\ and\ \citenamefont
  {Pekola}(2016)}]{karimi2016}%
  \BibitemOpen
  \bibfield  {author} {\bibinfo {author} {\bibfnamefont {B.}~\bibnamefont
  {Karimi}}\ and\ \bibinfo {author} {\bibfnamefont {J.~P.}\ \bibnamefont
  {Pekola}},\ }\href {\doibase 10.1103/PhysRevB.94.184503} {\bibfield
  {journal} {\bibinfo  {journal} {Phys. Rev. B}\ }\textbf {\bibinfo {volume}
  {94}},\ \bibinfo {pages} {184503} (\bibinfo {year} {2016})}\BibitemShut
  {NoStop}%
\bibitem [{\citenamefont {Hardal}\ \emph {et~al.}(2017)\citenamefont {Hardal},
  \citenamefont {Aslan}, \citenamefont {Wilson},\ and\ \citenamefont
  {M\"ustecapl\ifmmode \imath \else \i \fi{}o\ifmmode~\breve{g}\else
  \u{g}\fi{}lu}}]{hardal2017}%
  \BibitemOpen
  \bibfield  {author} {\bibinfo {author} {\bibfnamefont {A.~U.~C.}\
  \bibnamefont {Hardal}}, \bibinfo {author} {\bibfnamefont {N.}~\bibnamefont
  {Aslan}}, \bibinfo {author} {\bibfnamefont {C.~M.}\ \bibnamefont {Wilson}}, \
  and\ \bibinfo {author} {\bibfnamefont {O.~E.}\ \bibnamefont
  {M\"ustecapl\ifmmode \imath \else \i \fi{}o\ifmmode~\breve{g}\else
  \u{g}\fi{}lu}},\ }\href {\doibase 10.1103/PhysRevE.96.062120} {\bibfield
  {journal} {\bibinfo  {journal} {Phys. Rev. E}\ }\textbf {\bibinfo {volume}
  {96}},\ \bibinfo {pages} {062120} (\bibinfo {year} {2017})}\BibitemShut
  {NoStop}%
\bibitem [{\citenamefont {Manikandan}\ \emph {et~al.}(2019)\citenamefont
  {Manikandan}, \citenamefont {Giazotto},\ and\ \citenamefont
  {Jordan}}]{manikandan2019}%
  \BibitemOpen
  \bibfield  {author} {\bibinfo {author} {\bibfnamefont {S.~K.}\ \bibnamefont
  {Manikandan}}, \bibinfo {author} {\bibfnamefont {F.}~\bibnamefont
  {Giazotto}}, \ and\ \bibinfo {author} {\bibfnamefont {A.~N.}\ \bibnamefont
  {Jordan}},\ }\href {\doibase 10.1103/PhysRevApplied.11.054034} {\bibfield
  {journal} {\bibinfo  {journal} {Phys. Rev. Applied}\ }\textbf {\bibinfo
  {volume} {11}},\ \bibinfo {pages} {054034} (\bibinfo {year}
  {2019})}\BibitemShut {NoStop}%
\bibitem [{\citenamefont {Giazotto}\ \emph {et~al.}(2006)\citenamefont
  {Giazotto}, \citenamefont {Heikkil\"a}, \citenamefont {Luukanen},
  \citenamefont {Savin},\ and\ \citenamefont {Pekola}}]{giazotto2006}%
  \BibitemOpen
  \bibfield  {author} {\bibinfo {author} {\bibfnamefont {F.}~\bibnamefont
  {Giazotto}}, \bibinfo {author} {\bibfnamefont {T.~T.}\ \bibnamefont
  {Heikkil\"a}}, \bibinfo {author} {\bibfnamefont {A.}~\bibnamefont
  {Luukanen}}, \bibinfo {author} {\bibfnamefont {A.~M.}\ \bibnamefont {Savin}},
  \ and\ \bibinfo {author} {\bibfnamefont {J.~P.}\ \bibnamefont {Pekola}},\
  }\href {\doibase 10.1103/RevModPhys.78.217} {\bibfield  {journal} {\bibinfo
  {journal} {Rev. Mod. Phys.}\ }\textbf {\bibinfo {volume} {78}},\ \bibinfo
  {pages} {217} (\bibinfo {year} {2006})}\BibitemShut {NoStop}%
\bibitem [{\citenamefont {Huang}\ \emph {et~al.}(2024)\citenamefont {Huang},
  \citenamefont {Xi}, \citenamefont {Long}, \citenamefont {Liu}, \citenamefont
  {Fan}, \citenamefont {Wang}, \citenamefont {Zheng}, \citenamefont {Feng},
  \citenamefont {Nie},\ and\ \citenamefont {Lu}}]{huang2024}%
  \BibitemOpen
  \bibfield  {author} {\bibinfo {author} {\bibfnamefont {K.}~\bibnamefont
  {Huang}}, \bibinfo {author} {\bibfnamefont {C.}~\bibnamefont {Xi}}, \bibinfo
  {author} {\bibfnamefont {X.}~\bibnamefont {Long}}, \bibinfo {author}
  {\bibfnamefont {H.}~\bibnamefont {Liu}}, \bibinfo {author} {\bibfnamefont
  {Y.-a.}\ \bibnamefont {Fan}}, \bibinfo {author} {\bibfnamefont
  {X.}~\bibnamefont {Wang}}, \bibinfo {author} {\bibfnamefont {Y.}~\bibnamefont
  {Zheng}}, \bibinfo {author} {\bibfnamefont {Y.}~\bibnamefont {Feng}},
  \bibinfo {author} {\bibfnamefont {X.}~\bibnamefont {Nie}}, \ and\ \bibinfo
  {author} {\bibfnamefont {D.}~\bibnamefont {Lu}},\ }\href {\doibase
  10.1103/PhysRevLett.132.210403} {\bibfield  {journal} {\bibinfo  {journal}
  {Phys. Rev. Lett.}\ }\textbf {\bibinfo {volume} {132}},\ \bibinfo {pages}
  {210403} (\bibinfo {year} {2024})}\BibitemShut {NoStop}%
\bibitem [{\citenamefont {Maslennikov}\ \emph {et~al.}(2019)\citenamefont
  {Maslennikov}, \citenamefont {Ding}, \citenamefont {Habl{\"u}tzel},
  \citenamefont {Gan}, \citenamefont {Roulet}, \citenamefont {Nimmrichter},
  \citenamefont {Dai}, \citenamefont {Scarani},\ and\ \citenamefont
  {Matsukevich}}]{maslennikov2019}%
  \BibitemOpen
  \bibfield  {author} {\bibinfo {author} {\bibfnamefont {G.}~\bibnamefont
  {Maslennikov}}, \bibinfo {author} {\bibfnamefont {S.}~\bibnamefont {Ding}},
  \bibinfo {author} {\bibfnamefont {R.}~\bibnamefont {Habl{\"u}tzel}}, \bibinfo
  {author} {\bibfnamefont {J.}~\bibnamefont {Gan}}, \bibinfo {author}
  {\bibfnamefont {A.}~\bibnamefont {Roulet}}, \bibinfo {author} {\bibfnamefont
  {S.}~\bibnamefont {Nimmrichter}}, \bibinfo {author} {\bibfnamefont
  {J.}~\bibnamefont {Dai}}, \bibinfo {author} {\bibfnamefont {V.}~\bibnamefont
  {Scarani}}, \ and\ \bibinfo {author} {\bibfnamefont {D.}~\bibnamefont
  {Matsukevich}},\ }\href {\doibase 10.1038/s41467-018-08090-0} {\bibfield
  {journal} {\bibinfo  {journal} {Nature Communications}\ }\textbf {\bibinfo
  {volume} {10}},\ \bibinfo {pages} {202} (\bibinfo {year} {2019})}\BibitemShut
  {NoStop}%
\bibitem [{\citenamefont {Chen}\ and\ \citenamefont {Li}(2012)}]{chen2012}%
  \BibitemOpen
  \bibfield  {author} {\bibinfo {author} {\bibfnamefont {Y.-X.}\ \bibnamefont
  {Chen}}\ and\ \bibinfo {author} {\bibfnamefont {S.-W.}\ \bibnamefont {Li}},\
  }\href {\doibase 10.1209/0295-5075/97/40003} {\bibfield  {journal} {\bibinfo
  {journal} {{EPL} (Europhysics Letters)}\ }\textbf {\bibinfo {volume} {97}},\
  \bibinfo {pages} {40003} (\bibinfo {year} {2012})}\BibitemShut {NoStop}%
\bibitem [{\citenamefont {Venturelli}\ \emph {et~al.}(2013)\citenamefont
  {Venturelli}, \citenamefont {Fazio},\ and\ \citenamefont
  {Giovannetti}}]{venturelli2013}%
  \BibitemOpen
  \bibfield  {author} {\bibinfo {author} {\bibfnamefont {D.}~\bibnamefont
  {Venturelli}}, \bibinfo {author} {\bibfnamefont {R.}~\bibnamefont {Fazio}}, \
  and\ \bibinfo {author} {\bibfnamefont {V.}~\bibnamefont {Giovannetti}},\
  }\href {\doibase 10.1103/PhysRevLett.110.256801} {\bibfield  {journal}
  {\bibinfo  {journal} {Phys. Rev. Lett.}\ }\textbf {\bibinfo {volume} {110}},\
  \bibinfo {pages} {256801} (\bibinfo {year} {2013})}\BibitemShut {NoStop}%
\bibitem [{\citenamefont {Hofer}\ \emph {et~al.}(2016)\citenamefont {Hofer},
  \citenamefont {Perarnau-Llobet}, \citenamefont {Brask}, \citenamefont
  {Silva}, \citenamefont {Huber},\ and\ \citenamefont {Brunner}}]{hofer2016}%
  \BibitemOpen
  \bibfield  {author} {\bibinfo {author} {\bibfnamefont {P.~P.}\ \bibnamefont
  {Hofer}}, \bibinfo {author} {\bibfnamefont {M.}~\bibnamefont
  {Perarnau-Llobet}}, \bibinfo {author} {\bibfnamefont {J.~B.}\ \bibnamefont
  {Brask}}, \bibinfo {author} {\bibfnamefont {R.}~\bibnamefont {Silva}},
  \bibinfo {author} {\bibfnamefont {M.}~\bibnamefont {Huber}}, \ and\ \bibinfo
  {author} {\bibfnamefont {N.}~\bibnamefont {Brunner}},\ }\href {\doibase
  10.1103/PhysRevB.94.235420} {\bibfield  {journal} {\bibinfo  {journal} {Phys.
  Rev. B}\ }\textbf {\bibinfo {volume} {94}},\ \bibinfo {pages} {235420}
  (\bibinfo {year} {2016})}\BibitemShut {NoStop}%
\bibitem [{\citenamefont {Mitchison}\ \emph {et~al.}(2016)\citenamefont
  {Mitchison}, \citenamefont {Huber}, \citenamefont {Prior}, \citenamefont
  {Woods},\ and\ \citenamefont {Plenio}}]{mitchison2016}%
  \BibitemOpen
  \bibfield  {author} {\bibinfo {author} {\bibfnamefont {M.~T.}\ \bibnamefont
  {Mitchison}}, \bibinfo {author} {\bibfnamefont {M.}~\bibnamefont {Huber}},
  \bibinfo {author} {\bibfnamefont {J.}~\bibnamefont {Prior}}, \bibinfo
  {author} {\bibfnamefont {M.~P.}\ \bibnamefont {Woods}}, \ and\ \bibinfo
  {author} {\bibfnamefont {M.~B.}\ \bibnamefont {Plenio}},\ }\href {\doibase
  10.1088/2058-9565/1/1/015001} {\bibfield  {journal} {\bibinfo  {journal}
  {Quantum Science and Technology}\ }\textbf {\bibinfo {volume} {1}},\ \bibinfo
  {pages} {015001} (\bibinfo {year} {2016})}\BibitemShut {NoStop}%
\bibitem [{\citenamefont {Ghanavati}\ and\ \citenamefont
  {Movahhedian}(2014)}]{Ghanavati2014}%
  \BibitemOpen
  \bibfield  {author} {\bibinfo {author} {\bibfnamefont {M.}~\bibnamefont
  {Ghanavati}}\ and\ \bibinfo {author} {\bibfnamefont {H.}~\bibnamefont
  {Movahhedian}},\ }\href {\doibase 10.1142/S021974991450018X} {\bibfield
  {journal} {\bibinfo  {journal} {Int. J. Quant. Info.}\ }\textbf {\bibinfo
  {volume} {12}},\ \bibinfo {pages} {1450018} (\bibinfo {year}
  {2014})}\BibitemShut {NoStop}%
\bibitem [{\citenamefont {Arısoy}\ and\ \citenamefont {M{\"u}stecaplıo{\u
  g}lu}(2021)}]{Arisoy2021}%
  \BibitemOpen
  \bibfield  {author} {\bibinfo {author} {\bibfnamefont {O.}~\bibnamefont
  {Arısoy}}\ and\ \bibinfo {author} {\bibfnamefont {{\"O}.~E.}\ \bibnamefont
  {M{\"u}stecaplıo{\u g}lu}},\ }\href {\doibase 10.1038/s41598-021-92258-0}
  {\bibfield  {journal} {\bibinfo  {journal} {Scientific Reports}\ }\textbf
  {\bibinfo {volume} {11}},\ \bibinfo {pages} {12981} (\bibinfo {year}
  {2021})}\BibitemShut {NoStop}%
\bibitem [{\citenamefont {Landi}\ \emph {et~al.}(2022)\citenamefont {Landi},
  \citenamefont {Poletti},\ and\ \citenamefont {Schaller}}]{Landi2022}%
  \BibitemOpen
  \bibfield  {author} {\bibinfo {author} {\bibfnamefont {G.~T.}\ \bibnamefont
  {Landi}}, \bibinfo {author} {\bibfnamefont {D.}~\bibnamefont {Poletti}}, \
  and\ \bibinfo {author} {\bibfnamefont {G.}~\bibnamefont {Schaller}},\ }\href
  {\doibase 10.1103/RevModPhys.94.045006} {\bibfield  {journal} {\bibinfo
  {journal} {Rev. Mod. Phys.}\ }\textbf {\bibinfo {volume} {94}},\ \bibinfo
  {pages} {045006} (\bibinfo {year} {2022})}\BibitemShut {NoStop}%
\bibitem [{\citenamefont {Wichterich}\ \emph {et~al.}(2007)\citenamefont
  {Wichterich}, \citenamefont {Henrich}, \citenamefont {Breuer}, \citenamefont
  {Gemmer},\ and\ \citenamefont {Michel}}]{Wichterich2007}%
  \BibitemOpen
  \bibfield  {author} {\bibinfo {author} {\bibfnamefont {H.}~\bibnamefont
  {Wichterich}}, \bibinfo {author} {\bibfnamefont {M.~J.}\ \bibnamefont
  {Henrich}}, \bibinfo {author} {\bibfnamefont {H.-P.}\ \bibnamefont {Breuer}},
  \bibinfo {author} {\bibfnamefont {J.}~\bibnamefont {Gemmer}}, \ and\ \bibinfo
  {author} {\bibfnamefont {M.}~\bibnamefont {Michel}},\ }\href {\doibase
  10.1103/PhysRevE.76.031115} {\bibfield  {journal} {\bibinfo  {journal} {Phys.
  Rev. E}\ }\textbf {\bibinfo {volume} {76}},\ \bibinfo {pages} {031115}
  (\bibinfo {year} {2007})}\BibitemShut {NoStop}%
\bibitem [{\citenamefont {Barra}(2015)}]{Barra2015}%
  \BibitemOpen
  \bibfield  {author} {\bibinfo {author} {\bibfnamefont {F.}~\bibnamefont
  {Barra}},\ }\href {\doibase 10.1038/srep14873} {\bibfield  {journal}
  {\bibinfo  {journal} {Scientific Reports}\ }\textbf {\bibinfo {volume} {5}},\
  \bibinfo {pages} {14873} (\bibinfo {year} {2015})}\BibitemShut {NoStop}%
\bibitem [{\citenamefont {Strasberg}\ \emph {et~al.}(2017)\citenamefont
  {Strasberg}, \citenamefont {Schaller}, \citenamefont {Brandes},\ and\
  \citenamefont {Esposito}}]{Strasberg2017}%
  \BibitemOpen
  \bibfield  {author} {\bibinfo {author} {\bibfnamefont {P.}~\bibnamefont
  {Strasberg}}, \bibinfo {author} {\bibfnamefont {G.}~\bibnamefont {Schaller}},
  \bibinfo {author} {\bibfnamefont {T.}~\bibnamefont {Brandes}}, \ and\
  \bibinfo {author} {\bibfnamefont {M.}~\bibnamefont {Esposito}},\ }\href
  {\doibase 10.1103/PhysRevX.7.021003} {\bibfield  {journal} {\bibinfo
  {journal} {Phys. Rev. X}\ }\textbf {\bibinfo {volume} {7}},\ \bibinfo {pages}
  {021003} (\bibinfo {year} {2017})}\BibitemShut {NoStop}%
\bibitem [{\citenamefont {Stockburger}\ and\ \citenamefont
  {Motz}(2017)}]{Motz2017}%
  \BibitemOpen
  \bibfield  {author} {\bibinfo {author} {\bibfnamefont {J.~T.}\ \bibnamefont
  {Stockburger}}\ and\ \bibinfo {author} {\bibfnamefont {T.}~\bibnamefont
  {Motz}},\ }\href {\doibase https://doi.org/10.1002/prop.201600067} {\bibfield
   {journal} {\bibinfo  {journal} {Fort. der Phys.}\ }\textbf {\bibinfo
  {volume} {65}},\ \bibinfo {pages} {1600067} (\bibinfo {year}
  {2017})}\BibitemShut {NoStop}%
\bibitem [{\citenamefont {Korepin}\ \emph {et~al.}(1993)\citenamefont
  {Korepin}, \citenamefont {Bogoliubov},\ and\ \citenamefont
  {Izergin}}]{korepin_bogoliubov_izergin_1993}%
  \BibitemOpen
  \bibfield  {author} {\bibinfo {author} {\bibfnamefont {V.~E.}\ \bibnamefont
  {Korepin}}, \bibinfo {author} {\bibfnamefont {N.~M.}\ \bibnamefont
  {Bogoliubov}}, \ and\ \bibinfo {author} {\bibfnamefont {A.~G.}\ \bibnamefont
  {Izergin}},\ }\href {\doibase 10.1017/CBO9780511628832} {\emph {\bibinfo
  {title} {Quantum Inverse Scattering Method and Correlation Functions}}},\
  Cambridge Monographs on Mathematical Physics\ (\bibinfo  {publisher}
  {Cambridge University Press},\ \bibinfo {year} {1993})\BibitemShut {NoStop}%
\bibitem [{\citenamefont {Giamarchi}(2004)}]{Giamarchi2004}%
  \BibitemOpen
  \bibfield  {author} {\bibinfo {author} {\bibfnamefont {T.}~\bibnamefont
  {Giamarchi}},\ }\href {\doibase 10.1093/acprof:oso/9780198525004.001.0001}
  {\emph {\bibinfo {title} {{Quantum physics in one dimension}}}},\
  International series of monographs on physics\ (\bibinfo  {publisher}
  {Clarendon Press},\ \bibinfo {address} {Oxford},\ \bibinfo {year}
  {2004})\BibitemShut {NoStop}%
\bibitem [{\citenamefont {Franchini}(2017)}]{Franchini2017}%
  \BibitemOpen
  \bibfield  {author} {\bibinfo {author} {\bibfnamefont {F.}~\bibnamefont
  {Franchini}},\ }\href {\doibase 10.1007/978-3-319-48487-7} {\emph {\bibinfo
  {title} {{An Introduction to Integrable Techniques for One-Dimensional
  Quantum Systems}}}},\ Lecture Notes in Physics\ (\bibinfo  {publisher}
  {Springer Cham},\ \bibinfo {address} {Switzerland},\ \bibinfo {year}
  {2017})\BibitemShut {NoStop}%
\bibitem [{\citenamefont {Fisher}(1964)}]{fisher1964}%
  \BibitemOpen
  \bibfield  {author} {\bibinfo {author} {\bibfnamefont {M.~E.}\ \bibnamefont
  {Fisher}},\ }\href {\doibase 10.1119/1.1970340} {\bibfield  {journal}
  {\bibinfo  {journal} {American Journal of Physics}\ }\textbf {\bibinfo
  {volume} {32}},\ \bibinfo {pages} {343} (\bibinfo {year} {1964})}\BibitemShut
  {NoStop}%
\bibitem [{\citenamefont {Richter}\ and\ \citenamefont
  {Voigt}(1994)}]{Richter_1994}%
  \BibitemOpen
  \bibfield  {author} {\bibinfo {author} {\bibfnamefont {J.}~\bibnamefont
  {Richter}}\ and\ \bibinfo {author} {\bibfnamefont {A.}~\bibnamefont
  {Voigt}},\ }\href {\doibase 10.1088/0305-4470/27/4/010} {\bibfield  {journal}
  {\bibinfo  {journal} {Journal of Physics A: Mathematical and General}\
  }\textbf {\bibinfo {volume} {27}},\ \bibinfo {pages} {1139} (\bibinfo {year}
  {1994})}\BibitemShut {NoStop}%
\bibitem [{\citenamefont {Hutton}\ and\ \citenamefont
  {Bose}(2004)}]{Hutton2004}%
  \BibitemOpen
  \bibfield  {author} {\bibinfo {author} {\bibfnamefont {A.}~\bibnamefont
  {Hutton}}\ and\ \bibinfo {author} {\bibfnamefont {S.}~\bibnamefont {Bose}},\
  }\href {\doibase 10.1103/PhysRevA.69.042312} {\bibfield  {journal} {\bibinfo
  {journal} {Phys. Rev. A}\ }\textbf {\bibinfo {volume} {69}},\ \bibinfo
  {pages} {042312} (\bibinfo {year} {2004})}\BibitemShut {NoStop}%
\bibitem [{\citenamefont {Anzà}\ \emph {et~al.}(2010)\citenamefont {Anzà},
  \citenamefont {Militello},\ and\ \citenamefont {Messina}}]{Anza_2010}%
  \BibitemOpen
  \bibfield  {author} {\bibinfo {author} {\bibfnamefont {F.}~\bibnamefont
  {Anzà}}, \bibinfo {author} {\bibfnamefont {B.}~\bibnamefont {Militello}}, \
  and\ \bibinfo {author} {\bibfnamefont {A.}~\bibnamefont {Messina}},\ }\href
  {\doibase 10.1088/0953-4075/43/20/205501} {\bibfield  {journal} {\bibinfo
  {journal} {Journal of Physics B: Atomic, Molecular and Optical Physics}\
  }\textbf {\bibinfo {volume} {43}},\ \bibinfo {pages} {205501} (\bibinfo
  {year} {2010})}\BibitemShut {NoStop}%
\bibitem [{\citenamefont {Militello}\ and\ \citenamefont
  {Messina}(2011)}]{Militello2011}%
  \BibitemOpen
  \bibfield  {author} {\bibinfo {author} {\bibfnamefont {B.}~\bibnamefont
  {Militello}}\ and\ \bibinfo {author} {\bibfnamefont {A.}~\bibnamefont
  {Messina}},\ }\href {\doibase 10.1103/PhysRevA.83.042305} {\bibfield
  {journal} {\bibinfo  {journal} {Phys. Rev. A}\ }\textbf {\bibinfo {volume}
  {83}},\ \bibinfo {pages} {042305} (\bibinfo {year} {2011})}\BibitemShut
  {NoStop}%
\bibitem [{\citenamefont {Haddadi}\ \emph {et~al.}(2021)\citenamefont
  {Haddadi}, \citenamefont {Ghominejad}, \citenamefont {Akhound},\ and\
  \citenamefont {Pourkarimi}}]{haddadi2021}%
  \BibitemOpen
  \bibfield  {author} {\bibinfo {author} {\bibfnamefont {S.}~\bibnamefont
  {Haddadi}}, \bibinfo {author} {\bibfnamefont {M.}~\bibnamefont {Ghominejad}},
  \bibinfo {author} {\bibfnamefont {A.}~\bibnamefont {Akhound}}, \ and\
  \bibinfo {author} {\bibfnamefont {M.~R.}\ \bibnamefont {Pourkarimi}},\ }\href
  {\doibase 10.1038/s41598-021-02045-0} {\bibfield  {journal} {\bibinfo
  {journal} {Scientific Reports}\ }\textbf {\bibinfo {volume} {11}},\ \bibinfo
  {pages} {22691} (\bibinfo {year} {2021})}\BibitemShut {NoStop}%
\bibitem [{\citenamefont {Karl\'{o}v\'{a}'}\ and\ \citenamefont
  {Stre\u{c}ka}(2023)}]{Karlova2023}%
  \BibitemOpen
  \bibfield  {author} {\bibinfo {author} {\bibfnamefont {K.}~\bibnamefont
  {Karl\'{o}v\'{a}'}}\ and\ \bibinfo {author} {\bibfnamefont {J.}~\bibnamefont
  {Stre\u{c}ka}},\ }\href {https://www.mdpi.com/1420-3049/28/10/4037}
  {\bibfield  {journal} {\bibinfo  {journal} {Molecules}\ }\textbf {\bibinfo
  {volume} {28}} (\bibinfo {year} {2023})}\BibitemShut {NoStop}%
\bibitem [{\citenamefont {Dicke}(1954)}]{Dicke1954}%
  \BibitemOpen
  \bibfield  {author} {\bibinfo {author} {\bibfnamefont {R.~H.}\ \bibnamefont
  {Dicke}},\ }\href {\doibase 10.1103/PhysRev.93.99} {\bibfield  {journal}
  {\bibinfo  {journal} {Phys. Rev.}\ }\textbf {\bibinfo {volume} {93}},\
  \bibinfo {pages} {99} (\bibinfo {year} {1954})}\BibitemShut {NoStop}%
\bibitem [{\citenamefont {Breuer}\ and\ \citenamefont
  {Petruccione}(2002)}]{breuer2002}%
  \BibitemOpen
  \bibfield  {author} {\bibinfo {author} {\bibfnamefont {H.~P.}\ \bibnamefont
  {Breuer}}\ and\ \bibinfo {author} {\bibfnamefont {F.}~\bibnamefont
  {Petruccione}},\ }\href@noop {} {\emph {\bibinfo {title} {The Theory of Open
  Quantum Systems}}}\ (\bibinfo  {publisher} {Oxford University Press,
  Oxford},\ \bibinfo {year} {2002})\BibitemShut {NoStop}%
\bibitem [{\citenamefont {Ángel Rivas}\ \emph {et~al.}(2010)\citenamefont
  {Ángel Rivas}, \citenamefont {Plato}, \citenamefont {Huelga},\ and\
  \citenamefont {Plenio}}]{Rivas2010}%
  \BibitemOpen
  \bibfield  {author} {\bibinfo {author} {\bibnamefont {Ángel Rivas}},
  \bibinfo {author} {\bibfnamefont {A.~D.~K.}\ \bibnamefont {Plato}}, \bibinfo
  {author} {\bibfnamefont {S.~F.}\ \bibnamefont {Huelga}}, \ and\ \bibinfo
  {author} {\bibfnamefont {M.~B.}\ \bibnamefont {Plenio}},\ }\href {\doibase
  10.1088/1367-2630/12/11/113032} {\bibfield  {journal} {\bibinfo  {journal}
  {New Journal of Physics}\ }\textbf {\bibinfo {volume} {12}},\ \bibinfo
  {pages} {113032} (\bibinfo {year} {2010})}\BibitemShut {NoStop}%
\bibitem [{\citenamefont {Potts}(2019)}]{potts2019}%
  \BibitemOpen
  \bibfield  {author} {\bibinfo {author} {\bibfnamefont {P.~P.}\ \bibnamefont
  {Potts}},\ }\href {https://arxiv.org/abs/1906.07439} {\bibfield  {journal}
  {\bibinfo  {journal} {arXiv:1906.07439}\ } (\bibinfo {year}
  {2019})}\BibitemShut {NoStop}%
\bibitem [{\citenamefont {Konar}\ \emph
  {et~al.}(2023{\natexlab{b}})\citenamefont {Konar}, \citenamefont {Ghosh},
  \citenamefont {Pal},\ and\ \citenamefont {Sen(De)}}]{Tanoy2023}%
  \BibitemOpen
  \bibfield  {author} {\bibinfo {author} {\bibfnamefont {T.~K.}\ \bibnamefont
  {Konar}}, \bibinfo {author} {\bibfnamefont {S.}~\bibnamefont {Ghosh}},
  \bibinfo {author} {\bibfnamefont {A.~K.}\ \bibnamefont {Pal}}, \ and\
  \bibinfo {author} {\bibfnamefont {A.}~\bibnamefont {Sen(De)}},\ }\href
  {\doibase 10.1103/PhysRevA.107.032602} {\bibfield  {journal} {\bibinfo
  {journal} {Phys. Rev. A}\ }\textbf {\bibinfo {volume} {107}},\ \bibinfo
  {pages} {032602} (\bibinfo {year} {2023}{\natexlab{b}})}\BibitemShut
  {NoStop}%
\bibitem [{\citenamefont {Nielsen}\ and\ \citenamefont
  {Chuang}(2010)}]{nielsen2010}%
  \BibitemOpen
  \bibfield  {author} {\bibinfo {author} {\bibfnamefont {M.~A.}\ \bibnamefont
  {Nielsen}}\ and\ \bibinfo {author} {\bibfnamefont {I.~L.}\ \bibnamefont
  {Chuang}},\ }\href@noop {} {\emph {\bibinfo {title} {Quantum Computation and
  Quantum Information}}}\ (\bibinfo  {publisher} {Cambridge University Press},\
  \bibinfo {year} {2010})\BibitemShut {NoStop}%
\bibitem [{\citenamefont {Lidar}(2020)}]{lidar2020lecture}%
  \BibitemOpen
  \bibfield  {author} {\bibinfo {author} {\bibfnamefont {D.~A.}\ \bibnamefont
  {Lidar}},\ }\href@noop {} {\enquote {\bibinfo {title} {Lecture notes on the
  theory of open quantum systems},}\ } (\bibinfo {year} {2020}),\ \Eprint
  {http://arxiv.org/abs/1902.00967} {arXiv:1902.00967 [quant-ph]} \BibitemShut
  {NoStop}%
\bibitem [{\citenamefont {Weiss}(2012)}]{wiess2012}%
  \BibitemOpen
  \bibfield  {author} {\bibinfo {author} {\bibfnamefont {U.}~\bibnamefont
  {Weiss}},\ }\href {\doibase 10.1142/8334} {\emph {\bibinfo {title} {Quantum
  Dissipative Systems}}},\ \bibinfo {edition} {4th}\ ed.\ (\bibinfo
  {publisher} {World Scientific},\ \bibinfo {year} {2012})\BibitemShut
  {NoStop}%
\bibitem [{\citenamefont {Liao}\ \emph {et~al.}(2011)\citenamefont {Liao},
  \citenamefont {Huang},\ and\ \citenamefont {Kuang}}]{Liao2011}%
  \BibitemOpen
  \bibfield  {author} {\bibinfo {author} {\bibfnamefont {J.-Q.}\ \bibnamefont
  {Liao}}, \bibinfo {author} {\bibfnamefont {J.-F.}\ \bibnamefont {Huang}}, \
  and\ \bibinfo {author} {\bibfnamefont {L.-M.}\ \bibnamefont {Kuang}},\ }\href
  {\doibase 10.1103/PhysRevA.83.052110} {\bibfield  {journal} {\bibinfo
  {journal} {Phys. Rev. A}\ }\textbf {\bibinfo {volume} {83}},\ \bibinfo
  {pages} {052110} (\bibinfo {year} {2011})}\BibitemShut {NoStop}%
\end{thebibliography}%

\end{document}